\documentclass[aip,jcp,preprint,superscriptaddress,noshowkeys]{revtex4-2}

\usepackage{media9}

\usepackage[vcentermath]{youngtab}
\usepackage{young}
\usepackage{amsmath}
\usepackage{amssymb}
\usepackage{braket}
\usepackage{bm}
\usepackage{tikz}
\usetikzlibrary{decorations.pathmorphing}

\usepackage{graphicx}
\usepackage{footnote}
\usepackage{booktabs}

\usepackage{subcaption}
\usepackage{colortbl}

\usepackage{hyperref}
\hypersetup{
	colorlinks,
	linkcolor={red!50!black},
	citecolor={red!70!black},
	urlcolor={red!80!black}
}

\usepackage{listings}
\definecolor{codegreen}{rgb}{0.58,0.4,0.2}
\definecolor{codegray}{rgb}{0.5,0.5,0.5}
\definecolor{codepurple}{rgb}{0.25,0.35,0.55}
\definecolor{codeblue}{rgb}{0.30,0.60,0.8}
\definecolor{backcolour}{rgb}{0.98,0.98,0.98}
\definecolor{mygray}{rgb}{0.5,0.5,0.5}

\definecolor{sqred}{rgb}{0.85,0.1,0.1}
\definecolor{sqgreen}{rgb}{0.25,0.65,0.15}
\definecolor{sqorange}{rgb}{0.90,0.50,0.15}
\definecolor{sqblue}{rgb}{0.10,0.3,0.60}

\lstdefinestyle{mystyle}{
	backgroundcolor=\color{backcolour},
	commentstyle=\color{codegreen},
	keywordstyle=\color{codeblue},
	numberstyle=\tiny\color{codegray},
	stringstyle=\color{codepurple},
	basicstyle=\ttfamily\footnotesize,
	breakatwhitespace=false,
	breaklines=true,
	captionpos=b,
	keepspaces=true,
	numbers=left,
	numbersep=5pt,
	numberstyle=\ttfamily\tiny\color{mygray},
	showspaces=false,
	showstringspaces=false,
	showtabs=false,
	tabsize=2
}

\newcommand{\hH}{\Hat{H}}

\newcommand{\cre}[1]{\Hat{a}^{\dagger}_{#1}}
\newcommand{\ani}[1]{\Hat{a}_{#1}}
\newcommand{\Sp}[1]{\Hat{P}^{+}_{#1}}
\newcommand{\Sm}[1]{\Hat{P}^{-}_{#1}}
\newcommand{\Spm}[1]{\Hat{P}^{\pm}_{#1}}

\newcommand{\Az}[1]{\Hat{A}^{0}_{#1}}
\newcommand{\Ap}[1]{\Hat{A}^{+}_{#1}}
\newcommand{\Am}[1]{\Hat{A}^{-}_{#1}}

\newcommand{\n}[1]{\Hat{n}_{#1}}

\newcommand{\PP}{\boldsymbol{\omega}}
\newcommand{\RG}{\bold{u}}
\newcommand{\PPE}{\Psi}

\newcommand{\ger}[1]{#1_{0}}
\newcommand{\ung}[1]{#1_{1}}

\newcommand{\SpG}[1]{\Hat{P}^{+}(#1)}
\newcommand{\SmG}[1]{\Hat{P}^{-}(#1)}

\newcommand{\fone}[2]{4^{#1}_{#2}}
\newcommand{\ftwo}[2]{\bar{4}^{#1}_{#2}}

\newcommand{\stwo}[2]{ \hat{\varphi}^+_{#1,#2} }

\newcommand{\gzer}[2]{ \textbf{g}_{#1 #2}}
\newcommand{\gone}[2]{ \textbf{g}^-_{#1 #2} }
\newcommand{\gtwo}[2]{ \textbf{g}^{--}_{#1 #2} }

\newcommand{\eg}[1]{\eta_{#1}}

\newcommand{\twamp}[1]{t_{#1}}

\newcommand{\sgn}[1]{\text{sgn}(#1)}

\newcommand{\kd}{\mathbf{d}}

\newcommand{\vac}[1]{\theta_{#1}}

\begin{document}

\title{Richardson-Gaudin states of non-zero seniority III: The Perfect-Pairing limit}
\author{Paul A. Johnson}
 \email{paul.johnson@chm.ulaval.ca}
 \affiliation{D\'{e}partement de chimie, Universit\'{e} Laval, Qu\'{e}bec, Qu\'{e}bec, G1V 0A6, Canada}

\date{\today}

\begin{abstract}
Strongly correlated electrons can be treated with a configuration interaction of Slater determinants grouped by their number of unpaired electrons with exponential cost. The first two papers in this series demonstrated that single reference methods built from Richardson-Gaudin states gave results of similar quality at polynomial cost. In this contribution, the states are simplified substantially yielding the perfect-pairing state as a reference along with its low-lying excitations. The states are much simpler, the computational cost is substantially reduced, and there is no sacrifice in numerical accuracy. Second-order Epstein-Nesbet perturbative corrections for the valence electrons are similar in quality to the complete active space self-consistent field.
\end{abstract}

\maketitle

\section{Introduction}
Weak correlation is understood in terms of Slater determinants. The Hartree-Fock (HF) Slater determinant provides a good first approximation upon which corrections from low-lying excitations may be added systematically. Strong correlation, on the other hand, is not well understood in terms of Slater determinants. One Slater determinant is a poor starting point. Many are required even for a qualitative first approximation. The complete active space self-consistent field (CASSCF),\cite{roos:1980a,siegbahn:1980,siegbahn:1981,roos:1987} the accepted approach for strong correlation, relies on the ability to identify the important Slater determinants. If the required active space is sufficiently small this approach is simple. However, as the active space grows this quickly becomes difficult for the user and unfeasible for the computer. The density matrix renormalization group (DMRG) algorithm has proven itself\cite{chan:2002,ghosh:2008,yanai:2009,wouters:2014,sun:2017,ma:2017} in larger systems even if it is less simple to understand conceptually.

Grouping Slater determinants by their number of unpaired electrons, their \emph{seniority}, is productive and leads to systematic improvement with increasing seniority.\cite{bytautas:2011,bytautas:2015, wahlen:2018} In particular, the configuration interaction (CI) of all seniority-zero Slater determinants is qualitatively correct for bond-breaking processes, being nearly parallel to full CI:\cite{weinhold:1967a,weinhold:1967b,veillard:1967,clementi:1967,cook:1975} strong correlation manifests in the seniority-zero sector.\cite{gus_sanibel} However, even seniority-zero CI scales faster than exponentially. In addition, as orbital rotations do not conserve seniority, the orbitals must be optimized for seniority-zero CI. A cheap method which reproduces seniority-zero CI quantitatively was quickly discovered and reported as the antisymmetric product of 1 reference orbital geminals (AP1roG)\cite{limacher:2013, limacher:2014a, limacher:2014b, boguslawski:2014a, boguslawski:2014b, boguslawski:2014c,tecmer:2014} or pair coupled cluster doubles (pCCD).\cite{stein:2014, henderson:2014a, henderson:2014b, boguslawski:2015, boguslawski:2016a, boguslawski:2016b, boguslawski:2017, boguslawski:2019, boguslawski:2021, marie:2021, kossoski:2021, baran:2021} While these methods are successful, it is not obvious how to add the outstanding effects from non-zero seniorities, though different avenues have been tested.\cite{cassam:2006, parkhill:2009, cassam:2010, lehtola:2016, lehtola:2018, vu:2019, kossoski:2022, kossoski:2023, cassam:2023, johnson:2025a, calero-osorio:2026a, calero-osorio:2026b}

An alternative basis for the Hilbert space is provided by the eigenvectors of the reduced Bardeen-Cooper-Schrieffer (BCS) Hamiltonian,\cite{cooper:1956, bardeen:1957a, bardeen:1957b, schrieffer_book} the so-called Richardson\cite{richardson:1963, richardson:1964, richardson:1965}-Gaudin\cite{gaudin:1976, gaudin_book} (RG) states. A single RG state is very close to seniority-zero CI, while a second-order Epstein\cite{epstein:1926}-Nesbet\cite{nesbet:1955} perturbative correction (EN2) makes the agreement quantitative.\cite{fecteau:2022, johnson:2024b} For RG states, it is a tedious but straightforward exercise to include higher seniorities. One simply needs to add the effects of the low-lying RG excitations in seniorities two and four. The first paper in this series,\cite{johnson:2025b} hereafter referred to as Part I, reported in exhaustive detail the matrix elements required between RG states of seniorities zero, two, and four. The second paper in this series,\cite{johnson:2025c} hereafter Part II, demonstrated numerically that a short CI expansion in terms of RG states was a very good approximation to the full Slater determinant CI in seniorities zero, two, and four. A much cheaper EN2 correction was computed and found to be a reasonable, but not quantitative, alternative. These approaches were expensive, but the cost was polynomial. The next step is to reduce the cost to be competitive. 

Recently, the connection between generalized valence-bond/perfect-pairing (PP) and RG states has been reported:\cite{johnson:2025d} PP is a particular limit of the RG state construction. PP is an established method,\cite{goddard:1967, hunt:1972, hay:1972, goddard:1973, goddard:1978, dykstra:1980, small:2009, lawler:2010, small:2011} though it is understood as either a multireference wavefunction written in Slater determinants,\cite{dunning:2016} or as a particular coupled cluster state.\cite{cullen:1996, cullen:1999, vanvoorhis:2000, vanvoorhis:2000b, vanvoorhis:2001, small:2012, cullen:2007} PP is a particular case of the antisymmetric product of strongly-orthogonal geminals (APSG), an ansatz first proposed by Fock,\cite{fock:1950} and studied thoroughly by others.\cite{hurley:1953, kutzelnigg:1964, surjan_book, surjan:2012} In this manuscript, the low-lying excitations of PP will be constructed in seniorities zero, two, and four to be employed in an EN2 correction. For RG states it was possible to write general expressions for most matrix elements as the RG states are structurally identical. As will be seen, the PP excited states can be broken into several different classes, so that the present contribution will focus on the couplings with the PP reference to compute an EN2 correction. One would expect that this would make a cheaper alternative to the EN2 correction built from RG states with a tradeoff in numerical accuracy. Indeed, the method is substantially cheaper, but there is no loss in the quality of the results.

This manuscript is organized as follows. Section \ref{sec:prelim} fixes the notation and summarizes the pairing Lie algebras su(2) and sp(N). Section \ref{sec:pp} presents PP as a limiting case of an RG state. The low-lying excitations are constructed in Section \ref{sec:en2}, along with their excitation energies and couplings to PP through the Coulomb Hamiltonian. The resulting EN2 correction is computed numerically in Section \ref{sec:numbers}.

\section{Preliminaries} \label{sec:prelim}
This section will summarize the algebraic structure of electron-pairs necessary to construct PP along with its weak excitations. For a more complete description see ref. \citenum{johnson:2025b}. 

A requirement of the PP construction is the separation of orbitals into $M_c$ doubly-occupied core orbitals, a valence space of $2M$ partially occupied orbitals, and $M_v$ empty virtual orbitals. That such a separation is possible is well-established, even procedural.\cite{wang:2019} Core orbitals are labelled $i,j,k,l$ while virtual orbitals are labelled $a,b,c,d$. This choice is essentially standard for HF, where there are only cores and virtuals.\cite{helgaker_book} The valence consists of $M$ sets of bonding and antibonding orbitals. Orbitals in the valence will be labelled with Greek letters: the letters $\alpha,\beta,\gamma,\delta$ will denote the sets of orbitals while the letters $\mu,\nu,\lambda,\kappa$ are used to distinguish the bonding and antibonding orbitals. The letters $p,q,r,s$ are used to denote orbitals that could be in any of the three spaces. These are the choices made for ref. \citenum{johnson:2025d}.

Operators for electron pairs are built from second-quantized operators which create/remove electrons with the usual structure
\begin{align}
	\left[ \cre{p\sigma} \ani{q\tau} \right]_+ = \kd_{pq} \kd_{\sigma\tau}.
\end{align}
As the letter $\delta$ is reserved as an index for the valence, the Kronecker delta is denoted $\kd_{pq}$. Seniority-zero pairs in each orbital $p$ require the three operators
\begin{align} \label{eq:pair_su2}
	\Sp{p} & = \cre{p\uparrow} \cre{p\downarrow} 
	&
	\Sm{p} & = \ani{p\downarrow} \ani{p\uparrow}
	&
	\n{p} & = \cre{p\uparrow} \ani{p\uparrow} + \cre{p\downarrow}\ani{p\downarrow}
\end{align}
which have the structure of su(2)
\begin{align}
	[\Sp{p},\Sm{q}] &= \kd_{pq} \left( \n{p} -1 \right) 
	&
	[\n{p},\Spm{q}] &= \pm 2 \kd_{pq} \Spm{p}.
\end{align}
$\Sp{p}$ creates a pair in the orbital $p$ while $\Sm{p}$ removes a pair from orbital $p$. The third object, $\n{p}$, is just the usual number operator.  Seniority-two singlet pairs in spatial orbitals $p$ and $q$ require the operators
\begin{subequations}
	\begin{align}
		\Ap{pq} &= \frac{1}{\sqrt{2}} \left( \cre{p\uparrow} \cre{q\downarrow}
			- \cre{p\downarrow} \cre{q\uparrow}
		\right) \\
		\Am{pq} &= \frac{1}{\sqrt{2}} \left( \ani{q\downarrow} \ani{p\uparrow}
			- \ani{q\uparrow} \ani{p\downarrow}
		\right) \\
		\Az{pq} &= \cre{p\uparrow} \ani{q\uparrow} + \cre{p\downarrow} \ani{q\downarrow}.
	\end{align}	
\end{subequations}
For any $p\neq q$, there are four operators as $\Ap{pq} = \Ap{qp}$ and $\Am{pq} = \Am{qp}$, but $\Az{pq} \neq \Az{qp}$. The adjoint of the pair creator is the annihilator $\left(\Ap{pq}\right)^{\dagger}=\Am{pq}$, while $\left(\Az{pq}\right)^{\dagger}=\Az{qp}$. In the chosen convention the two-electron state created by $\Ap{pq}$ is normalized, while the operator $\Az{pq}$ is traditionally known as a singlet excitation operator.\cite{helgaker_book} These objects have the structure of the Lie algebra sp(N), with the relevant structure
\begin{subequations}
	\begin{align}
		[ \Az{pq} , \Az{rs} ] &= \kd_{qr} \Az{ps} - \kd_{ps} \Az{rq} \\
		[ \Az{pq} , \Ap{rs} ] &= \kd_{qr} \Ap{ps} + \kd_{qs} \Ap{pr} \\
		[ \Az{pq} , \Sp{r}  ] &= \sqrt{2} \kd_{qr} \Ap{pq}.
	\end{align}
\end{subequations}
Diagonal elements are not defined, but it is easily understood that
\begin{align}
	\hat{A}^{\pm}_{pp} &= \sqrt{2} \Spm{p}
	&
	\Az{pp} &= \n{p}.
\end{align}
Further structure amongst these objects are consequences of the Pauli principle. In particular, a seniority-zero pair cannot be created more than once
\begin{align}
	\Sp{p} \Sp{p} = 0,
\end{align}
nor can a seniority-zero pair be created in partially occupied spatial orbitals
\begin{align}
	\Sp{p} \Ap{pq} = 0.
\end{align}
Seniority-two pairs can be created more than once however
\begin{subequations}
\begin{align}
	\Ap{pq} \Ap{pq} &= - \Sp{p} \Sp{q} \\
	\Ap{pq} \Ap{pr} &= - \frac{1}{\sqrt{2}} \Sp{p} \Ap{qr}.
\end{align}
\end{subequations}
Singlet-excitations with shared indices are pair excitations
\begin{subequations}
\begin{align}
	\Az{pq} \Az{pq} &= 2 \Sp{p} \Sm{q} \\
	\Az{pq} \Az{pr} &= \sqrt{2} \Sp{p} \Am{qr} \\
	\Az{pr} \Az{qr} &= \sqrt{2} \Ap{pq} \Sm{r}.
\end{align}
\end{subequations}
Finally, for any choice of four indices,
\begin{align}
	\Ap{pq} \Ap{rs} + \Ap{pr} \Ap{qs} + \Ap{ps} \Ap{qr} = 0
\end{align}
which reflects the fact there are only two linearly independent seniority-four singlets. This is a property of fermion pairing, not of sp(N). Different choices of an orthogonal basis are possible, as one could use Gram-Schmidt or L\"{o}wdin schemes to arrive at distinct answers. Clebsch-Gordan coupling leads to the states $\Ap{pq}\Ap{rs} \ket{\theta}$ and
\begin{align}
	\stwo{pq}{rs}  \ket{\theta} = \frac{1}{\sqrt{3}} \left(\Ap{pr} \Ap{qs} - \Ap{ps} \Ap{qr}\right) \ket{\theta}
\end{align}
which imposes a choice of ordering on the indices $p,q,r,s$. $\ket{\theta}$ is the empty state. For RG states, all the labels were general, and this caused an enormous headache.\cite{johnson:2025b} A complete and unambiguous solution was possible, but was incredibly tedious. In the present case, all of the seniority-four states have obvious groupings of indices so that the choice is simple. This choice of basis does not affect a CI expansion, but will have an impact in perturbation theory.

The Coulomb Hamiltonian has an explicit expression in terms of the singlet-excitation operators
\begin{align}
	\hat{H}_C = \sum_{pq} h_{pq} \Az{pq} + \frac{1}{2} \sum_{pqrs} V_{pqrs} 
	\left( \Az{pq}\Az{rs}  - \delta_{qr} \Az{ps}  \right) + \hat{V}_{NN}
\end{align}
whose one- and two-electron integrals
\begin{align}
	h_{pq} &= \int d\textbf{x} \phi^*_p (\textbf{x})
	\left( - \frac{\nabla^2}{2} - \sum_I \frac{Z_I}{\left\vert\textbf{x} - \textbf{R}_I\right\vert} \right) \phi_q(\textbf{x}), \\
	V_{pqrs} &= \int d\textbf{x}_1 d\textbf{x}_2 
	\frac{\phi^*_p (\textbf{x}_1) \phi_q(\textbf{x}_1) \phi^*_r (\textbf{x}_2) \phi_s (\textbf{x}_2)}
	{\left\vert \textbf{x}_1 - \textbf{x}_2 \right\vert},
\end{align}
have been pre-computed in a basis of orbitals $\{\phi\}$, and \emph{Chemists'} notation is employed for the two-electron integrals. PP and its excited states all have defined seniority. This is not unusual, Slater determinants have this property as well. However, the Coulomb Hamiltonian does not conserve seniority so that its eigenvectors do not have defined seniority. The Coulomb Hamiltonian can be regrouped
\begin{align}
	\hH_C = \hH_C^{(0)} + \hH_C^{(2)} + \hH_C^{(4)},
\end{align}
into a channel that conserves seniority $\hH_C^{(0)}$, a channel that changes seniority by two $\hH_C^{(2)}$, and a channel that changes seniority by four $\hH_C^{(4)}$. The whole point in this line of research is that strong correlation manifests in the seniority-conserving channel:\cite{bytautas:2011,bytautas:2015,wahlen:2018,gus_sanibel} the orbitals can always be chosen to maximize the contribution of the seniority-conserving piece of the Hamiltonian. Explicitly the seniority-conserving channel is
\begin{equation} \label{eq:sen0channel}
	\begin{split}
		\hH_C^{(0)} &=
		\sum_p h_{pp} \Az{pp}
		+ \frac{1}{2} \sum_p L_{pp} \left( \Az{pp} \Az{pp} - \Az{pp} \right) \\
		&+ \sum_{p<q} J_{pq} \Az{pp} \Az{qq} + \sum_{p<q} K_{pq} \left( \Az{pq} \Az{qp} - \Az{pp} \right) 
		+ \frac{1}{2} \sum_{p \neq q} L_{pq} \Az{pq} \Az{pq}
	\end{split}	
\end{equation}
where, the only two-body integrals that show up
\begin{align}
		J_{pq} &= V_{ppqq}
		&
		K_{pq} &= V_{pqqp}
		&
		L_{pq} &= V_{pqpq}
\end{align}
are \emph{direct}, \emph{exchange}, and \emph{pair-transfer} integrals. For real orbitals $\{\phi\}$, $K_{pq} = L_{pq}$. It is convenient to abbreviate
\begin{align}
	G_{pq} = 2 J_{pq} - K_{pq}.
\end{align}
Notice that in eq. \eqref{eq:sen0channel} the diagonal two-electron integral is chosen to be written as $L_{pp}$. Quickly, the other channels are
\begin{equation}
	\begin{split}
		\hH_C^{(2)} &= \sum_{p\neq q} h_{pq} \Az{pq} 
		+   \sum_p \sum_{q \neq r} V_{prqp} \Az{pr} \Az{qp} 
		+ \sum_p \sum_{q \neq r (\neq p)} V_{ppqr} \Az{pp} \Az{qr} \\
		& + \sum_p \sum_{q < r (\neq p)} \left(
		V_{pqpr} \Az{pq} \Az{pr} + V_{qprp} \Az{qp} \Az{rp}
		\right)
	\end{split}
\end{equation}
and
\begin{equation}
	\begin{split}
		\hH_C^{(4)} &=
		\sum_{p < q < r < s} V_{pqrs}
		\left(
		\Az{pq} \Az{rs} + \Az{qp} \Az{rs} + \Az{pq} \Az{sr} + \Az{qp} \Az{sr}
		\right) \\
		&+ \sum_{p < q < r < s} V_{prqs}
		\left(
		\Az{pr} \Az{qs} + \Az{rp} \Az{qs} + \Az{pr} \Az{sq} + \Az{rp} \Az{sq}
		\right) \\
		&+ \sum_{p < q < r < s} V_{psqr}
		\left(
		\Az{ps} \Az{qr} + \Az{sp} \Az{qr} + \Az{ps} \Az{rq} + \Az{sp} \Az{rq}
		\right).
	\end{split}
\end{equation}
With the algebraic structure introduced, PP and its excitations will be discussed.

\section{Perfect-Pairing} \label{sec:pp}
RG states are the eigenvectors of the reduced BCS Hamiltonian
\begin{align} \label{eq:hbcs}
	\hH_{\text{BCS}} = \frac{1}{2} \sum_p \xi_p \n{p} + \frac{1}{2} \sum_{pq} \Sp{p} \Sm{q},
\end{align}
with an arbitrary set of single-particle energies $\{\xi\}$ and a constant pairing strength.\footnote{The constant pairing strength can also be varied, though doing so will amplify all the single-particle energies by the same factor.} They are products of $N_p$ pairs delocalized over $N$ spatial orbitals
\begin{align}
	\label{eq:rg_states}
	\ket{\RG} = \prod^{N_p}_{\alpha=1} \sum^N_{p=1} \frac{\Sp{p}}{u_{\alpha} - \xi_p} \ket{\theta}
\end{align}
weighted by a set of numbers called \emph{rapidities} $\{u\}$. Acting with $\hH_{\text{BCS}}$ on the RG state \eqref{eq:rg_states} gives one term proportional to it, and a collection of $N_p$ terms which are not. The unwanted terms are eliminated by solving Richardson's equations
\begin{align}
	\label{eq:rich_eq}
	2 + \sum^N_{p=1} \frac{1}{u_{\alpha} - \xi_p} + \sum^{N_p}_{\beta (\neq \alpha)=1} \frac{2}{u_{\beta} - u_{\alpha}} = 0,\quad \forall \; \alpha=1,\dots,N_p
\end{align}
for the rapidities. Richardson's equations are non-linear, and are difficult to solve directly. The correct approach is to change variables so that the equations are stable and easy to solve.\cite{faribault:2011,elaraby:2012} The $\binom{N}{N_p}$ distinct solutions of Richardson's equations define the complete set of seniority-zero eigenvectors of \eqref{eq:hbcs}. RG states of seniorities two and four are obtained from Richardson's equations with particular orbitals omitted.\cite{johnson:2025b} Each RG state requires a complete solution of Richardson's equations so that none of the rapidities are common among different RG states.

RG states can thus be employed in variational approximations by minimizing their expectation value of the Coulomb Hamiltonian with respect to $\{\xi\}$. When the pairing strength is zero, this approach is equivalent to RHF. This general structure is tractable: the complete set of eigenvectors of \eqref{eq:hbcs} is polynomially computable, and their correlation functions are sums of cofactors of the corresponding effective overlap matrix. However, this construction relies on the solution of non-linear equations for intermediate quantities which makes it rather opaque: it is difficult to interpret the ground state and its low-lying excitations qualitatively.

Several studies have indicated how the single-particle energies arrange themselves.\cite{fecteau:2022,johnson:2023,johnson:2024a,johnson:2024b} In short, they adopt a core-valence-virtual pattern: isolated strongly-negative values (core), pairs of near-degenerate values (valence), and isolated strongly-positive values (virtual). For the valence, each set of near-degenerate $\{\xi_p\}$ and their coresponding spatial orbitals will be referred to as valence-bond subsystems (VBS). The orbitals in one VBS are disjoint from the others, which is to say that they are strongly orthogonal. In the APSG literature, VBS are Arai subspaces with only two elements. Each VBS contains a bonding orbital $\ket{\ger{\alpha}}$ and an antibonding orbital $\ket{\ung{\alpha}}$. Effectively, the reduced BCS Hamiltonian becomes
\begin{align}
	\hH_{\text{PP}}' &= \frac{1}{2}\sum_{p} \xi_p \n{p} + \frac{1}{2} \sum_{\alpha}
	\left( \Sp{\ger{\alpha}} \Sm{\ung{\alpha}} + \Sp{\ung{\alpha}} \Sm{\ger{\alpha}} \right),
\end{align}
a model in which transfer of electron pairs only occurs within each VBS. This is the structure of the optimal reduced BCS Hamiltonians found for dissociation of linear hydrogen chains for example. Now, this model has many symmetries that allow for simplfication: the number of electrons in each core orbital $\n{i}$, in each virtual orbital $\n{a}$, and in each VBS $\n{\alpha} = \n{\ger{\alpha}}  + \n{\ung{\alpha}}$ all commute with $\hH_{\text{PP}}'$. Thus, this model can be reduced to
\begin{align} \label{eq:hgvb}
	\hH_{\text{PP}} = \frac{1}{2} \sum_{\alpha} \omega_{\alpha} \n{\ung{\alpha}} 
	+ \frac{1}{2} \sum_{\alpha} \left( \Sp{\ger{\alpha}} \Sm{\ung{\alpha}} + \Sp{\ung{\alpha}} \Sm{\ger{\alpha}} \right),
\end{align}
where $\omega_{\alpha} = \xi_{\alpha_1} - \xi_{\alpha_0}$. In each VBS, there is competition between the \emph{aufbau} behaviour and the Coulomb repulsion. Here, the Coulomb repulsion has been rotated into a seniority-zero picture and is represented by transferring electrons between the bonding and antibonding orbitals. The VBS gap $\omega_{\alpha}$ is a measure of the relative contributions of each effect: the aufbau filling dominates (weak correlation) when the gap is large while the Coulomb repulsion dominates (strong correlation) when the gap is small. These two effects are in balance when the gap is equal to one, which marks the transition from weak to strong correlation.

The important point is that the Hamiltonian \eqref{eq:hgvb} acts locally in each VBS with no coupling between them. It has \emph{no} action on the core or the virtual orbitals. Its eigenvectors thus factor into contributions from the core orbitals, each VBS, and the virtual orbitals. The core orbitals are either empty $\ket{\vac{i}}$ or full $\Sp{i} \ket{\vac{i}}$, and the same for the virtual orbitals. Each VBS has four possible states. The VBS can be empty $\ket{\vac{\alpha}}$, full $\Sp{\ger{\alpha}}\Sp{\ung{\alpha}} \ket{\vac{\alpha}}$ or the two orbitals $\ket{\alpha_0} = \Sp{\ger{\alpha}} \ket{\vac{\alpha}}$ and $\ket{\ung{\alpha}} = \Sp{\ung{\alpha}} \ket{\vac{\alpha}}$ can be partially occupied in two ways
\begin{align}
	\ket{\alpha_{\pm} } = \ket{\ger{\alpha}} + (\omega_{\alpha} \pm \eg{\alpha}) \ket{\ung{\alpha}},
\end{align}
with the \emph{pair-amplitude}
\begin{align}
	\eg{\alpha} = \sqrt{\omega^2_{\alpha} + 1}
\end{align}
which will occur so frequently that it is worth abbreviating. The last two states are not normalized, though it is easily verified that
\begin{align}
	\braket{\alpha_{\pm} | \alpha_{\pm}} 
	= 2 \eg{\alpha} \left(\eg{\alpha} \pm \omega_{\alpha}\right).
\end{align}
Both states $\ket{\alpha_{\pm} }$ represent partial occupation of both orbitals in VBS $\alpha$ based on the gap $\omega_{\alpha}$. In $\ket{\alpha_-}$, the bonding orbital is more strongly occupied while the antibonding orbital is more strongly occupied in $\ket{\alpha_+}$. Thus, $\ket{\alpha_{-}}$ will be called a \emph{bond} while $\ket{\alpha_{+}}$ will be called an \emph{antibond}. When the gap $\omega_{\alpha}$ goes to zero, the orbitals are equally occupied in both states, which themselves become
\begin{align}
	\lim_{\omega_{\alpha} \rightarrow 0} \ket{\alpha_{\pm} } = \frac{1}{\sqrt{2}} \left( \ket{\alpha_0} \pm \ket{\alpha_1} \right).
\end{align}
In this limit, $\ket{\alpha_-}$ is a ``pair singlet'', while the three states $\ket{\vac{\alpha}}$, $\ket{\alpha_+}$ and $\Sp{\ger{\alpha}}\Sp{\ung{\alpha}} \ket{\vac{\alpha}}$ form a ``pair triplet''. The precise nature of the ``pair-spin'' coupling will be treated thoroughly elsewhere.

With the pair creators
\begin{align} \label{eq:PP_pair_creators}
	\SpG{\alpha_{\pm}} = \frac{\Sp{\ger{\alpha}} + (\omega_{\alpha} \pm \eg{\alpha}) \Sp{\ung{\alpha}} }
		{\sqrt{ 2 \eg{\alpha} (\omega_{\alpha} \pm \eg{\alpha}) }}
\end{align}
and a core of doubly-occupied orbitals
\begin{align}
	\ket{\Phi} = \Sp{1} \dots \Sp{M_c} \ket{\theta},
\end{align}
the PP state with $M_c$ core pairs and $M$ bonds is
\begin{align} \label{eq:PP_ref}
	\ket{\PP} = \prod_{\alpha} \SpG{\alpha_-} \ket{\Phi}
\end{align}
a product of fully-occupied core orbitals, $\ket{\alpha_-}$ in each VBS, and empty virtual orbitals. It is an eigenvector of $\eqref{eq:hgvb}$ with eigenvalue
\begin{align} \label{eq:pp_eig}
	E = \frac{1}{2} \sum_{\alpha} \left( \omega_{\alpha} - \eg{\alpha}  \right).
\end{align}
The notation $\ket{\PP}$ serves to remind that PP \emph{is} an RG state, whose rapidities are replaced with the VBS gaps $\{\omega\}$. The major simplification is that Richardson's equations have an explicit solution, and as a result the same set of values $\{\omega\}$ will define \emph{all} the eigenvectors of \eqref{eq:hgvb}. While \eqref{eq:PP_ref} is not necessarily the ground state of \eqref{eq:hgvb}, what matters is that $\ket{\PP}$ is an eigenvector. The other eigenvectors are a basis for its orthogonal complement, of which only a small number will couple meaningfully through a two-body operator.  As the core and virtual orbitals give no contribution to the eigenvalue \eqref{eq:pp_eig}, it is $\binom{M_c + M_v}{M_c}$-fold degenerate.

PP and its excitations all have defined seniority. As a result, only RDM elements which conserve seniority 
\begin{subequations}
\begin{align}
	n_p    &= \braket{\PP | \Az{pp}                   | \PP} \\
	D_{pq} &= \braket{\PP | \Az{pp} \Az{qq}           | \PP} \\
	X_{pq} &= \braket{\PP | \Az{pq} \Az{qp} - \Az{pp} | \PP} \\
	P_{pq} &= \braket{\PP | \Az{pq} \Az{pq}           | \PP} 
\end{align}
\end{subequations}
will be non-zero. It must be emphasized that the convention is different from that in ref. \citenum{johnson:2025d}: here the occupation numbers have range $0 < n_p < 2$. The \emph{direct} 2-RDM elements $D_{pq}$ measure the simultaneous occupation of orbitals $p$ and $q$, For a seniority-zero state the \emph{exchange} 2-RDM elements $X_{pq}$ are strictly proportional to the direct elements
\begin{align} \label{eq:exc_rdm}
	X_{pq} = - \frac{1}{2} D_{pq},
\end{align}
whereas they \emph{are not always} for states of higher seniority. The \emph{pair} 2-RDM elements $P_{pq}$ measure the probability of the transfer of a pair of electrons from orbital $p$ to orbital $q$. The diagonal element is ambiguously defined as either a direct or a pair 2-RDM element. By convention, it is assigned
\begin{subequations}
\begin{align}
	D_{pp} &= 0 \\
	P_{pp} &= \braket{\PP | \Az{pp} \Az{pp} - \Az{pp} | \PP}.
\end{align}
\end{subequations}
The PP occupation numbers
\begin{subequations}
\begin{align}
	n_i &= 2 \\
	n_{\alpha_{\mu}} &=  1+ (-1)^{\mu}  \frac{ \omega_{\alpha}}{\eg{\alpha}} \\
	n_a &= 0,
\end{align}
\end{subequations}
are usually considered the fundamental quantities. The direct 2-RDM elements are products of the individual occupation numbers
\begin{align}
	D_{pq} = n_p n_q
\end{align}
\emph{except} if $p$ and $q$ are in the same VBS, in which case $D_{\ger{\alpha}\ung{\alpha}} = 0$.
Notice that this includes the possibilities of core and virtual orbitals. For PP, the exchange elements are \eqref{eq:exc_rdm}. The only non-vanishing pair elements occur within a VBS:
\begin{align}
	P_{\alpha_0 \alpha_1} = - \sqrt{n_{\alpha_0} n_{\alpha_1}} = - \frac{1}{\eg{\alpha}}
\end{align}
and $P_{\alpha_1 \alpha_0} = P_{\alpha_0 \alpha_1}$. 

It is convenient to define orbital energies\cite{johnson:2025d} 
\begin{subequations}
\begin{align}
	\varepsilon_i &= h_{ii} + \frac{1}{2} L_{ii} + \sum_{j (\neq i)} G_{ij} + \frac{1}{2} \sum_{\gamma} \sum_{\lambda} G_{i \gamma_{\lambda}} n_{\gamma_{\lambda}} \\
	\varepsilon_{\alpha_{\mu}} &= h_{\alpha_{\mu} \alpha_{\mu}} + \frac{1}{2} L_{\alpha_{\mu} \alpha_{\mu}} 
		+ \sum_i G_{i \alpha_{\mu}} + \frac{1}{2} \sum_{\gamma (\neq \alpha)} \sum_{\lambda} G_{\alpha_{\mu} \gamma_{\lambda}} n_{\gamma_{\lambda}} \\
	\varepsilon_a &= h_{aa} + \frac{1}{2} L_{aa} + \sum_i G_{ia} + \frac{1}{2} \sum_{\gamma} \sum_{\lambda} G_{\gamma_{\lambda} a} n_{\gamma_{\lambda}},
\end{align}
\end{subequations}
which lead to a HF-like energy expression:
\begin{align} \label{eq:pp_occ_energy}
	E[\PP] = \sum_p \varepsilon_p n_p 
	- \sum_{\alpha} L_{\ger{\alpha} \ung{\alpha}} \sqrt{n_{\ger{\alpha}} n_{\ung{\alpha}}} 
	- \frac{1}{2} \sum_{p<q}{}' G_{pq} n_p n_q
\end{align}
where the final summation is restricted so that pairs of indices in the same VBS are excluded. The energy is written entirely in terms of the occupation numbers, emphasizing the well-known fact that the PP energy is a functional of the 1-RDM.\cite{piris:2011, pernal:2013} The energy must be made stationary with respect to both electronic and orbital degrees of freedom, like in multiconfigurational self-consistent-field (MCSCF) type approaches. The electronic degrees of freedom are the VBS gaps $\{\omega\}$ with gradient
\begin{align}
	\eg{\alpha}^3 \frac{\partial E}{\partial \omega_{\alpha}} = \omega_{\alpha} L_{\ger{\alpha}\ung{\alpha}}
		+ \varepsilon_{\ger{\alpha}} - \varepsilon_{\ung{\alpha}},
\end{align}
which, when stationary gives a direct meaning to the VBS gaps as the ratio of the difference of the orbital energies to the Coulomb repulsion pushing the pair of electrons from the bonding to the antibonding orbital
\begin{align} \label{eq:stat_grad_omega}
	\omega_{\alpha} = \frac{ \varepsilon_{\ung{\alpha}} - \varepsilon_{\ger{\alpha}}}{L_{\ger{\alpha}\ung{\alpha}}}.
\end{align}
The elements of the electronic Hessian are
\begin{subequations}
	\begin{align}
		\eg{\alpha}^3 \frac{\partial^2 E}{\partial \omega_{\alpha}^2} &= L_{\ger{\alpha}\ung{\alpha}}
		- 3 \eg{\alpha} \omega_{\alpha} \frac{\partial E}{\partial \omega_{\alpha}} \\
		\eg{\alpha}^3 \eg{\beta}^3 \frac{\partial^2 E}{\partial \omega_{\alpha} \partial \omega_{\beta}}
		&= \frac{1}{2} \sum_{\mu \nu} (-1)^{\mu + \nu} G_{\alpha_{\mu} \beta_{\nu}}.
	\end{align}
\end{subequations}
For a vanishing electronic gradient, the diagonal elements of the electronic Hessian are just the pair-transfer integrals which are non-negative for real orbitals. The off-diagonal elements however can be negative if there is substantial coupling between the bonding orbital of one VBS and the antibonding orbital of another.

The stationary condition \eqref{eq:stat_grad_omega} allows for different expressions of the PP energy. The VBS gaps could be eliminated but that does not lead an informative result. A more useful expression
\begin{equation} \label{eq:pp_pair_energy}
\begin{split}
	E[\PP] &= \sum_{\alpha} \left[
	\varepsilon_{\ger{\alpha}} + \varepsilon_{\ung{\alpha}} - \eg{\alpha} L_{\ger{\alpha} \ung{\alpha}}
	\right] 
	+ 2 \sum_i \varepsilon_i
	- \frac{1}{2} \sum_{p<q}{}' G_{pq} n_p n_q
\end{split}
\end{equation}
is obtained from the valence occupation numbers, and replacing the difference in valence orbital energies using \eqref{eq:stat_grad_omega}. This energy is the same as a state of singly-occupied valence orbitals, stabilised by a pairing interaction $\eg{\alpha} L_{\ger{\alpha} \ung{\alpha}}$. This is a clear indication of a direct connection to UHF. It is understood that UHF and PP are similar, though not equivalent: UHF has a Coulson-Fischer point whereas PP does not for example. UHF orbitals serve as an excellent guess for PP orbitals.\cite{wang:2019} This viewpoint will be useful when computing PP excitation energies.

While the pair amplitudes $\{\eg{} \}$ are clear functions of the VBS gaps, the energy expression \eqref{eq:pp_pair_energy} emphasizes that they are fundamental quantities. The last term in eq. \eqref{eq:pp_pair_energy} removes the terms that have been doubly-counted in the summation of orbital energies, as in HF. It will be convenient computationally to rewrite this summation
\begin{equation}
\begin{split}
	\frac{1}{2} \sum_{p<q}{}' G_{pq} n_p n_q &= 2 \sum_{i<j} G_{ij} + \sum_i \sum_{\alpha} \sum_{\mu} G_{i \alpha_{\mu}}
		+ 2 \sum_i \sum_{\alpha} \gone{\alpha}{i} \\
		&+ \sum_{\alpha < \beta} \left(
			  \gzer{\alpha}{\beta} + \gone{\alpha}{ \ger{\beta} } + \gone{\alpha}{ \ung{\beta} }
			+ \gone{{\beta}}{ \ger{\alpha} } + \gone{\beta}{ \ung{\alpha} } + \gtwo{\alpha}{\beta}
		\right)
\end{split}
\end{equation}
in terms of quantities reduced over $\{\omega\}$ and $\{\eg{}\}$
\begin{subequations}
\begin{align}
	\gzer{\alpha}{\beta} &= \frac{1}{2}
		( G_{\ger{\alpha} \ger{\beta} } + G_{ \ger{\alpha} \ung{\beta} }
		+ G_{ \ung{\alpha} \ger{\beta} } + G_{ \ung{\alpha} \ung{\alpha} } ) \\
	\gone{\alpha}{p}     &= \frac{1}{2} \frac{ \omega_{\alpha} }{ \eg{\alpha} }
		( G_{\ger{\alpha} p } - G_{\ung{\alpha} p}  ) \\
	\gtwo{\alpha}{\beta} &= \frac{1}{2} \frac{\omega_{\alpha} \omega_{\beta}}{ \eg{\alpha} \eg{\beta} }
	( G_{\ger{\alpha}\ger{\beta}} - G_{\ger{\alpha}\ung{\beta}}
	- G_{\ung{\alpha}\ger{\beta}} + G_{\ung{\alpha}\ung{\beta}}).
\end{align}
\end{subequations}
If the goal were the PP energy, then these quantities would serve no purpose, and the energy would be best understood as \eqref{eq:pp_occ_energy}. As post-PP computations are the point of the manuscript, these quantities will be reused several times in the computation of matrix elements, in particular the energies of the individual states. Computing them once and storing them is thus productive.

Orbitals must also be optimized through sequential unitary transformations
\begin{align}
	\hat{U} = \exp \left[ \sum_{p<q} \kappa_{pq} \left( \Az{pq} - \Az{qp} \right)  \right],
\end{align}
with $\boldsymbol{\kappa}$ set to zero after each step.
The orbital gradient may be written in terms of a generalized Fock matrix,\cite{helgaker_book}
\begin{align}
	\frac{\partial E}{\partial \kappa_{pq}} = 2 \left( f_{pq} - f_{qp} \right)
\end{align}
for any state $\ket{\Psi}$
\begin{align}
	f_{pq} = \sum_{\sigma} \braket{\Psi | a^{\dagger}_{p\sigma}[a_{q\sigma},\hat{H}_C] | \Psi}.
\end{align}
If $\ket{\Psi}$ has zero seniority, this reduces to
\begin{align} \label{eq:genfock}
	f_{pq} = h_{pq} n_p + \frac{1}{2} \sum_{r} (2V_{rrpq} - V_{rqpr}) D_{rp}
	+ \sum_r V_{rprq}  P_{rp}.
\end{align}
Again, the diagonal elements are $D_{pp}=0$ are $P_{pp} = n_p$. For PP, there are core, valence, and virtual orbitals, and the generalized Fock matrix is different in each case. For a core orbital $i$, and $q$ arbitrary,
\begin{align}
	f_{iq} = 2h_{iq} + 2 \sum_{j} \left(2 V_{jjiq} - V_{jqij}\right) 
	+ \sum_{\gamma}\sum_{\lambda} \left( 2V_{\gamma_{\lambda} \gamma_{\lambda} iq} - V_{ \gamma_{\lambda} qi \gamma_{\lambda} } \right) n_{\gamma_{\lambda}},
\end{align}
for a valence orbital $\alpha_{\mu}$
\begin{align}
	\begin{split}
	f_{\alpha_{\mu}q} &= \left( h_{\alpha_{\mu}q} + V_{ \alpha_{\mu}\alpha_{\mu}\alpha_{\mu}q } \right)n_{\alpha_{\mu} }
	- \frac{1}{\eg{\alpha}} V_{\alpha_{\mu} \alpha_{1 - \mu} q \alpha_{1 - \mu}} \\
	&+ \sum_{i} \left( 2V_{\alpha_{\mu} qii } - V_{ \alpha_{\mu} iiq } \right) n_{\alpha_{\mu}}
	+ \sum_{\gamma (\neq \alpha)} \sum_{\lambda} \left( V_{ \alpha_{\mu} q \gamma_{\lambda} \gamma_{\lambda} }  
	- \frac{1}{2} V_{ \alpha_{\mu} \gamma_{\lambda} \gamma_{\lambda} q } \right) n_{\alpha_{\mu}} n_{\gamma_{\lambda}},
	\end{split}
\end{align}
while for a virtual orbital $a$, the generalized Fock matrix is strictly $f_{aq} = 0$. In HF, the orbital energies \emph{are} the diagonal elements of the Fock matrix, but for PP they are different enough that their notations will be kept separate. For a state with zero seniority the orbital Hessian reduces to
\begin{equation}
\begin{split}
	\frac{\partial^2 E}{\partial \kappa_{pq} \partial \kappa_{rs}} &= \;
		2        ( P_{pr} + P_{qs} - P_{ps} - P_{qr} ) (   V_{prqs} + V_{psqr} ) \\
		&\quad + ( D_{pr} + D_{qs} - D_{ps} - D_{qr} ) ( 4 V_{pqrs} - V_{psqr} - V_{prqs} ) \\
		&\quad + \kd_{pr} [ ( 2 n_p - n_q - n_s ) h_{qs} + W_{pqs} ] \\
		&\quad + \kd_{qs} [ ( 2 n_q - n_p - n_r ) h_{pr} + W_{qpr} ] \\
		&\quad - \kd_{ps} [ ( 2 n_p - n_q - n_r ) h_{qr} + W_{pqr} ] \\
		&\quad - \kd_{qr} [ ( 2 n_q - n_p - n_s ) h_{ps} + W_{qps} ]
\end{split}
\end{equation}
in terms of intermediates
\begin{align}
	W_{pqr} &= \sum_s ( 2 P_{ps} - P_{qs} - P_{rs} ) V_{qsrs}
		+ \frac{1}{2} \sum_s  ( 2 D_{ps} - D_{qs} - D_{rs} ) ( 2 V_{qrss} - V_{qssr} ).
\end{align}
The orbital Hessian is thus built with $\mathcal{O}(N^4)$ cost. Even more efficient computation is possible.\cite{limacher:2026} The electronic and orbital degrees of freedom are however coupled, and the choice has been made to build the complete Hessian. The mixed elements are not complicated. First, for all $p$ and $q$, 
\begin{align}
	Q_{pq} = h_{pq} + \frac{1}{2} \sum_r ( 2 V_{rrpq} - V_{rqpr} ).
\end{align}
If neither $p$ nor $q$ are in the $\alpha$th VBS,
\begin{align}
	\eg{\alpha}^3  \frac{\partial^2 E}{ \partial \omega_{\alpha} \partial \kappa_{pq} } &=
		\sum_{\lambda} (-1)^{\lambda} ( 2 V_{\alpha_{\lambda} \alpha_{\lambda} pq } - V_{ \alpha_{\lambda} qp \alpha_{\lambda} } ) (n_p - n_q),
\end{align}
whereas if $p$ is in the $\alpha$th VBS but $q$ is not,
\begin{equation}
\begin{split}
	\eg{\alpha}^3  \frac{\partial^2 E}{ \partial \omega_{\alpha} \partial \kappa_{ \alpha_{\mu} q} } &=
		2 \omega_{\alpha} V_{ \alpha_{1 - \mu} \alpha_{\mu} \alpha_{1 - \mu} q } 
		+ 2 (-1)^{\mu} ( Q_{ \alpha_{\mu} q } + V_{ \alpha_{\mu} \alpha_{\mu} \alpha_{\mu} q } )  \\
		& \quad - \sum_{\lambda} ( 2 V_{ \alpha_{\lambda} \alpha_{\lambda} \alpha_{\mu} q } - V_{ \alpha_{\lambda} q \alpha_{\mu} \alpha_{ \lambda } } )
		( (-1)^{\mu} n_{\alpha_{ \lambda } } + (-1)^{\lambda} n_q ).
\end{split}
\end{equation}
If $q$ is in the $\alpha$th VBS but $p$ is not, the element is the same with the roles of $p$ and $q$ exchanged, and an overall negative sign. Finally, if both $p$ and $q$ are in the $\alpha$th VBS,
\begin{align}
	\eg{\alpha}^3  \frac{\partial^2 E}{ \partial \omega_{\alpha} \partial \kappa_{ \ger{\alpha} \ung{\alpha} } } &=
		4 Q_{\ger{\alpha} \ung{\alpha}} + 2 \omega_{\alpha} \left( 1 + \frac{1}{\eg{\alpha}} \right)
		( V_{\ung{\alpha} \ung{\alpha} \ung{\alpha} \ger{\alpha}} - V_{ \ger{\alpha} \ger{\alpha} \ger{\alpha} \ung{\alpha} } ).
\end{align}
With the PP state summarized in the language of RG states, we move on to compute post-PP corrections.

\section{Epstein-Nesbet Correction}
\label{sec:en2}

\subsection{Qualitative summary of excitations}
The low-lying excitations will now be included to compute a second-order perturbative correction. An EN partitioning is chosen as it is the simplest possible approach leading to a reasonable result. Choosing \eqref{eq:hbcs} as $\hH_0$ led to positive perturbative corrections as the chosen RG reference was not its ground state.\cite{johnson:2024b} The EN partitioning was the choice made in Parts I and II, which allows for a direct comparison of PP and RG results. For RG states it was possible to treat many cases at once, leading to a few difficult developments. Here it is productive to separate the individual types of states, and evaluate \emph{many} simple matrix elements. The number of off-diagonal elements to evaluate is incredibly large, so this contribution will focus only on a perturbative correction, leaving a CI construction to the next contribution.

The reference Hamiltonian is the Coulomb Hamiltonian represented in terms of the PP reference $\ket{\PP}$ and its excited states $\{ \ket{\PPE}\}$
\begin{align}
	\hH_0 = \ket{\PP} \bra{\PP} \hH_C \ket{\PP} \bra{\PP} 
	+ \sum_{\{\PPE\} } \ket{\PPE} \bra{\PPE} \hH_C \ket{\PPE} \bra{\PPE}.
\end{align}
The EN2 correction is
\begin{align}
	E^{(2)}_{\text{EN}} = - \sum_{\{\PPE\}} \frac{\vert \braket{\PPE | \hH_C | \PP} \vert^2}{E[\PPE] - E[\PP]},
\end{align}
where the denominators are ordered to denote excitation energies. This convention is adopted as they are the natural quantities to compute: for each state $\ket{\PPE}$, the energy is the energy of the reference $E[\PP]$ plus an update particular to the state. 

Before discussing the states, their energies, and their couplings to the reference in detail, the taxonomy and notation will be summarized. Generally, there are three groups of excitations: \emph{singles}, \emph{doubles}, and \emph{pair-transfers}. With Slater determinants, pair-transfers are equivalent to particular doubles. Here this is not the case: the excitations are quite different and must be treated separately. In HF, singles are one-electron excitations while doubles are two simultaneous one-electron excitations. This will also not be the case here, so a justification of the terms singles and doubles is warranted. A single is an excitation whose coupling to the reference should be related to the first derivative of the energy with respect to one of the variational parameters. With one caveat to be discussed in the following section, this is always the case. A double is an excitation that is obtained through two simultaneous single excitations. Their coupling to the reference is generally related to a second derivative of the energy. Here these excitations are correlated so the corresponding energy will be the sum of the individual excitation energies along with an update. Singles couple to the PP reference through one- and two-electron integrals, whereas doubles and pair-transfers only couple to the PP reference through two-electron integrals. 

\begin{figure}[htbp]
	\centering
	\includegraphics[
	width=0.5\textwidth  
	]{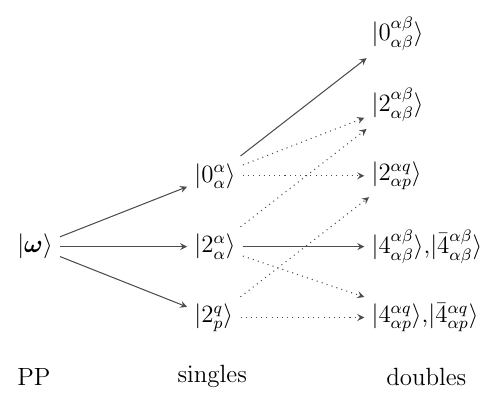}
	\caption{Single and double excitations from the PP reference $\ket{\PP}$: solid edges represent unique excitation channels. The asymmetry emphasizes that there are no double electron-transfers in the usual sense.}
	\label{fig:singles_doubles}
\end{figure}

The singles and doubles are summarised in Figure \ref{fig:singles_doubles}.
In the reference $\ket{\PP}$, each VBS $\alpha$ is in the bond $\ket{\alpha_{-}}$ state, in which a pair of electrons is delocalized across the bonding and antibonding orbitals, though the bonding orbital is more strongly occupied. Choosing $\ket{\alpha_{+}}$ in the VBS $\alpha$ reverses the occupations and does not change the seniority. These excitations are named \emph{swaps} and denoted $\ket{0^{\alpha}_{\alpha}}$ emphasizing that they change the seniority by zero, and excite from the $\alpha$th VBS to the $\alpha$th VBS. The pair of electrons in VBS $\alpha$ can be \emph{split}, leaving one electron in the bonding orbital and one electron in the antibonding orbital. These excitations also act locally in the VBS $\alpha$ but augment the seniority by two, thus are labelled $\ket{2^{\alpha}_{\alpha}}$. The last type of single is the one expected from Slater determinants, $\ket{2^q_p}$ corresponding to the \emph{electron-transfer} from one spatial orbital $p$ to another $q$, increasing the seniority by two. Minor restrictions are required on the indices: $p$ must be either in the core or valence while $q$ must be in the valence or virtual. In addition, $p$ and $q$ cannot both be in the same VBS: splits are different from single electron-transfers. As expected, the single excitations do not couple to the reference $\ket{\PP}$ through $\hH_C$, except for valence-valence electron-transfers, as will be discussed below. Briefly, the states $\ket{2^{\beta_{\nu}}_{\alpha_{\mu}} }$ and $\ket{2_{\beta_{\nu}}^{\alpha_{\mu}} }$ are distinct excitations. A specific linear combination of the two states decouples from $\ket{\PP}$ while its orthogonal complement does not.

With the singles understood, the doubles are simple to summarize. Two swaps can occur $\ket{0^{\alpha\beta}_{\alpha\beta}}$ or a swap may be combined with a split $\ket{2^{\alpha\beta}_{\alpha\beta}}$. In the latter case, to avoid ambiguity in the notation, the swap occurs in the first set of indices, while the split occurs in the second. A double-split is the first example of a seniority-four excitation, with the two resulting states labelled $\ket{4^{\alpha\beta}_{\alpha\beta}}$ and $\ket{\bar{4}^{\alpha\beta}_{\alpha\beta}}$. All of these doubles occur only in the valence.

The remaining doubles can involve the core and virtual orbitals: swaps may be combined with single electron-transfers $\ket{2^{\alpha q}_{\alpha p}}$ as can splits $\ket{4^{\alpha q}_{\alpha p}}$ and $\ket{\bar{4}^{\alpha q}_{\alpha p}}$. Double electron-transfers, the excitations one would expect from Slater determinants, \emph{are not present}. They are instead replaced with correlated two-electron excitations, \emph{pair-transfers}.

Pair-transfers are analogous to particle-hole type excitations, but the excitations are correlated. Interpretation of these excitations in terms of second quantization is ambiguous, \emph{particularly in the valence}. First, there is $\ket{0^{qq}_{pp}}$, where a seniority-zero pair-particle scatters to a seniority-zero pair-hole. This can also happen with a seniority-two pair-particle $\ket{2^{pp}_{qr}}$, a seniority-two pair-hole $\ket{2^{qr}_{pp}}$ or both $\ket{4^{qs}_{pr}}$ (and $\ket{ \bar{4}^{qs}_{pr} }$). For a seniority-zero pair in the valence, only the VBS label is kept.

With the excitations introduced, their excitation energies and couplings will be computed. The excitations in the valence will first be treated at length, while those involving the core or virtual orbitals will be briefly summarised. The valence excitations are more complicated while the external excitations are simplified versions of them. Everything will be reduced to a set of common quantities \emph{already computed for the PP mean-field}. Excitation energies can be written in different ways. In terms of occupation numbers it is possible to write each excitation energy as
\begin{equation} \label{eq:ON_EE}
	\begin{split}
		E[\PPE] - E[\PP] &=
		\sum_p \varepsilon_p \left(
			n[\PPE]_p - n_p
		\right)
		+ \frac{1}{2} \sum_{p<q}{}' G_{pq}
		\left( n[\PPE]_p - n_p \right) \left( n[\PPE]_q - n_q \right) 
		+ \Delta[\PPE]
	\end{split}
\end{equation}
with a quantity $\Delta[\PPE]$ particular to each excited state. While this expression involves complete summations, it is always easy to identify precisely which occupation numbers change so that \eqref{eq:ON_EE} may be evaluated with a fixed number of floating point operations. $\Delta[\PPE]$ is a cumulant for the energy: it is the two-body information that does not factor into one-body information. 

It is also possible to reduce the excitation energies to expressions involving only the VBS gaps $\omega_{\alpha}$ and pair-amplitudes $\eg{\alpha}$. This provides a direct physical interpretation to each excitation and is thus the chosen approach. Transition elements $\braket{\PP | \hH_C | \PPE}$ are reducible to the generalized Fock matrix $f_{pq}$ and roots of PP occupation numbers. The physically relevant quantities, the transition probabilities $\vert \braket{\PP | \hH_C | \PPE} \vert^2$, \emph{are} expressible directly in terms of $\omega_{\alpha}$ and $\eg{\alpha}$. However, this is less important to understand, and as the resulting expressions are lengthy they are omitted. 

Finally, it is convenient to abbreviate sums of particular 2-body integrals that will recur in seniorities two and four. For two indices that are both core, virtual, or within the same VBS, define the intermediates
\begin{subequations}
\begin{align}
	\twamp{ij} &= J_{ij} + K_{ij} - \frac{1}{2} L_{ii} - \frac{1}{2} L_{jj} \\
	\twamp{ab} &= J_{ab} + K_{ab} - \frac{1}{2} L_{aa} - \frac{1}{2} L_{bb} \\
	\twamp{\ger{\alpha} \ung{\alpha}} &= J_{\ger{\alpha} \ung{\alpha}} + K_{\ger{\alpha} \ung{\alpha}} 
	- \frac{1}{2} L_{\ger{\alpha} \ger{\alpha}} - \frac{1}{2} L_{\ung{\alpha} \ung{\alpha}}
\end{align}
\end{subequations}
while for two indices belonging to different subspaces, or different VBS:
\begin{subequations}
\begin{align}
	\twamp{\alpha_{\mu} \beta_{\nu}} &= J_{\alpha_{\mu} \beta_{\nu}} + K_{\alpha_{\mu} \beta_{\nu}}
	- \frac{1}{2} L_{\alpha_{\mu} \alpha_{\mu}} - \frac{1}{2} L_{\beta_{\nu} \beta_{\nu}}
	- \frac{1}{2} G_{\alpha_{\mu} \beta_{\nu}} \\
	\twamp{ia} &= J_{ia} + K_{ia} - \frac{1}{2} L_{ii} - \frac{1}{2} L_{aa} - \frac{1}{2} G_{ia} \\
	\twamp{i \alpha_{\mu}} &= J_{i \alpha_{\mu}} + K_{i \alpha_{\mu}} 
	- \frac{1}{2} L_{ii} - \frac{1}{2} L_{\alpha_{\mu} \alpha_{\mu}}
	- \frac{1}{2} G_{i \alpha_{\mu}} \\
	\twamp{\alpha_{\mu} a} &= J_{\alpha_{\mu} a} + K_{\alpha_{\mu} a} 
	- \frac{1}{2} L_{\alpha_{\mu} \alpha_{\mu}} - \frac{1}{2} L_{aa} 
	- \frac{1}{2} G_{\alpha_{\mu} a}.
\end{align}
\end{subequations}
The valence excitations can now be discussed in complete detail. The reader interested only in the numerical results is directed to Section \ref{sec:numbers}.

\subsection{Valence Excitations}
\label{sec:val_pp_states}

\subsubsection{Singles}
The first type of singles are swaps in the VBS $\alpha$. Concretely, the bond $\ket{\alpha_-}$ is replaced with the antibond $\ket{\alpha_+}$
\begin{align}
	\ket{0^{\alpha}_{\alpha}} = \SpG{\alpha_+} \ket{\PP_{\alpha}},
\end{align}
where the state $\ket{\PP_{\alpha}}$ is $\ket{\PP}$ with $\SpG{\alpha_-}$ omitted. This notation is inherited from RG states. It bears emphasis that excitations, \emph{particularly those in the valence}, are poorly described with second quantization. Dual to the PP pair creators in eq. \eqref{eq:PP_pair_creators}, one could define 
\begin{align} \label{eq:PP_pair_anni}
	\SmG{\alpha_{\pm}} = \frac{\Sm{\ger{\alpha}} + (\omega_{\alpha} \pm \eg{\alpha}) \Sm{\ung{\alpha}} }
		{\sqrt{ 2 \eg{\alpha} (\omega_{\alpha} \pm \eg{\alpha}) }}
\end{align}
which behave as desired,
\begin{align}
	\SpG{\alpha_{+}} \SmG{\alpha_{-}} \ket{\alpha_{-}} &= \ket{\alpha_+}.
\end{align}
\emph{But that is the only desired property it gives}, and using these objects is dangerous as repeated actions \emph{do not} behave as they should: the pair-creators are not nilpotent 
\begin{align}
	\SpG{\alpha_{\pm}} \SpG{\alpha_{\pm}} &= \pm \frac{1}{\eg{\alpha}} \Sp{\ger{\alpha}} \Sp{\ung{\alpha}},
\end{align}
and have vanishing action as $\omega_{\alpha}\rightarrow 0$
\begin{align}
	\SpG{\alpha_{+}  } \SpG{\alpha_{-  }} &= \frac{\omega_{\alpha}}{\eg{\alpha}} \Sp{\ger{\alpha}} \Sp{\ung{\alpha}}.
\end{align}
To be clear, no physically spurious results will occur, but we will always have to work back to the original objects $\Sp{}$ and $\Sm{}$ so that there is no benefit. A correct understanding of the PP excited states \emph{cannot} rely on second-quantized operators acting on the reference $\ket{\PP}$. The states will therefore be reported as individual objects, and presented so that they are orthonormal. When possible, second-quantized excitations will be presented, but the message is that \emph{this is the incorrect way to think about them}.

For RG states, the RDM elements of each state were strictly different. Each state required the solution of non-linear equations, their norms were obtained as the determinant of the Jacobian of the same non-linear equations, while their RDM elements were computed from sums involving first and second cofactors of the Jacobian. Here, the RDM elements of $\ket{0^{\alpha}_{\alpha}}$ are \emph{the same as the reference} $\ket{\PP}$ except that the occupations of the VBS $\alpha$ have been swapped
\begin{align}
	n[0^{\alpha}_{\alpha}]_{\alpha_{\mu}} = \braket{0^{\alpha}_{\alpha} | \hat{n}_{\alpha_{\mu}} | 0^{\alpha}_{\alpha}} = n_{\alpha_{1-\mu}},
\end{align}
and the pair-transfer elements in the VBS $\alpha$
\begin{align}
	P[0^{\alpha}_{\alpha}]_{\alpha_0 \alpha_1} = + \sqrt{n_{\alpha_0} n_{\alpha_1}} = \frac{1}{\eg{\alpha}} 
\end{align}
have the opposite phase. The direct/exchange elements do not change symbolically: they still factor into products of occupation numbers, but those in VBS $\alpha$ have been modified. The transition
\begin{align}
	\eg{\alpha} \braket{\PP | \hH_C | 0^{\alpha}_{\alpha}} = 
	\varepsilon_{\ger{\alpha}} - \varepsilon_{\ung{\alpha}} + \omega_{\alpha} L_{\ger{\alpha} \ung{\alpha}}
\end{align}
vanishes provided that the electronic gradient is zero. This is a Brillouin condition for PP. Other conditions will appear for seniority-two excitations provided the generalized Fock matrix is symmetric, which is the usual Brillouin condition for HF. To be well defined, the excitation energy cannot also vanish, and it is easily verified that
\begin{align}
	E[0^{\alpha}_{\alpha}] - E[\PP] = 2 \eg{\alpha} L_{\ger{\alpha} \ung{\alpha}}.
\end{align}
This energy is large when $\omega_{\alpha}$ is large, and tends to the pair-transfer integral $L_{\ger{\alpha}\ung{\alpha}}$ as $\omega_{\alpha}\rightarrow0$. 

Next, a split is a valence excitation occuring locally within one VBS, leaving both the bonding and antibonding orbitals singly-occupied
\begin{align}
	\ket{2^{\alpha}_{\alpha}} &= \Ap{\ger{\alpha} \ung{\alpha}} \ket{\PP_{\alpha}}
	= \frac{(-1)^{\mu}}{ \sqrt{n_{\alpha_{\mu}}} } \Az{\alpha_{1-\mu} \alpha_{\mu}} \ket{\PP}.
\end{align}
Here there are two ways of accessing this state ($\mu = 0,1$) through one-electron excitations on the reference. Both excitations lead to the same state. The RDM elements of this state are the same as PP, except in the $\alpha$th VBS, where both orbitals are singly-occupied $n[2^{\alpha}_{\alpha}]_{\ger{\alpha}} = n[2^{\alpha}_{\alpha}]_{\ung{\alpha}} = 1$. It is convenient to refer to the singly-occupied orbitals as \emph{blocked}. The direct elements do not change symbolically, but the exchange elements in the blocked levels are $X[2^{\alpha}_{\alpha}]_{\ger{\alpha} \ung{\alpha}} = 1$. All the pair elements \emph{including the diagonal} are zero within the blocked orbitals. Its excitation energy
\begin{equation}
	\begin{split}
		E[2^{\alpha}_{\alpha}] - E[\PP] &= \eg{\alpha} L_{\ger{\alpha}\ung{\alpha} } + \twamp{\ger{\alpha} \ung{\alpha}}
	\end{split}
\end{equation}
bears discussion. It is again useful to consider the interpretation that in $\ket{\PP}$ both the bonding and antibonding levels are singly-occupied and stabilised by the energy $\eg{\alpha}L_{\ger{\alpha}\ung{\alpha} }$. In $\ket{2^{\alpha}_{\alpha}}$, both levels \emph{are} singly-occupied but no longer stabilised by the pairing energy. As a result, a difference of orbital energies does not appear. The quantity $\twamp{\ger{\alpha} \ung{\alpha}}$ is the energy required to place a seniority-two pair in the blocked orbitals. One can see that its coupling to the PP reference
\begin{align}
	\braket{\PP | \hH_C | 2^{\alpha}_{\alpha}} = \frac{1}{ \sqrt{n_{\ger{\alpha}}} + \sqrt{n_{\ung{\alpha}}}  }
	\left(f_{\ger{\alpha}\ung{\alpha} } - f_{ \ung{\alpha} \ger{\alpha} } \right) =0
\end{align}
should vanish since the orbital optimization forces the generalized Fock matrix to be symmetric. This is no surprise, and corresponds to the Brillouin theorem for HF.

Single electron-transfers occuring between different VBS
\begin{align} \label{eq:state_single}
	\ket{2^{\beta_{\nu}}_{\alpha_{\mu}}} &= \Ap{\alpha_{\mu} \beta_{\nu}} \Sp{\beta_{1-\nu}} \ket{\PP_{\alpha\beta}}
	= (-1)^{\mu + \nu +1} \sqrt{\frac{2}{n_{\alpha_{\mu}} n_{\beta_{1-\nu}} } } \Az{\beta_{\nu} \alpha_{\mu}} \ket{\PP}
\end{align}
lead to 2 different states. Their excitation energy is more involved than for splits
\begin{equation} \label{eq:exc_single}
	\begin{split}
		E[2^{\beta_{\nu}}_{\alpha_{\mu}}] - E[\PP] &= 
			  \eg{\alpha} L_{\ger{\alpha}\ung{\alpha} } + \eg{\beta} L_{\ger{\beta}\ung{\beta} }
			+ \twamp{\alpha_{\mu} \beta_{\nu}}
			+ \varepsilon_{\beta_{1-\nu}} - \varepsilon_{\alpha_{1 - \mu}} 
			+G_{ \ger{\beta} \ung{\beta} } \\
		&\quad+ \gtwo{\alpha}{\beta}
			- \gone{\alpha}{\beta_{1-\nu}}
			+ \gone{\beta}{\alpha_{1 - \mu}}
			- \frac{1}{2} G_{\alpha_{1-\mu} \beta_{1-\nu} }.
	\end{split}
\end{equation}
The shape of this result will recur for all of the pair-transfer excitations, thus will be discussed further. The first two terms of \eqref{eq:exc_single} summarize the energetic penalties from breaking the bonds $\alpha$ and $\beta$. The next term $\twamp{\alpha_{\mu} \beta_{\nu}}$ is the energy required to put an open-shell singlet in the orbitals $\alpha_{\mu}$ and $\beta_{\nu}$, arising from the fact that the exchange elements are $X[2^{\beta_{\nu}}_{\alpha_{\mu}}]_{\alpha_{\mu} \beta_{\nu}} = D[2^{\beta_{\nu}}_{\alpha_{\mu}}]_{\alpha_{\mu} \beta_{\nu}} = 1$. The difference in orbital energies reflects the nature of the excitation, while $G_{ \ger{\beta} \ung{\beta} }$ appears as the the $\beta$th VBS now has $n[2^{\beta_{\nu}}_{\alpha_{\mu}}]_{\beta_{\nu}} =1$ and $n[2^{\beta_{\nu}}_{\alpha_{\mu}}]_{\beta_{1 - \nu}} =2$. The terms on the second line correct the double-counting present in the first line.

Notice that in the state \eqref{eq:state_single}, there is for the first time an unambiguous action of one singlet-excitation operator $\Az{\beta_{\nu} \alpha_{\mu}}$. Acting on a Slater determinant, this would be understood as exciting an electron from orbital $\ket{\alpha_{\mu}}$ to orbital $\ket{\beta_{\nu}}$. However, looking at the excitation energy \eqref{eq:exc_single}, it seems that this excitation instead occurs from $\ket{\alpha_{1 - \mu}}$ to $\ket{\beta_{1 - \nu}}$. Why does this happen? In $\ket{\alpha_{-}}$, there are configurations with either $\Sp{\alpha_{\mu}}$ or $\Sp{ \alpha_{1 - \mu} }$ but not both. The singlet-excitation promotes $\Sp{\alpha_{\mu}}$ to $\Ap{\beta_{\nu} \alpha_{\mu}}$ and destroys the configurations including $\Sp{\alpha_{1 - \mu}}$. From the $\beta$th VBS, configurations involving $\Sp{\beta_{\nu}}$ are disallowed by the Pauli principle, so the only surviving configurations must include $\Sp{\beta_{1 - \nu}}$. As a result, the partial occupation of $\ket{\alpha_{1 - \mu}}$ present in $\ket{\PP}$ is totally suppressed, while the spatial orbital $\ket{ \beta_{1 - \nu} }$ has been completely filled. Again, the action of the singlet-excitation is unambiguous, but does not produce the result one would expect from Slater determinants. It is most productive to consider orthonormal states directly.

Now, the single electron-transfers $\ket{2^{\beta_{\nu}}_{\alpha_{\mu}}}$ and $\ket{2_{\beta_{\nu}}^{\alpha_{\mu}}}$ are distinct excitations. These two states are orthonormal, but \emph{both couple to the reference} $\ket{\PP}$ through $\hH_C$
\begin{align}
		(-1)^{\mu + \nu +1} \sqrt{\frac{ 2 n_{\alpha_{\mu}} }{ n_{\beta_{1 - \nu}} }} 
		\braket{\PP | \hH_C | 2^{\beta_{\nu}}_{\alpha_{\mu}}} 
		&=  f_{\alpha_{\mu} \beta_{\nu} } 
		+ n_{\alpha_{\mu}} \sqrt{ \frac{n_{\beta_{\nu}}}{ n_{\beta_{1 - \nu}} } } 
		V_{\beta_{\nu} \beta_{1 - \nu} \alpha_{\mu} \beta_{1 - \nu} } \\
		&\quad + \frac{1}{2} \left( 2 V_{\beta_{1 - \nu} \beta_{1 - \nu} \alpha_{\mu} \beta_{\nu}} 
		- V_{\beta_{1 - \nu} \beta_{\nu} \alpha_{\mu} \beta_{1 - \nu}}
		- V_{\beta_{\nu} \beta_{\nu} \alpha_{\mu} \beta_{\nu}}
		\right) n_{\alpha_{\mu}} n_{\beta_{\nu}}. \nonumber
\end{align}
The orbital optimization does not decouple single excitations, rather it decouples the action of $\Az{pq} - \Az{qp}$. When both indices are in the valence, these two excitations are distinct. To get a state that decouples from the reference, define
\begin{align}
	\ket{2^-_{\alpha_{\mu} \beta_{\nu} }} = \frac{(-1)^{\mu + \nu + 1}}{\sqrt{2}}
	\left(
	\sqrt{ n_{\alpha_{\mu}} n_{\beta_{1-\nu}} } \ket{2^{\beta_{\nu}}_{\alpha_{\mu}}}
	- \sqrt{ n_{\alpha_{1-\mu}} n_{\beta_{\nu}} } \ket{2_{\beta_{\nu}}^{\alpha_{\mu}}}
	\right),
\end{align}
which gives
\begin{align}
	\braket{\PP | \hH_C | 2^-_{\alpha_{\mu} \beta_{\nu} }} = 
	f_{ \alpha_{\mu} \beta_{\nu} } - f_{\beta_{\nu} \alpha_{\mu}} = 0.
\end{align}
However, the orthogonal complement to $\ket{ 2^-_{\alpha_{\mu} \beta_{\nu} } }$,
\begin{align}
	\ket{2^+_{\alpha_{\mu} \beta_{\nu} }} = \frac{1}{\sqrt{2}} \left(
	\sqrt{ n_{\alpha_{1-\mu}} n_{\beta_{\nu}} } \ket{2^{\beta_{\nu}}_{\alpha_{\mu}}}
	+ \sqrt{ n_{\alpha_{\mu}} n_{\beta_{1-\nu}} } \ket{2_{\beta_{\nu}}^{\alpha_{\mu}}}
	\right)
\end{align}
\emph{does} couple to the reference
\begin{equation}
	\begin{split}
		2 (-1)^{\mu + \nu + 1} \eg{\alpha} \eg{\beta} \braket{\PP | \hH_C | 2^+_{\alpha_{\mu} \beta_{\nu} } }
		&= \frac{1}{ n_{\alpha_{\mu}} } f_{ \alpha_{\mu} \beta_{\nu} }
		 + \frac{1}{ n_{\beta_{\nu}} }  f_{ \beta_{\nu}  \alpha_{\mu}} \\
		&+ \frac{1}{2} \left(
		2 V_{ \alpha_{1 - \mu} \alpha_{1 - \mu} \beta_{\nu} \alpha_{\mu} }
		- V_{ \alpha_{1 - \mu} \alpha_{\mu} \beta_{\nu} \alpha_{1 - \mu} } \right. \\
		&- \left. V_{ \alpha_{\mu} \alpha_{\mu} \beta_{\nu} \alpha_{\mu} }
		+ 2 \eg{\alpha} V_{ \alpha_{\mu} \alpha_{1 - \mu} \beta_{\nu} \alpha_{1 - \mu} } 
		\right) n_{\alpha_{\mu}} \\
		&+ \frac{1}{2} \left(
		2 V_{ \beta_{1 - \nu} \beta_{1 - \nu} \alpha_{\mu} \beta_{\nu} }
		- V_{ \beta_{1 - \nu} \beta_{\nu} \alpha_{\mu} \beta_{1 - \nu} } \right. \\
		&- \left. V_{ \beta_{\nu} \beta_{\nu} \alpha_{\mu} \beta_{\nu} }
		+ 2 \eg{\beta} V_{ \beta_{\nu} \beta_{1 - \nu} \alpha_{\mu} \beta_{1 - \nu} }
		\right) n_{\beta_{\nu}}.
	\end{split}
\end{equation}
Both of these states have the same norm
\begin{align}
	\braket{ 2^{\pm}_{\alpha_{\mu} \beta_{\nu} } | 2^{\pm}_{\alpha_{\mu} \beta_{\nu} }} = 
	1 + (-1)^{\mu + \nu + 1} \frac{\omega_{\alpha} \omega_{\beta}}{\eg{\alpha} \eg{\beta}},
\end{align}
while their energies can be computed
\begin{subequations}
\begin{align}
	\begin{split}
		2 \braket{2^-_{\alpha_{\mu} \beta_{\nu} } | \hH_C | 2^-_{\alpha_{\mu} \beta_{\nu} } } &= 
		n_{\alpha_{\mu}} n_{\beta_{1 - \nu}} E[2^{\beta_{\nu}}_{\alpha_{\mu}} ] 
		+ n_{\alpha_{1 - \mu}} n_{\beta_{\nu}} E[2^{\alpha_{\mu}}_{\beta_{\nu}}] \\
		&\quad - \frac{1}{\eg{\alpha} \eg{\beta}} ( L_{\alpha_{1 - \mu} \beta_{1 - \nu}} + L_{\beta_{1 - \nu} \alpha_{1 - \mu}} )
	\end{split} \\
	\begin{split}
		2 \braket{2^+_{\alpha_{\mu} \beta_{\nu} } | \hH_C | 2^+_{\alpha_{\mu} \beta_{\nu} } } &= 
		n_{\alpha_{1 - \mu}} n_{\beta_{\nu}} E[2^{\beta_{\nu}}_{\alpha_{\mu}} ] 
		+ n_{\alpha_{\mu}} n_{\beta_{1 - \nu}} E[2^{\alpha_{\mu}}_{\beta_{\nu}}] \\
		&\quad + \frac{1}{\eg{\alpha} \eg{\beta}} ( L_{\alpha_{1 - \mu} \beta_{1 - \nu}} + L_{\beta_{1 - \nu} \alpha_{1 - \mu}} ).
	\end{split}
\end{align}
\end{subequations}
In HF, it is understood that the orbital optimization decouples the reference from its single excitations. Here, a linear combination of these two states decouples from the reference, while its orthogonal complement does not. The computational effort would be reduced by using the states $\ket{ 2^{\pm}_{\alpha_{\mu}\beta_{\nu}} }$, but it would not affect the scaling, and makes the treatment less clean. Presently, the choice doesn't really matter. When building the Hamiltonian matrix in these states to perform a CI, it will become clear which choice is to be preferred.

\subsubsection{Doubles}

The only seniority-zero doubles are the double-swaps
\begin{align}
	\ket{0^{\alpha\beta}_{\alpha\beta}} = \SpG{\alpha_+} \SpG{\beta_+} \ket{\PP_{\alpha\beta}}
\end{align}
where the occupations have been swapped in two distinct VBS: $n[0^{\alpha\beta}_{\alpha\beta}]_{\alpha_{\mu}} = n_{\alpha_{1-\mu}}$ and $n[0^{\alpha\beta}_{\alpha\beta}]_{\beta_{\nu}} = n_{\beta_{1-\nu}}$. The corresponding pair-transfer 2-RDM elements likewise have changed phases: $P[0^{\alpha\beta}_{\alpha\beta}]_{\ger{\alpha} \ung{\alpha} } = + \frac{1}{\eg{\alpha}}$ and 
$P[0^{\alpha\beta}_{\alpha\beta}]_{\ger{\beta} \ung{\beta} } = + \frac{1}{\eg{\beta}}$. The direct/exchange 2-RDM elements do not change symbolically. The excitation energy may be written in a couple of different ways
\begin{subequations}
\begin{align}
	E[0^{\alpha\beta}_{\alpha\beta}] - E[\PP] &= 2 \eg{\alpha} L_{\ger{\alpha} \ung{\alpha}}
	+ 2 \eg{\beta} L_{\ger{\beta} \ung{\beta}} 
	+ 4 \gtwo{\alpha}{\beta} \\
	&= 2 \eg{\alpha}^4 \frac{\partial^2 E}{\partial \omega_{\alpha}^2}
	+  2 \eg{\beta}^4  \frac{\partial^2 E}{\partial \omega_{\beta}^2} 
	+  4 \omega_{\alpha} \omega_{\beta} \eg{\alpha}^2 \eg{\beta}^2 \frac{\partial^2 E}{\partial \omega_{\alpha} \partial \omega_{\beta}},
\end{align}
\end{subequations}
or as one might expect
\begin{align}
	E[0^{\alpha\beta}_{\alpha\beta}] - E[\PP] &= (E[0^{\alpha}_{\alpha}] - E[\PP])
		+ (E[0^{\beta}_{\beta}] - E[\PP])
		+ 4 \gtwo{\alpha}{\beta}.
\end{align}
The first and second terms are the excitation energies from swaps in the $\alpha$th and $\beta$th VBS, while the last term removes the double-counting. Such a correction is also present for Slater determinants in an EN scheme. The mixed second derivatives \emph{may be negative} but the diagonal second derivatives should dominate. The transitions
\begin{subequations}
\begin{align}
	\braket{\PP | \hH_C | 0^{\alpha\beta}_{\alpha\beta} } &= 
	\frac{1}{2} \frac{1}{\eg{\alpha}\eg{\beta}} \sum_{\mu\nu} (-1)^{\mu+\nu} G_{\alpha_{\mu} \beta_{\nu}} \\
	&= \eg{\alpha}^2 \eg{\beta}^2 \frac{\partial^2 E}{\partial \omega_{\alpha} \partial \omega_{\beta}}
\end{align}
\end{subequations}
are small, but not necessarily positive.

Next, a swap in $\alpha$ can accompany a split in $\beta$
\begin{align}
	\ket{2^{\alpha \beta}_{\alpha \beta}} &= \Ap{\ger{\beta} \ung{\beta}} \SpG{\alpha_{+}} \ket{\PP_{\alpha\beta}}
	=   \frac{(-1)^{\nu}}{\sqrt{n_{\beta_{\nu}}}} \Az{ \beta_{1-\nu} \beta_{\nu} } \ket{0^{\alpha}_{\alpha}}
\end{align}
for either choice of $\nu = 0,1$. The notation is ambiguous, so it is chosen that the first subscript/superscript corresponds to the swap while the second indices correspond to the split. The excitation energy is
\begin{subequations}
\begin{align}
	\begin{split}
		E[2^{\alpha\beta}_{\alpha\beta}] - E[\PP] &=
		2 \eg{\alpha} L_{\ger{\alpha} \ung{\alpha}} + \eg{\beta} L_{\ger{\beta} \ung{\beta}}
			+ \twamp{\ger{\beta} \ung{\beta}} + 2\gtwo{\alpha}{\beta},
	\end{split}
	\\
	&= (E[0^{\alpha}_{\alpha}] - E[\PP])
	+ (E[2^{\beta}_{\beta}] - E[\PP])
	+ 2 \gtwo{\alpha}{\beta}.
\end{align}
\end{subequations}
The transition is
\begin{subequations}
\begin{align}
	\braket{\PP | \hH_C | 2^{\alpha \beta}_{\alpha \beta}} &=
	\frac{1}{2 \eg{\alpha}} \left( \sqrt{n_{\ger{\beta}}} - \sqrt{ n_{\ung{\beta}} } \right)
	\sum_{\mu} (-1)^{\mu} \left(
	2 V_{ \alpha_{\mu} \alpha_{\mu} \ger{\beta} \ung{\beta} }
	- V_{ \alpha_{\mu} \ung{\beta} \ger{\beta} \alpha_{\mu} }
	\right) \\
	&= \frac{1}{2} \frac{\eg{\alpha}^2}{ \sqrt{n_{\ger{\beta}}} + \sqrt{n_{\ung{\beta}}} } 
	\frac{\partial^2 E}{ \partial \omega_{\alpha} \partial \kappa_{\ger{\beta} \ung{\beta}} }.
\end{align}
\end{subequations}

Splits can occur in two different VBS, leading to the two excited states
\begin{subequations}
	\begin{align}
		\ket{\fone{\alpha\beta}{\alpha\beta}} &= \Ap{\ger{\alpha} \ung{\alpha}} \Ap{\ger{\beta} \ung{\beta}} \ket{\PP_{\alpha\beta}} \\
		\ket{\ftwo{\alpha\beta}{\alpha\beta}} &= \stwo{\ger{\alpha} \ung{\alpha}}{\ger{\beta} \ung{\beta}} \ket{\PP_{\alpha\beta}}.
	\end{align}
\end{subequations}
This is the first seniority-four state discussed, so its RDM elements will be highlighted. First, the occupation numbers in each of the blocked orbitals are
\begin{align}
	n[\fone{\alpha\beta}{\alpha\beta}]_{ \alpha_{\mu} } = 
	n[\fone{\alpha\beta}{\alpha\beta}]_{ \beta_{\nu}  } = 1,
\end{align}
with the same values for the second state. The direct elements in the blocked orbitals are all 1 for both states. The exchange elements on the other hand are
\begin{align}
	  X[\fone{\alpha \beta}{\alpha \beta}]_{\ger{\alpha} \ung{\alpha}} = 
	  X[\fone{\alpha \beta}{\alpha \beta}]_{\ger{\beta}  \ung{\beta}}  =
	- X[\ftwo{\alpha \beta}{\alpha \beta}]_{\ger{\alpha} \ung{\alpha}} = 
	- X[\ftwo{\alpha \beta}{\alpha \beta}]_{\ger{\beta}  \ung{\beta}}  = 1
\end{align}
and
\begin{align}
	  X[\fone{\alpha \beta}{\alpha \beta}]_{ \alpha_{\mu} \beta_{\nu} } = 
	- X[\ftwo{\alpha \beta}{\alpha \beta}]_{ \alpha_{\mu} \beta_{\nu} } = 
	- \frac{1}{2}.
\end{align}
The pair elements are all zero between blocked levels, including the diagonal elements.

The excitation energy of the first state is
\begin{subequations}
\begin{align}
\begin{split}
	E[4^{\alpha\beta}_{\alpha\beta}] - E[\PP] &= 
		\eg{\alpha} L_{\ger{\alpha} \ung{\alpha}} + \eg{\beta} L_{\ger{\beta} \ung{\beta}}
			+ \twamp{\ger{\alpha} \ung{\alpha}} + \twamp{\ger{\beta}  \ung{\beta} }
			+ \gtwo{\alpha}{\beta}
\end{split}
	\\
	&=   (E[2^{\alpha}_{\alpha}] - E[\PP])
	+ (E[2^{\beta }_{\beta }] - E[\PP])
	+ \gtwo{\alpha}{\beta}.
\end{align}
\end{subequations}
The excitation energy of the second state is likewise computed, but it is more efficient to view it as an update to the excitation energy of the first state. The RDM elements of the two seniority-four states differ only in their exchange elements, so that
\begin{equation}
\begin{split}
	E[\bar{4}^{\alpha\beta}_{\alpha\beta}] - E[\PP] &= E[4^{\alpha\beta}_{\alpha\beta}] - E[\PP] \\
		&\quad + K_{\ger{\alpha} \ger{\beta}}  + K_{\ger{\alpha} \ung{\beta}} 
		       + K_{\ung{\alpha} \ger{\beta}}  + K_{\ung{\alpha} \ung{\beta}} \\
		&\quad -2K_{\ger{\alpha} \ung{\alpha}} -2K_{\ger{\beta}  \ung{\beta}}.
\end{split}
\end{equation}
Such a result holds between any of the pairs of seniority-four states, though as the notation is imperfect, this will be highlighted in each case. The transitions are
\begin{subequations}
\begin{align}
	\begin{split}
	\braket{ \PP | \hH_C | \fone{\alpha\beta}{\alpha\beta}} &=
	\left( \sqrt{n_{\ger{\alpha}}} - \sqrt{ n_{\ung{\alpha}} } \right)
	\left( \sqrt{n_{\ger{\beta}}}  - \sqrt{ n_{\ung{\beta}} } \right)
	V_{ \ger{\alpha} \ung{\alpha} \ger{\beta} \ung{\beta} } \\
	&\quad - \frac{1}{2} \left( \sqrt{ n_{\ger{\alpha}} n_{\ger{\beta}} } + \sqrt{ n_{\ung{\alpha}} n_{\ung{\beta}} } \right)
	V_{ \ger{\alpha} \ung{\beta} \ger{\beta} \ung{\alpha} } \\
	&\quad + \frac{1}{2} \left( \sqrt{ n_{\ung{\alpha}} n_{\ger{\beta}} } + \sqrt{ n_{\ger{\alpha}} n_{\ung{\beta}} } \right)
	V_{ \ung{\alpha} \ung{\beta} \ger{\beta} \ger{\alpha} }
	\end{split} \\
	\begin{split} \label{eq:conjugate_dble_split}
	\braket{ \PP | \hH_C | \ftwo{\alpha\beta}{\alpha\beta}} &=
	- \frac{\sqrt{3}}{2} \left( \sqrt{ n_{\ger{\alpha}} n_{\ger{\beta}} } + \sqrt{ n_{\ung{\alpha}} n_{ \ung{\beta} } } \right)
	V_{ \ger{\alpha} \ung{\beta} \ger{\beta} \ung{\alpha} } \\
	&\quad
	- \frac{\sqrt{3}}{2} \left( \sqrt{ n_{\ung{\alpha}} n_{\ger{\beta}} } + \sqrt{ n_{\ger{\alpha}} n_{ \ung{\beta} } } \right)
	V_{ \ung{\alpha} \ung{\beta} \ger{\beta} \ger{\alpha} }.
	\end{split}
\end{align}
\end{subequations}
It will be convenient to refer to these states individually. The first state $\ket{\fone{\alpha\beta}{\alpha\beta}}$ will be called a double-split while the second state $\ket{\ftwo{\alpha\beta}{\alpha\beta}}$ will be called a \emph{complementary double-split}.

A swap coupled with an electron-transfer leads to the seniority-two double excitation
\begin{align}
	\ket{2^{\alpha \gamma_{\lambda}}_{\alpha \beta_{\nu}} } &= 
	\Ap{\beta_{\nu} \gamma_{\lambda}} \Sp{\gamma_{1-\lambda}} \SpG{\alpha_+} \ket{\PP_{\alpha\beta\gamma}}
	= (-1)^{\nu + \lambda +1} \sqrt{\frac{ 2 }{ n_{\beta_{\nu}} n_{\gamma_{1-\lambda}} }} \Az{\gamma_{\lambda} \beta_{\nu}}
	\ket{0^{\alpha}_{\alpha}},
\end{align}
whose occupation numbers have been swapped in the $\alpha$th VBS $n[2^{\alpha \gamma_{\lambda}}_{\alpha \beta_{\nu}}]_{ \alpha_{\mu} } = n_{\alpha_{1 - \mu}}$, and otherwise modified $n[ 2^{\alpha \gamma_{\lambda}}_{\alpha \beta_{\nu}} ]_{\beta_{\nu}} = n[2^{\alpha \gamma_{\lambda}}_{\alpha \beta_{\nu}} ]_{\gamma_{\lambda}} =1$, $n[2^{\alpha \gamma_{\lambda}}_{\alpha \beta_{\nu}} ]_{\beta_{1 - \nu}} = 0$, and $n[2^{\alpha \gamma_{\lambda}}_{\alpha \beta_{\nu}} ]_{\gamma_{1 - \lambda}} = 2$. The exchange element between the blocked levels $X[2^{\alpha \gamma_{\lambda}}_{\alpha \beta_{\nu}} ]_{\beta_{\nu} \gamma_{\lambda}} = 1$, whereas the rest of the direct/exchange elements remain products of the corresponding occupation numbers. Pair elements in the $\alpha$th VBS have the opposite phase $P[2^{\alpha \gamma_{\lambda}}_{\alpha \beta_{\nu}} ]_{\ger{\alpha} \ung{\alpha}} = + \frac{1}{\eg{\alpha}}$, pair elements in the $\beta$th are all zero, while there is one non-zero pair element in the $\gamma$th VBS $P[2^{\alpha \gamma_{\lambda}}_{\alpha \beta_{\nu}} ]_{\gamma_{1 - \lambda} \gamma_{1 - \lambda}} = 2$. This gives the excitation energy
\begin{subequations}
\begin{align}
\begin{split}
	E[ 2^{\alpha \gamma_{\lambda}}_{\alpha \beta_{\nu}} ] - E[\PP] &=
		2 \eg{\alpha} L_{\ger{\alpha} \ung{\alpha}} 
		+ \eg{\beta}  L_{\ger{\beta}  \ung{\beta} }
		+ \eg{\gamma} L_{\ger{\gamma} \ung{\gamma}}
		+ \twamp{\beta_{\nu} \gamma_{\lambda}}
		+ \varepsilon_{\gamma_{1-\lambda}}
		- \varepsilon_{\beta_{1-\nu}}
		+ G_{\ger{\gamma} \ung{\gamma}} \\
		&\quad 
		+ 2 \gtwo{\alpha}{\beta} 
		+ 2 \gtwo{\alpha}{\gamma}
		+ \gtwo{\beta}{\gamma} 
		- 2 \gone{\alpha}{\gamma_{1-\lambda}}
		- \gone{\beta}{\gamma_{1-\lambda}}
		+ 2 \gone{\alpha}{\beta_{1-\nu}}
		+ \gone{\gamma}{\beta_{1-\nu}}
		\\
		&\quad 
		- \frac{1}{2} G_{\beta_{1-\nu} \gamma_{1-\lambda}} 
\end{split}
\\
\begin{split}
&= (E[ 0^{\alpha}_{\alpha}] - E[\PP]) +  ( E[2^{\gamma_{\lambda}}_{\beta_{\nu}} ] -E[\PP] ) \\
&\quad + 2 \gtwo{\alpha}{\beta} + 2 \gtwo{\alpha}{\gamma}
	+ 2 \gone{\alpha}{\beta_{1-\nu}} - 2 \gone{\alpha}{\gamma_{1-\lambda}}.
\end{split}
\end{align}
\end{subequations}
The transition
\begin{align}
	\braket{\PP | \hH_C | 2^{\alpha \gamma_{\lambda}}_{\alpha \beta_{\nu}} } &=
	\frac{(-1)^{\nu + \lambda +1}}{2 \sqrt{2} \eg{\alpha}} \sqrt{n_{\beta_{\nu}} n_{\gamma_{1 - \lambda}} }
	\sum_{\mu} (-1)^{\mu} \sqrt{n_{\alpha_{\mu}}} ( 2V_{\beta_{\nu} \gamma_{\lambda} \alpha_{\mu} \alpha_{\mu}} - V_{\alpha_{\mu} \gamma_{\lambda} \beta_{\nu} \alpha_{\mu}} )
\end{align}
is in principle related to a second derivative of the energy, but to get a clear relationship, the single electron-transfer must be treated as $\ket{2^-_{\beta_{\nu} \gamma_{\lambda}} }$. Frankly, it is more important to have a clean set of states than a clear connection to a second derivative.

The corresponding split may be combined with a single electron-transfer
\begin{align}
	\ket{ \fone{\alpha \gamma_{\lambda}}{\alpha \beta_{\nu}} } &= \Ap{\ger{\alpha} \ung{\alpha}} \Ap{\beta_{\nu} \gamma_{\lambda}}
	\Sp{\gamma_{1-\lambda}} \ket{\PP_{\alpha\beta\gamma}} 
	= \frac{(-1)^{\nu + \lambda + 1}}{ \sqrt{n_{\beta_{\nu}} n_{\gamma_{1-\lambda}} } } \Az{\gamma_{\lambda} \beta_{\nu}} \ket{2^{\alpha}_{\alpha}}
\end{align}
yielding the excitation energy
\begin{subequations}
\begin{align}
\begin{split}
	E[\fone{\alpha \gamma_{\lambda}}{\alpha \beta_{\nu}}] - E[\PP] &=
		  \eg{\alpha} L_{\ger{\alpha} \ung{\alpha}}
		+ \eg{\beta}  L_{\ger{\beta}  \ung{\beta} }
		+ \eg{\gamma} L_{\ger{\gamma} \ung{\gamma}}
		+ \twamp{\ger{\alpha} \ung{\alpha}}
		+ \twamp{\beta_{\nu} \gamma_{\lambda}} \\
		&\quad + \varepsilon_{\gamma_{1-\lambda}} 
		- \varepsilon_{\beta_{1-\nu}}
		+ G_{\ger{\gamma} \ung{\gamma}}
		\\
		&\quad + \gtwo{\alpha}{\beta} + \gtwo{\alpha}{\gamma} + \gtwo{\beta}{\gamma}
			- \gone{\alpha}{\gamma_{1-\lambda}}
			- \gone{\beta}{\gamma_{1-\lambda}}
			+ \gone{\alpha}{\beta_{1-\nu}}
			+ \gone{\gamma}{\beta_{1-\nu}}		
		\\	
		&\quad 
		- \frac{1}{2} G_{\beta_{1-\nu} \gamma_{1-\lambda}} 
\end{split}
	\\
\begin{split}
	&=
	(E[2^{\alpha}_{\alpha} ] - E[\PP] ) + (E[2^{\gamma_{\lambda}}_{\beta_{\nu}} ] - E[\PP])  \\
	&\quad
	+ \gtwo{\alpha}{\beta} + \gtwo{\alpha}{\gamma}
	+ \gone{\alpha}{\beta_{1-\nu}} - \gone{\alpha}{\gamma_{1-\lambda}}.	
\end{split}
\end{align}
\end{subequations}
and the transition
\begin{align}
	\braket{\PP | \hH_C | \fone{\alpha \gamma_{\lambda}}{\alpha \beta_{\nu}}} &=
	\frac{(-1)^{\nu + \lambda + 1}}{2 \sqrt{2}} \sqrt{n_{\beta_{\nu}} n_{\gamma_{1 - \lambda}} }
	\sum_{\mu} (-1)^{\mu} \sqrt{n_{\alpha_{\mu}}} 
	(2V_{\beta_{\nu} \gamma_{\lambda} \alpha_{\mu} \alpha_{1 - \mu}} - V_{\alpha_{\mu} \gamma_{\lambda} \beta_{\nu} \alpha_{1 - \mu}} ).
\end{align}
Again, this transition is in principle related to a second derivative of the energy, but a clean orthonormal basis is the priority. The simplest basis for seniority-four states has already been chosen.

The second seniority-four state corresponding to a split combined with an electron-transfer is
\begin{subequations}
	\begin{align}
		\ket{ \ftwo{\alpha \gamma_{\lambda}}{\alpha \beta_{\nu}} } &= \stwo{\ger{\alpha} \ung{\alpha}} { \beta_{\nu} \gamma_{\lambda}} \Sp{\gamma_{1-\lambda}}
		\ket{\PP_{\alpha\beta\gamma}}  \\
		&= (-1)^{\nu + \lambda} \sqrt{ \frac{2}{ 3 n_{\beta_{\nu}} n_{\gamma_{1-\lambda}} } } 
		\left(
		\frac{1}{\sqrt{n_{\ger{\alpha}}}} \Az{\gamma_{\lambda} \ger{\alpha}} \Az{\ung{\alpha} \beta_{\nu}} 
		+ \frac{1}{\sqrt{n_{\ung{\alpha}}}} \Az{\gamma_{\lambda} \ung{\alpha}} \Az{\ger{\alpha} \beta_{\nu}}
		\right) \ket{\PP}.
	\end{align}
\end{subequations}
The excitation energy is again an update to that of the first state
\begin{equation}
	\begin{split}
		E[\ftwo{\alpha \gamma_{\lambda}}{\alpha \beta_{\nu}}] - E[\PP] &=
		E[\fone{\alpha \gamma_{\lambda}}{\alpha \beta_{\nu}}] - E[\PP] \\
		&\quad + K_{\ger{\alpha} \beta_{\nu}} + K_{\ger{\alpha} \gamma_{\lambda} }
		       + K_{\ung{\alpha} \beta_{\nu}} + K_{\ung{\alpha} \gamma_{\lambda} } \\
		&\quad - 2 K_{\ger{\alpha} \ung{\alpha}} - 2 K_{\beta_{\nu} \gamma_{\lambda}}, 
	\end{split}
\end{equation}
while its coupling to PP through $\hH_C$ is different
\begin{align}
	\braket{\PP | \hH_C | \ftwo{\alpha \gamma_{\lambda}}{ \alpha \beta_{\nu} } } &=
	\frac{(-1)^{\nu + \lambda}}{2} \sqrt{\frac{3}{2}} \sqrt{ n_{\beta_{\nu}} n_{\gamma_{1 - \lambda}} }
	\sum_{\mu} \sqrt{ n_{\alpha_{\mu}} } V_{ \alpha_{\mu} \gamma_{\lambda} \beta_{\nu} \alpha_{1 - \mu} }.
\end{align}

The seniority-four state closest to a ``double electron-transfer'' is
\begin{subequations}
	\begin{align} \label{eq:s4_de}
		\ket{ \ftwo{\gamma_{\lambda} \delta_{\kappa}}{ \alpha_{\mu} \beta_{\nu} } } &= 
		\stwo{\gamma_{\lambda} \delta_{\kappa}}{ \alpha_{\mu} \beta_{\nu}} 
		\Sp{\gamma_{1-\lambda}} \Sp{\delta_{1-\kappa}} \ket{\PP_{\alpha\beta\gamma\delta}} \\
		&= \frac{2 (-1)^{\mu + \nu + \gamma + \lambda}}{ \sqrt{3 n_{\alpha_{\mu}} n_{\beta_{\nu}} n_{\gamma_{1-\lambda}} n_{\delta_{1-\kappa}} } }
		\left(
		\Az{\gamma_{\lambda} \alpha_{\mu}} \Az{\delta_{\kappa} \beta_{\nu}} - \Az{\gamma_{\lambda} \beta_{\nu}} \Az{\delta_{\kappa} \alpha_{\mu}}
		\right) \ket{\PP}
	\end{align}
\end{subequations}
as one can see the action of direct and exchange singlet excitations on top of PP. However, its excitation energy does not simplify to a sum of two single-excitation energies plus an update. It is better understood as an update to the corresponding seniority-four state, 
\begin{align}
	\begin{split}
		E[\ftwo{\gamma_{\lambda} \delta_{\kappa}}{\alpha_{\mu} \beta_{\nu}}] - E[\PP] &=
		E[\fone{\gamma_{\lambda} \delta_{\kappa}}{\alpha_{\mu} \beta_{\nu}}] - E[\PP] \\
		&\quad + K_{\alpha_{\mu} \gamma_{\lambda}} + K_{\alpha_{\mu} \delta_{\kappa}} 
		+ K_{\beta_{\nu}  \gamma_{\lambda}} + K_{\beta_{\nu}  \delta_{\kappa}} \\
		&\quad - 2 K_{\alpha_{\mu} \beta_{\nu}} - 2 K_{\gamma_{\lambda} \delta_{\kappa}}
	\end{split}
\end{align}
which is a pair-transfer treated in the next section.

\subsubsection{Pair-Transfers}
In seniority-zero, a pair can be transferred from one VBS to another
\begin{align}
	\ket{0^{\beta\beta}_{\alpha\alpha}} = \Sp{\ger{\beta}} \Sp{\ung{\beta}} \ket{\PP_{\alpha\beta}},
\end{align}
which modifies the occupation numbers $n[0^{\beta\beta}_{\alpha\alpha}]_{\alpha_{\mu}}=0$ and $n[0^{\beta\beta}_{\alpha\alpha}]_{\beta_{\nu}}=2$. The pair-transfer elements in both modified VBS are zero $P[0^{\beta\beta}_{\alpha\alpha}]_{\ger{\alpha} \ung{\alpha}} = P[0^{\beta\beta}_{\alpha\alpha}]_{\ger{\beta} \ung{\beta}} =0$. The direct/exchange elements do not change symbolically, except that the element $D[0^{\beta\beta}_{\alpha\alpha}]_{\ger{\beta} \ung{\beta}} = 4$ is not zero. These states can be accessed through several uncorrelated excitations ($\mu,\nu$ can be either 0 or 1),
\begin{align}
	\ket{0^{\beta\beta}_{\alpha\alpha}} &= \frac{2 (-1)^{\mu + \nu +1} }{\sqrt{ n_{\alpha_{\mu}} n_{\beta_{1-\nu}} }}
	\Sp{\beta_{\nu}} \Sm{\alpha_{\mu}} \ket{\PP}.
\end{align}
Why belabour this point? \emph{Because the idea that a pair of electrons is transferred from one orbital to another is insufficiently precise.} A pair of electrons has been transferred from one VBS to another, and the states should be understood as correlated. A simple description with second quantization is not good enough. 

The excitation energies has been computed in ref. \citenum{johnson:2025d}
\begin{equation}
	\begin{split}
		E[0^{\beta\beta}_{\alpha\alpha}] - E[\PP] &= 
		\eg{\alpha} L_{\ger{\alpha} \ung{\alpha}} 
		+ \eg{\beta} L_{\ger{\beta} \ung{\beta }}
		+ \varepsilon_{\ger{\beta}} + \varepsilon_{ \ung{\beta} } 
		- \varepsilon_{\ger{\alpha}} - \varepsilon_{ \ung{\alpha}}
		+ 2G_{\ger{\beta} \ung{\beta}}
		\\
		&\quad - \gzer{\alpha}{\beta} + \gtwo{\alpha}{\beta}
		- \gone{\alpha}{ \ger{\beta}  } - \gone{\alpha}{ \ung{\beta}  }
		+ \gone{\beta }{ \ger{\alpha} } + \gone{\beta }{ \ung{\alpha} },
	\end{split}
\end{equation}
which involves breaking the bonds in VBS $\alpha$ and $\beta$, and transferring two electrons from $\alpha$ to $\beta$. This results in direct/exchange integrals in VBS $\beta$, and there are corrections for double-counting. The transition was also computed in ref. \citenum{johnson:2025d}
\begin{align}
	\braket{ \PP | \hH_C | 0^{\beta\beta}_{\alpha\alpha} } = \frac{1}{2} \sum_{\mu\nu} (-1)^{\mu+\nu+1} L_{\alpha_{\mu} \beta_{\nu}}
	\sqrt{ n_{\alpha_{\mu}} n_{\beta_{1 - \nu}} }.
\end{align}

Seniority-two pair-transfers occur in two different ways (for each, $\lambda$ is either 0 or 1),
\begin{subequations}
	\begin{align}
		\ket{2^{\gamma \gamma}_{\alpha_{\mu} \beta_{\nu}}} &= \Ap{\alpha_{\mu} \beta_{\nu}} \Sp{\ger{\gamma}} \Sp{\ung{\gamma}}
		\ket{\PP_{\alpha\beta\gamma}} 
		= \frac{2 \sqrt{2} (-1)^{\mu+\nu+\lambda} }{\sqrt{ n_{\alpha_{\mu}} n_{\beta_{\nu}} n_{\gamma_{1-\lambda} } }} 
		\Sp{\gamma_{\lambda}} \Am{ \alpha_{\mu} \beta_{\nu} }  \ket{\PP} \\	
		\ket{2^{\alpha_{\mu} \beta_{\nu}}_{\gamma \gamma}} &= \Ap{\alpha_{\mu} \beta_{\nu}} \Sp{\alpha_{1-\mu}} \Sp{\beta_{1-\nu}}
		\ket{\PP_{\alpha\beta\gamma}}
		= \frac{2 \sqrt{2} (-1)^{\mu + \nu + \lambda} }{ \sqrt{n_{\alpha_{1 - \mu}}  n_{\beta_{1-\nu}} n_{\gamma_{\lambda}} } } 
		\Ap{ \alpha_{\mu} \beta_{\nu} } \Sm{\gamma_{\lambda}} 
		\ket{\PP}.
	\end{align}
\end{subequations}
With the quantities
\begin{subequations}
\begin{align}
	\begin{split}
	B[\alpha_{\mu}, \beta_{\nu}, \gamma] &= \eg{\alpha} L_{\ger{\alpha} \ung{\alpha}}
	+ \eg{\beta}  L_{\ger{\beta}  \ung{\beta}}
	+ \eg{\gamma} L_{\ger{\gamma} \ung{\gamma}}
	+ \twamp{\alpha_{\mu} \beta_{\nu}} \\
	&\quad 
		+ \gtwo{\alpha}{\beta} + \gtwo{\alpha}{\gamma} + \gtwo{\beta}{\gamma}
		+ \frac{1}{2} G_{\alpha_{1 - \mu} \beta_{1-\nu}} \\
	&\quad
		-\frac{1}{2} G_{\alpha_{1 - \mu} \ger{\gamma}}
		- \frac{1}{2} G_{\alpha_{1 - \mu} \ung{\gamma}}
		- \frac{1}{2} G_{ \beta_{1 - \nu} \ger{\gamma}}
		- \frac{1}{2} G_{ \beta_{1 - \nu} \ung{\gamma}}
	\end{split}
	\\
	\begin{split}
		C[\alpha_{\mu}, \beta_{\nu}, \gamma] &=
			  \varepsilon_{\ger{\gamma}} + \varepsilon_{\ung{\gamma}}
			- \varepsilon_{\alpha_{1 - \mu}} - \varepsilon_{\beta_{1 - \nu}} \\
		&\quad 
			- \gone{\alpha}{\ger{\gamma}} - \gone{\alpha}{\ung{\gamma}}
			- \gone{ \beta}{\ger{\gamma}} - \gone{ \beta}{\ung{\gamma}} \\
		&\quad
			+ \gone{\beta}{\alpha_{1 - \mu}} + \gone{\gamma}{\alpha_{1 - \mu}}
			+ \gone{\alpha}{\beta_{1 - \nu}} + \gone{\gamma}{ \beta_{1 - \nu}},
	\end{split}
\end{align}
\end{subequations}
the excitation energies of the two states are
\begin{subequations}
\begin{align}
	E[2^{\gamma \gamma}_{\alpha_{\mu} \beta_{\nu}}] - E[\PP] &=	
		B[\alpha_{\mu}, \beta_{\nu}, \gamma] + C[\alpha_{\mu}, \beta_{\nu}, \gamma] 
		+ 2 G_{\ger{\gamma} \ung{\gamma}} \\
	E[2_{\gamma \gamma}^{\alpha_{\mu} \beta_{\nu}}] - E[\PP] &=
		B[\alpha_{\mu}, \beta_{\nu}, \gamma] - C[\alpha_{\mu}, \beta_{\nu}, \gamma] 
		+ G_{\ger{\alpha} \ung{\alpha}} + G_{\ger{\beta} \ung{\beta}}.
\end{align}
\end{subequations}
The corresponding transitions are
\begin{subequations}
\begin{align}
	\braket{ \PP | \hH_C | 2^{\gamma\gamma}_{\alpha_{\mu} \beta_{\nu}} } &=
	\frac{1}{2} (-1)^{\mu + \nu +1} 
	\sqrt{ n_{\alpha_{\mu}} n_{\beta_{\nu}} } 
	\left(
	V_{ \alpha_{\mu} \ung{\gamma} \beta_{\nu} \ung{\gamma} } \sqrt{ n_{\ger{\gamma}} }
	- V_{ \alpha_{\mu} \ger{\gamma} \beta_{\nu} \ger{\gamma} } \sqrt{ n_{\ung{\gamma}} }
	\right) \\
	\braket{ \PP | \hH_C | 2^{\alpha_{\mu} \beta_{\nu}}_{\gamma\gamma}  } &=
	\frac{1}{2} (-1)^{\mu + \nu}
	\sqrt{ n_{\alpha_{1 - \mu}} n_{\beta_{1 - \nu}} }
	\left(
	V_{ \ger{\gamma} \alpha_{\mu} \ger{\gamma} \beta_{\nu} } \sqrt{ n_{\ger{\gamma}} }
	- V_{ \ung{\gamma} \alpha_{\mu} \ung{\gamma} \beta_{\nu} } \sqrt{ n_{\ung{\gamma}} }
	\right).
\end{align}
\end{subequations}

Finally, in seniority-four there is the excitation
\begin{align}
	\ket{ \fone{\gamma_{\lambda} \delta_{\kappa}}{ \alpha_{\mu} \beta_{\nu} } } &= 
	\Ap{\alpha_{\mu} \beta_{\nu}} \Ap{\gamma_{\lambda} \delta_{\kappa}}
	\ket{\PP_{\alpha\beta\gamma\delta}}
	= \frac{4 (-1)^{\mu + \nu + \gamma + \lambda + 1}}{\sqrt{ n_{\alpha_{\mu}} n_{\beta_{\nu}} n_{\gamma_{1-\lambda}} n_{\delta_{1-\kappa}} }}
	\Ap{\gamma_{\lambda} \delta_{\kappa}} \Am{\alpha_{\mu} \beta_{\nu}} \ket{\PP}.
\end{align}
The second seniority-four state is \eqref{eq:s4_de}. The excitation energy is
\begin{equation}
	\begin{split}
		E[\fone{\gamma_{\lambda} \delta_{\kappa}}{ \alpha_{\mu} \beta_{\nu} }] - E[\PP] &=
			  \eg{\alpha} L_{\ger{\alpha} \ung{\alpha}}
			+ \eg{\beta}  L_{\ger{\beta}  \ung{\beta}}
			+ \eg{\gamma} L_{\ger{\gamma} \ung{\gamma}}
			+ \eg{\delta} L_{\ger{\delta} \ung{\delta}}
			+\twamp{\alpha_{\mu} \beta_{\nu}}
			+\twamp{\gamma_{\lambda} \delta_{\kappa}}
		\\
		&\quad
			+ \varepsilon_{\gamma_{1 - \lambda}} 
			+ \varepsilon_{\delta_{1 - \kappa}}
			- \varepsilon_{\alpha_{1 - \mu}}
			- \varepsilon_{\beta_{1 - \nu}}
			+ G_{\ger{\gamma} \ung{\gamma}} + G_{\ger{\delta} \ung{\delta}}
		\\
		&\quad 
			+ \gtwo{\alpha}{ \beta} + \gtwo{\alpha}{\gamma} + \gtwo{\alpha}{\delta}
			+ \gtwo{ \beta}{\gamma} + \gtwo{ \beta}{\delta} + \gtwo{\gamma}{\delta} 
		\\
		&\quad
			+ \frac{1}{2} G_{\alpha_{1 - \mu} \beta_{1 - \nu}}
			+ \frac{1}{2} G_{\gamma_{1 - \lambda} \delta_{1 - \kappa}}	
		\\
		&\quad
			- \gone{\alpha}{\gamma_{1 - \lambda}}
			- \gone{ \beta}{\gamma_{1 - \lambda}}
			- \gone{\delta}{\gamma_{1 - \lambda}}
			- \gone{\alpha}{\delta_{1 - \kappa}}
			- \gone{ \beta}{\delta_{1 - \kappa}}
			- \gone{\gamma}{\delta_{1 - \kappa}}
		\\
		&\quad
			+ \gone{ \beta}{\alpha_{1 - \mu}}
			+ \gone{\gamma}{\alpha_{1 - \mu}}
			+ \gone{\delta}{\alpha_{1 - \mu}}
			+ \gone{\alpha}{\beta_{1 - \nu}}
			+ \gone{\gamma}{\beta_{1 - \nu}}
			+ \gone{\delta}{\beta_{1 - \nu}}
		\\
		&\quad	
			- \frac{1}{2} G_{\alpha_{1 - \mu} \gamma_{1 - \lambda}}
			- \frac{1}{2} G_{\alpha_{1 - \mu} \delta_{1 - \kappa}}
			- \frac{1}{2} G_{\beta_{1 - \nu} \gamma_{1 - \lambda}}
			- \frac{1}{2} G_{\beta_{1 - \nu} \delta_{1 - \kappa}},
	\end{split}
\end{equation}
and the transitions are
\begin{subequations}
	\begin{align}
		\braket{ \PP | \hH_C | \fone{\gamma_{\lambda} \delta_{\kappa}}{\alpha_{\mu} \beta_{\nu}} } &=
		\frac{1}{4} (-1)^{\mu + \nu + \lambda + \kappa +1} 
		\sqrt{ n_{\alpha_{\mu}} n_{\beta_{\nu}} n_{\gamma_{1 - \lambda}} n_{\delta_{1 - \kappa}} }
		\left(
		V_{ \alpha_{\mu} \gamma_{\lambda} \beta_{\nu} \delta_{\kappa}  }
		+ V_{ \alpha_{\mu} \delta_{\kappa}  \beta_{\nu} \gamma_{\lambda} }
		\right) \\
		\braket{ \PP | \hH_C | \ftwo{\gamma_{\lambda} \delta_{\kappa}}{ \alpha_{\mu} \beta_{\nu} } } &=
		\frac{\sqrt{3}}{4} (-1)^{\mu + \nu + \lambda + \kappa}
		\sqrt{ n_{\alpha_{\mu}} n_{\beta_{\nu}} n_{\gamma_{1 - \lambda}} n_{\delta_{1 - \kappa}} }
		\left(
		V_{ \alpha_{\mu} \gamma_{\lambda} \beta_{\nu} \delta_{\kappa}  }
		- V_{ \alpha_{\mu} \delta_{\kappa}  \beta_{\nu} \gamma_{\lambda} }
		\right).
	\end{align}
\end{subequations}
This completes the list of low-lying valence excitations which couple to $\ket{\PP}$. The remaining excitations will now be quickly summarized.

\subsection{External Excitations}
For the purposes of this manuscript, external excitations involve either the core or the virtual orbitals (or both). All of these states are simplified cases of those in the valence. As such, while there are a lot of them, they will be treated quickly in passing. The orthonormal basis for the singles and doubles are summarised in Table \ref{tab:ext_sing_doub}. 
\begin{center}
	\begin{table}[h] 
		
		\begin{tabular}{c|c|c|c|c}
			$p,q$ & $\ket{2^q_p}$ & $\ket{2^{\alpha q}_{\alpha p}}$ & $\ket{\fone{\alpha q}{\alpha p}}$ & $\ket{\ftwo{\alpha q}{\alpha p}}$ \\
			\toprule
			$i,a$ & $\Ap{ai} \ket{\PP_{i}}$ & $\Ap{ai} \SpG{\alpha_{+}} \ket{\PP_{\alpha i}} $ &
			$   \Ap{\ger{\alpha} \ung{\alpha}} \Ap{ai} \ket{\PP_{\alpha i}}$ & 
			$ \stwo{\ger{\alpha} \ung{\alpha}}{ai} \ket{\PP_{\alpha i}}$ \\
			$\beta_{\nu}, a$ & $\Ap{a \beta_{\nu}} \ket{\PP_{\beta}}$ & $ \Ap{a \beta_{\nu}} \SpG{\alpha_{+}} \ket{\PP_{\alpha\beta}}$ &
			$   \Ap{\ger{\alpha} \ung{\alpha}} \Ap{a \beta_{\nu}} \ket{\PP_{\alpha\beta}}$ &
			$ \stwo{\ger{\alpha} \ung{\alpha}}{a \beta_{\nu}} \ket{\PP_{\alpha\beta}}$ \\
			$ i, \beta_{\nu}$ & $\Ap{\beta_{\nu} i} \Sp{\beta_{1 - \nu}} \ket{ \PP_{\beta i} }$ &
			$ \Ap{\beta_{\nu} i} \Sp{\beta_{1 - \nu}} \SpG{\alpha_{+}} \ket{ \PP_{\alpha \beta i} } $ &
			$ \Ap{\ger{\alpha} \ung{\alpha}} \Ap{\beta_{\nu} i} \Sp{\beta_{1 - \nu}} \ket{ \PP_{\alpha \beta i} } $ &
			$ \stwo{\ger{\alpha} \ung{\alpha}}{\beta_{\nu} i} \Sp{\beta_{1 - \nu}} \ket{ \PP_{\alpha \beta i} } $
		\end{tabular}
		\caption{Orthonormal basis for external single and double PP excitations.}
		\label{tab:ext_sing_doub}
	\end{table}
\end{center}
As swaps and splits are only allowed in the valence, the only external singles are electron-transfers. Similarly, as double electron-transfers do not exist, the only doubles are swaps plus electron-transfers $\ket{2^{\alpha q}_{\alpha p}}$ or splits plus electron-transfers $\ket{\fone{\alpha q}{\alpha p}}$ and $\ket{\ftwo{\alpha q}{\alpha p}}$. General expressions can be written for these excitations with a sign: core and virtual orbitals never contribute to the sign, but the valence orbitals will. For $\sgn{p,q}$: if $p=\alpha_{\mu}$, it will contribute $(-1)^{\mu}$ to the sign, while if $q=\beta_{\nu}$ it will contribute $(-1)^{\nu+1}$. For $\sgn{pq,rs}$: $p$ and $q$ contribute the parity of their valence orbitals, while $r$ and $s$ contribute the opposite of the parity of their valence orbitals to the sign. The couplings for the external singles and doubles are
\begin{subequations}
	\begin{align}
		\braket{\PP | \hH_C | 2^{q}_{p}} &= \sgn{p,q} \sqrt{\frac{2}{n_p (2-n_q)}} (f_{pq} - f_{qp}) = 0 \\
		\braket{\PP | \hH_C | 2^{\alpha q}_{\alpha p }} &= \frac{\sgn{p,q}}{2 \eg{\alpha}} \sqrt{ \frac{n_p (2-n_q)}{2} }
			\sum_{\mu} (-1)^{\mu} ( 2 V_{pq \alpha_{\mu} \alpha_{\mu}} - V_{\alpha_{\mu} q p \alpha_{\mu}} ) \\
		\braket{\PP | \hH_C | \fone{\alpha q}{\alpha p}} &= \frac{\sgn{p,q}}{2} \sqrt{ \frac{n_p (2-n_q)}{2} } 
			\sum_{\mu} (-1)^{\mu} \sqrt{n_{\alpha_{\mu}}} ( 2 V_{ pq \alpha_{\mu} \alpha_{1 - \mu} } - V_{ \alpha_{\mu} q p \alpha_{1 - \mu} } ) \\
		\braket{\PP | \hH_C | \ftwo{\alpha q}{\alpha p}} &= - \frac{\sgn{p,q}}{2} \sqrt{ \frac{3 n_p (2-n_q)}{2} }
			\sum_{\mu} \sqrt{n_{\alpha_{\mu}}} V_{ \alpha_{\mu} q p \alpha_{1 - \mu} }.
	\end{align}
\end{subequations}
The excitation energies are less simple to summarize, so rather than trying to write an unreadable general case for each class, the explicit expressions are included in the appendix.

The seniority-four pair-transfers also fall into general expressions so they will be presented first. An orthonormal basis for the states is listed in Table \ref{tab:ext_s4_xfer}.
\begin{center}
	\begin{table}[h]

		\begin{tabular}{c|c|c}
			$pq,rs$ & $\ket{\fone{rs}{pq}}$ & $\ket{\ftwo{rs}{pq}}$ \\
			\toprule
			$i j, a b$ & 
			$\Ap{i j} \Ap{a b} \ket{\PP_{i j}}$ & 
			$ \stwo{i j}{a b} \ket{\PP_{i j}} $ \\
			
			$i \alpha_{\mu}, a b $ & 
			$  \Ap{i \alpha_{\mu}} \Ap{a b} \ket{\PP_{\alpha i} }$ &
			$\stwo{i \alpha_{\mu}}{a b} \ket{\PP_{\alpha i} }$ \\
			
			$ \alpha_{\mu} \beta_{\nu}, a b$ &
			$ \Ap{\alpha_{\mu} \beta_{\nu}} \Ap{a b} \ket{\PP_{\alpha \beta}} $  &
			$ \stwo{\alpha_{\mu \beta_{\nu}}}{a b}   \ket{\PP_{\alpha \beta}} $ \\
			
			$ i j, \alpha_{\mu} v$ &
			$   \Ap{i j} \Ap{\alpha_{\mu} a} \Sp{\alpha_{1 - \mu}} \ket{\PP_{\alpha i j}} $ &
			$ \stwo{i j}{\alpha_{\mu} a}     \Sp{\alpha_{1 - \mu}} \ket{\PP_{\alpha i j}} $ \\
			
			$ i \alpha_{\mu}, \beta_{\nu} a$ &
			$ \Ap{i \alpha_{\mu}} \Ap{\beta_{\nu} a} \Sp{\beta_{1 - \nu}} \ket{\PP_{\alpha \beta i}} $ &
			$ \stwo{i \alpha_{\mu}}{\beta_{\nu} a}   \Sp{\beta_{1 - \nu}} \ket{\PP_{\alpha \beta i}} $ \\
			
			$ \alpha_{\mu} \beta_{\nu}, \gamma_{\lambda} a$ &
			$ \Ap{\alpha_{\mu} \beta_{\nu}} \Ap{\gamma_{\lambda} a} \Sp{\gamma_{1 - \lambda}} \ket{\PP_{\alpha \beta \gamma}} $ &
			$ \stwo{\alpha_{\mu} \beta_{\nu}}{\gamma_{\lambda} a}   \Sp{\gamma_{1 - \lambda}} \ket{\PP_{\alpha \beta \gamma}} $\\
			
			$ i j, \alpha_{\mu} \beta_{\nu}$ & 
			$ \Ap{i j} \Ap{\alpha_{\mu} \beta_{\nu}} \Sp{\alpha_{1 - \mu}} \Sp{\beta_{1 - \nu}} \ket{\PP_{ \alpha \beta i j }} $ &
			$ \stwo{i j}{\alpha_{\mu}\beta_{\nu}}    \Sp{\alpha_{1 - \mu}} \Sp{\beta_{1 - \nu}} \ket{\PP_{ \alpha \beta i j }} $ \\
			
			$ i \alpha_{\mu}, \beta_{\nu} \gamma_{\lambda}$ &
			$   \Ap{i \alpha_{\mu}} \Ap{\beta_{\nu} \gamma_{\lambda}} \Sp{\beta_{1 - \nu}} \Sp{\gamma_{1 - \lambda}} \ket{\PP_{\alpha \beta \gamma i}} $ &
			$ \stwo{i \alpha_{\mu}}{\beta_{\nu}\gamma_{\lambda}}      \Sp{\beta_{1 - \nu}} \Sp{\gamma_{1 - \lambda}} \ket{\PP_{\alpha \beta \gamma i}} $
		\end{tabular}
		\caption{Orthonormal basis for external seniority-four pair transfer PP excitations.}
		\label{tab:ext_s4_xfer}
	\end{table}
\end{center}
Their couplings to the PP reference through the Coulomb Hamiltonian have general expressions
\begin{subequations}
	\begin{align}
		\braket{\PP | \hH_C | \fone{rs}{pq}} &= - \sgn{pq,rs} \frac{1}{4} \sqrt{ n_p n_q (2-n_r) (2-n_s) }  (V_{prqs} + V_{psqr} ) \\
		\braket{\PP | \hH_C | \ftwo{rs}{pq}} &= + \sgn{pq,rs} \frac{\sqrt{3}}{4} \sqrt{ n_p n_q (2-n_r) (2-n_s) }  (V_{prqs} - V_{psqr} ).
	\end{align}
\end{subequations}
However, the seniority-zero and seniority-two pair-transfers do not have clean general expressions, so the orthonormal basis along with their couplings to the PP reference are listed in Table \ref{tab:ext_s0_s2_xfer}.
\begin{center}
\begin{table}[h!]
\begin{tabular}{c|c|c}
	& $\ket{\Psi}$ & $\braket{\PP | \hH_C | \Psi}$ \\
	\toprule
	$\ket{0^{aa}_{ii}} $ & $\Sp{a} \ket{\PP_{i}}$ & $L_{ia}$ \\
	
	$\ket{0^{aa}_{\alpha \alpha}}$ & $ \Sp{a} \ket{\PP_{\alpha}}$ &
	$\frac{1}{ \sqrt{2} } \sum_{\mu} (-1)^{\mu} L_{a \alpha_{\mu}  } \sqrt{ n_{\alpha_{\mu}} }$ \\
	
	$\ket{0^{\alpha \alpha}_{ii}}$ & $\Sp{\ger{\alpha}} \Sp{\ung{\alpha}} \ket{\PP_{\alpha i}}$ &
	$ \frac{1}{ \sqrt{2} } \sum_{\mu} (-1)^{\mu +1} L_{i \alpha_{\mu}} \sqrt{ n_{\alpha_{1 - \mu}} } $ \\
	
	\midrule
	$\ket{2^{a b}_{ii}}$ & $ \Ap{a b} \ket{\PP_{i}}$ & $\sqrt{2} V_{i a i b}$ \\

	$\ket{2^{\alpha_{\mu}a}_{ii}}$ & $ \Ap{\alpha_{\mu} a} \Sp{\alpha_{1 - \mu}} \ket{\PP_{\alpha i}} $ &
	$ (-1)^{\mu +1} \sqrt{n_{\alpha_{1 - \mu}}} V_{i a i \alpha_{\mu}} $ \\
	
	$\ket{2^{\alpha_{\mu} \beta_{\nu}}_{ii}}$ & 
	$\Ap{\alpha_{\mu} \beta_{\nu}} \Sp{\alpha_{1 - \mu}} \Sp{\beta_{1 - \nu}} \ket{\PP_{\alpha \beta i}} $ &
	$ (-1)^{\mu + \nu} \sqrt{\frac{ n_{\alpha_{1 - \mu}} n_{\beta_{1 - \nu}} }{2}} V_{ i \alpha_{\mu} i \beta_{\nu} } $ \\
	
	$\ket{2^{a b}_{\alpha \alpha}}$ & $\Ap{a b} \ket{\PP_{\alpha}} $ &
	$ \sum_{\mu} (-1)^{\mu} \sqrt{n_{\alpha_{\mu}}} V_{ \alpha_{\mu} a \alpha_{\mu} b } $ \\
		
	$\ket{2^{\alpha_{\mu}a}_{\beta\beta}}$ & $ \Ap{\alpha_{\mu} a} \Sp{\alpha_{1 - \mu}} \ket{\PP_{\alpha \beta}} $ &
	$(-1)^{\mu + 1} \sqrt{\frac{n_{\alpha_{1 - \mu}}}{2}} 
	\sum_{\nu} (-1)^{\nu} \sqrt{n_{\beta_{\nu}} } V_{ \beta_{\nu} a \beta_{\nu} \alpha_{\mu}} $ \\

	\midrule
	
	$\ket{2^{aa}_{i j}}$ & $ \Ap{i j} \Sp{a} \ket{\PP_{i j} } $ & $-\sqrt{2} V_{i a j a}$ \\
	
	$\ket{2^{aa}_{i\alpha_{\mu}}}$ & $ \Ap{i \alpha_{\mu}} \Sp{v} \ket{\PP_{\alpha i}}$ &
	$(-1)^{\mu +1} \sqrt{n_{\alpha_{\mu}}} V_{i a \alpha_{\mu} a} $ \\
	
	$\ket{2^{aa}_{\alpha_{\mu} \beta_{\nu}}}$ & $\Ap{\alpha_{\mu} \beta_{\nu}} \Sp{a} \ket{\PP_{\alpha\beta}} $ &
	$ (-1)^{\mu + \nu +1} \sqrt{\frac{n_{\alpha_{\mu}} n_{\beta_{\nu}}}{2}} V_{ \alpha_{\mu} a \beta_{\nu} a } $ \\
	
	$\ket{2^{\alpha \alpha}_{i j}}$ & $\Ap{i j} \Sp{\ger{\alpha}} \Sp{\ung{\alpha}} \ket{\PP_{\alpha i j}} $ &
	$ \sum_{\mu} (-1)^{\mu} \sqrt{n_{\alpha_{1 - \mu}}} V_{ i \alpha_{\mu} j \alpha_{\mu} } $ \\
	
	$\ket{2^{\beta\beta}_{i\alpha_{\mu}}}$ &
	$ \Ap{c \alpha_{\mu}} \Sp{\ger{\beta}} \Sp{\ung{\beta}} \ket{\PP_{\alpha \beta i}} $ &
	$(-1)^{\mu +1} \sqrt{\frac{n_{\alpha_{\mu}}}{2}} 
	\sum_{\nu} (-1)^{\nu} \sqrt{n_{\beta_{\nu}}} V_{ i \beta_{1 - \nu} \alpha_{\mu} \beta_{1 - \nu} } $	
	
\end{tabular}
\caption{Transition energies for external pair transfers.}
\label{tab:ext_s0_s2_xfer}
\end{table}
\end{center}

\section{Numerical Results} \label{sec:numbers}
Orbital optimized PP results, hereafter referred to as simply PP, were computed by building the complete Hessian and using the Newton-Raphson (NR) procedure. A trust radius is required on the NR step near equilibrium geometries. EN2 corrections are computed from the expressions in Section \ref{sec:en2} implemented in Python. FCI results in the STO-6G basis were computed with PySCF.\cite{sun:2018} CASSCF results were computed with Gaussian 16 Revision B.01.\cite{gaussian_16}

\begin{figure} 
	\begin{subfigure}{\textwidth}
		\includegraphics[width=0.485\textwidth]{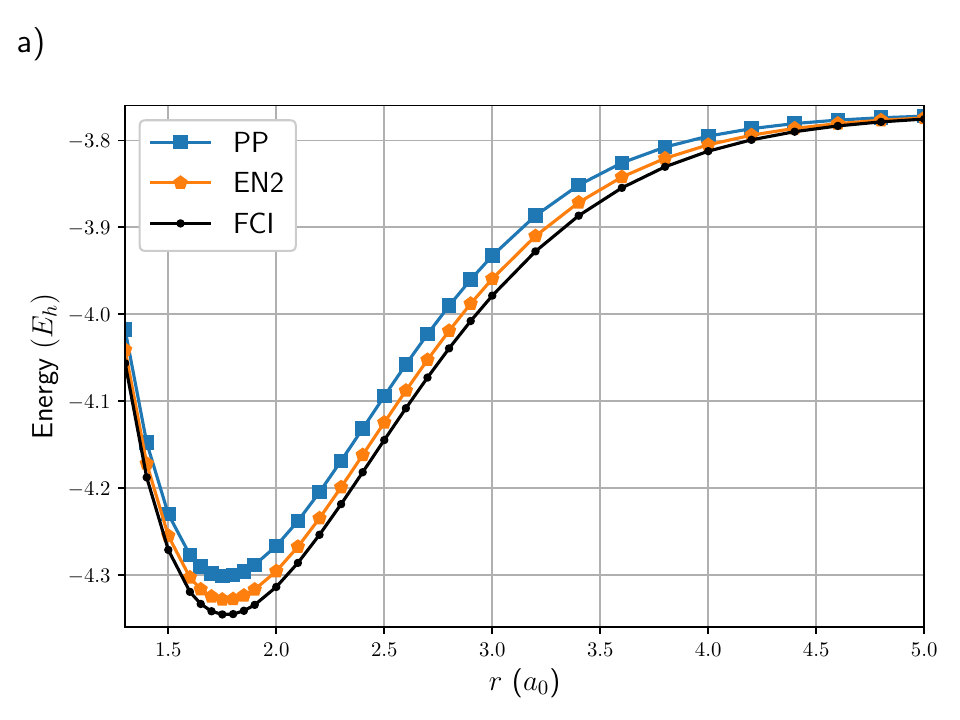}
		\hfill
		\includegraphics[width=0.485\textwidth]{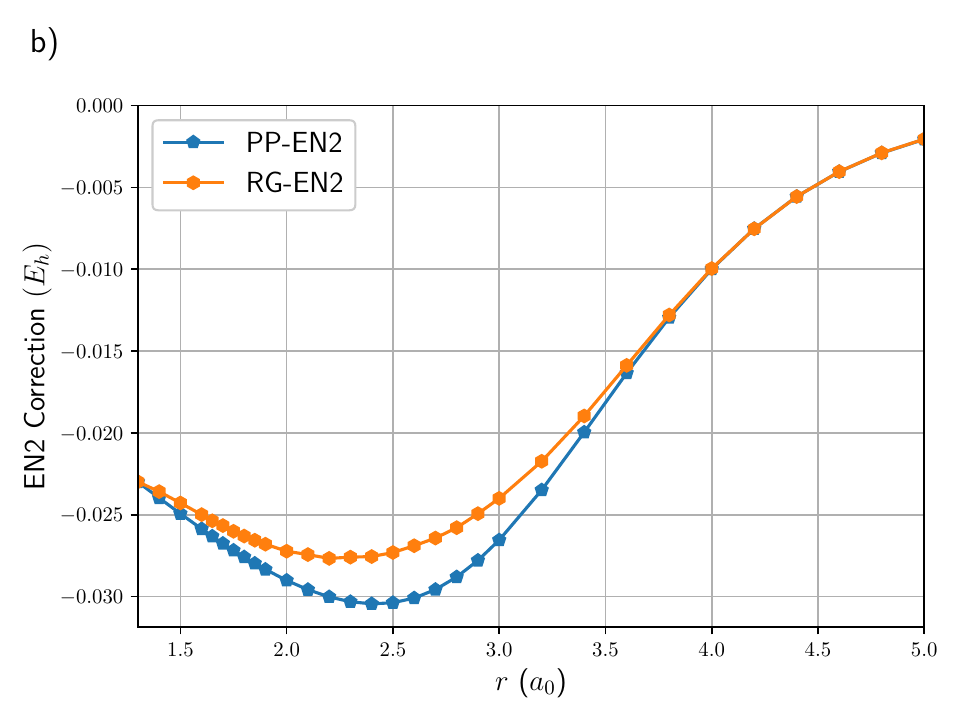}
	\end{subfigure}
	\caption[]{(a) Symmetric bond dissociation of linear H$_8$ in the STO-6G basis: EN2 computed for PP states, compared with optimized PP reference and FCI. PP and RG reference are difficult to discern, as are PP and RG EN2 results. (b) EN2 corrections computed for PP and RG states. RG and PP results are computed in the optimized PP orbitals in the STO-6G basis.}
	\label{fig:h8_rg_pp}
\end{figure}

The point of this contribution is to see how RG states of higher seniority degenerate to the PP case. The first test is thus to compare EN2 results computed with RG states and PP states. EN2 results with RG states were computed for linear equidistant H$_8$ in the minimal basis STO-6G in Part II. To have a clearly comparable set of results, the RG-EN2 correction was recomputed in the optimal PP orbitals. The energies of the optimal PP state and the optimal RG state are indiscernible so only the PP reference energy is shown in Figure \ref{fig:h8_rg_pp}.
The EN2 corrections are substantial improvements, but are still visually discernible from FCI. Treatments beyond EN2 should thus be explored. From Figure \ref{fig:h8_rg_pp} (b), one can see that the EN2 corrections in terms of PP and RG states are not the same, with the PP-EN2 correction out-performing the RG-EN2 correction. This happens for two reasons. First, the list of PP states in section \ref{sec:val_pp_states} is longer than the list of RG states considered in Parts I and II. The only seniority-two singles considered in Part I correspond to $\ket{2^{\ung{\beta}}_{\ger{\alpha}}}$, that is electron-transfers from the bonding orbital of one VBS to the antibonding orbital of another VBS. Part II included the states corresponding to $\ket{2^{\ger{\beta}}_{\ung{\alpha}}}$ which substantially improved the CI results. The additional PP states $\ket{2^{\ger{\beta}}_{\ger{\alpha}}}$ and $\ket{2^{\ung{\beta}}_{\ung{\alpha}}}$ have small contributions to EN2, as do their corresponding doubles $\ket{2^{\alpha \gamma_{\nu}}_{\alpha \beta_{\nu}}}$, $\ket{\fone{\alpha \gamma_{\nu}}{\alpha \beta_{\nu}}}$, and $\ket{\ftwo{\alpha \gamma_{\nu}}{\alpha \beta_{\nu}}}$. The second reason for the observed difference between PP-EN2 and RG-EN2 is the choice of seniority-four states. For RG states, the choice was made to split these states based only on the ordering of their blocked indices. As bonding orbitals are always listed before antibonding orbitals, double-splits are excluded and replaced with the states 
\begin{subequations}
	\begin{align}
		\ket{\fone{\ung{\alpha} \ung{\beta}}{\ger{\alpha} \ger{\beta}}} &= 
			\Ap{\ger{\alpha}\ger{\beta}} \Ap{\ung{\alpha}\ung{\beta}} \ket{\PP_{\alpha\beta}} \\
		\ket{\ftwo{\ung{\alpha} \ung{\beta}}{\ger{\alpha} \ger{\beta}}} &= 
			\stwo{\ger{\alpha}\ger{\beta}}{\ung{\alpha}\ung{\beta}} \ket{\PP_{\alpha\beta}}.
	\end{align}
\end{subequations}
The excitation energies of these states differ from double-splits
\begin{align}
	\begin{split}
	E[\fone{\ung{\alpha}\ung{\beta}}{\ger{\alpha} \ger{\beta}}] - E[\PP] &=
		\eg{\alpha} L_{\ger{\alpha}\ung{\alpha}} + \eg{\beta} L_{\ger{\beta}\ung{\beta}} 
		+ t_{\ger{\alpha}\ger{\beta}} + t_{\ung{\alpha}\ung{\beta}} \\
		&\quad + \frac{1}{2} G_{\ger{\alpha}\ung{\alpha}} + \frac{1}{2} G_{\ger{\beta}\ung{\beta}}
		+ \gtwo{\alpha}{\beta}
	\end{split}
\end{align}
and it is more difficult to consider them as doubles. As double-splits perform better in the EN2 correction, and have a clearer physical interpretation, they are the preferred choice. In CI treatments however, the choice of seniority-four states will have no impact on the results.

\begin{figure} 
	\begin{subfigure}{\textwidth}
		\includegraphics[width=0.32\textwidth]{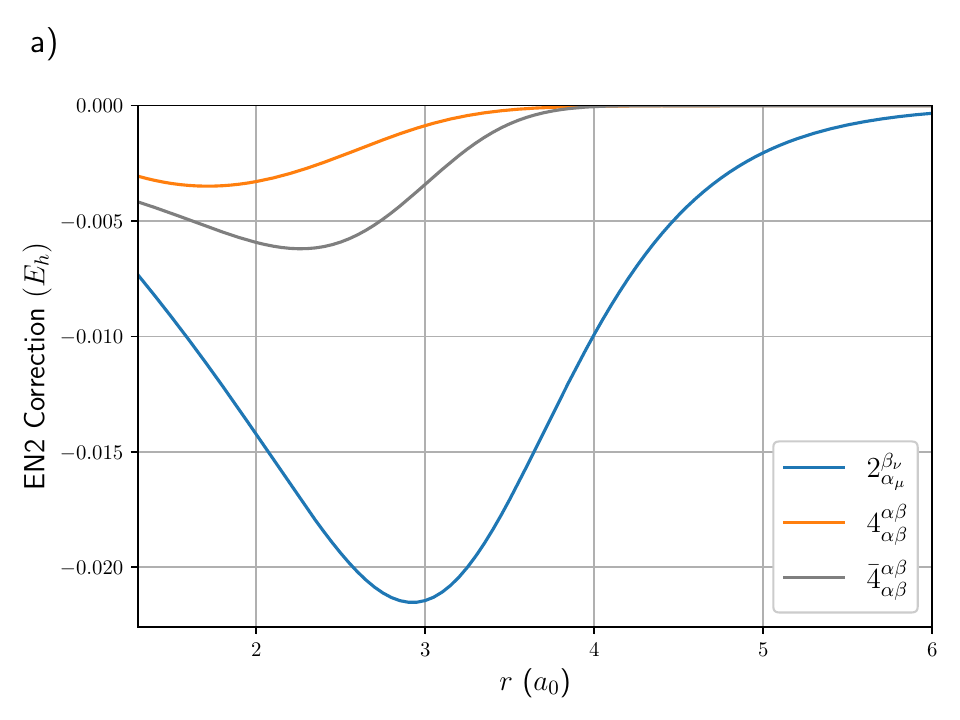}
		\hfill
		\includegraphics[width=0.32\textwidth]{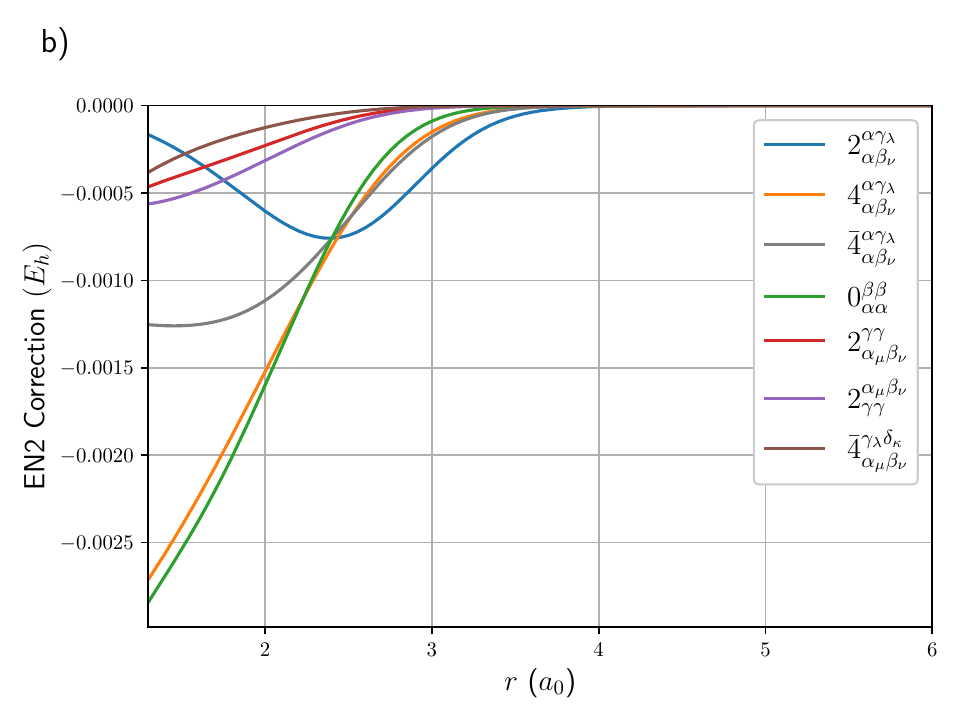}
		\hfill
		\includegraphics[width=0.32\textwidth]{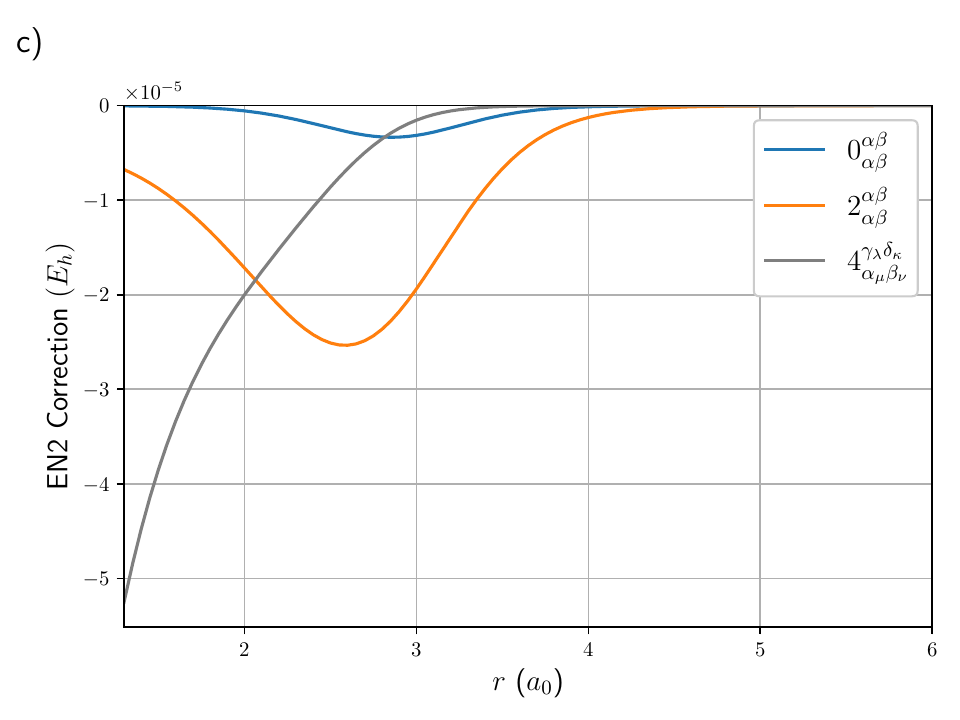}
	\end{subfigure}
	\caption[]{Symmetric bond dissociation of linear H$_8$ in the STO-6G basis: comparison of excitation channels. (a) Dominant contributions: single electron-transfers, double-splits and complementary double-splits. (b) Moderate contributions. (c) Irrelevant contributions: double swaps, swap plus splits, and seniority-four pair transfers. Single swaps and single splits are strictly zero everywhere.	Results are computed in the PP orbitals.}
	\label{fig:h8_channels}
\end{figure}

A breakdown of the contributions in each excitation channel is now in order. In short, the different PP excitations separate into a dominant small group, a less important larger group, and a totally irrelevant small group. This is shown in Figure \ref{fig:h8_channels}.
The dominant contributions to the EN2 correction are clearly the single electron-transfers, followed by the complementary double-splits, and the double-splits. All of the pair-transfer states (except $\ket{\fone{\gamma_{\lambda} \delta_{\kappa}}{\alpha_{\mu} \beta_{\nu}}}$) and the doubles involving electron-transfers give moderate contributions to EN2. The remaining states are irrelevant. Notice that each type of coupling goes to zero rapidly with increasing separation between the hydrogen centres. The optimal PP orbitals are localized\cite{limacher:2014a} so that most of the two-electron integrals will vanish. There is the possibility of non-zero integrals coupling distinct VBS, but in each transition element these are weighted by differences of roots of occupation numbers.

\begin{figure} 
	\begin{subfigure}{0.5\textwidth}
		\includegraphics[width=\textwidth]{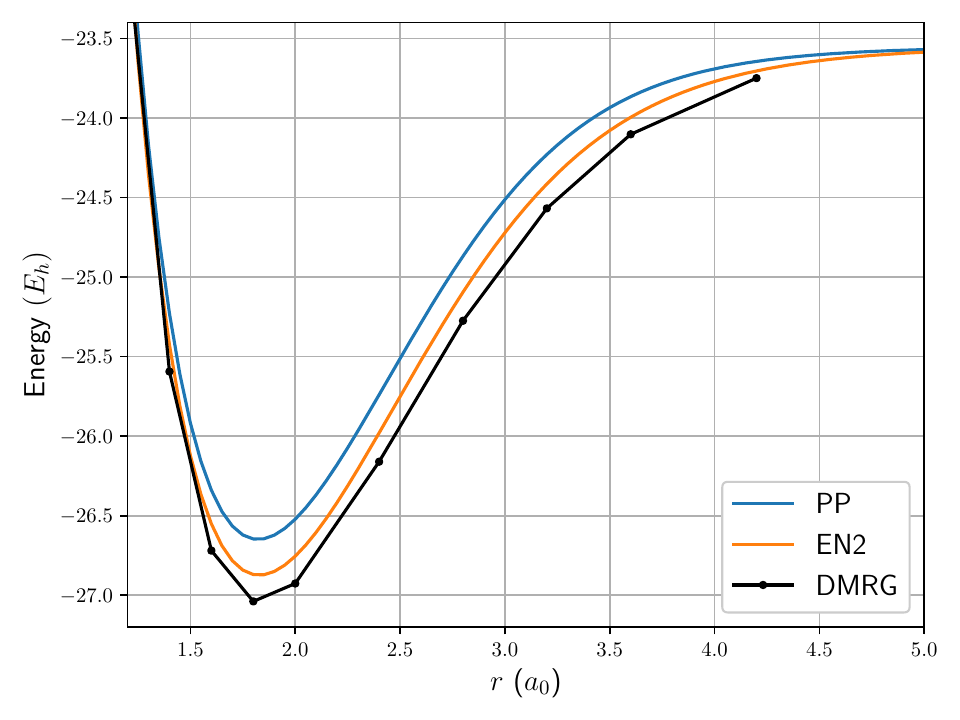}
	\end{subfigure}
	\caption[]{Symmetric bond dissociation of linear H$_{50}$ in the STO-6G basis: PP and EN2 results computed in the PP orbitals. DMRG results are taken from ref. \citenum{hachmann:2006}.}
	\label{fig:h50}
\end{figure}

PP scales favourably, and is thus easily extended to much larger systems. For example, H$_{50}$ is shown in Figure \ref{fig:h50}.
DMRG results obtained in ref. \citenum{hachmann:2006} are numerically exact, allowing for a comparison. PP and EN2 both dissociate correctly, and EN2 recovers about half of the weak correlation missed by PP. 

\begin{figure} 
	\begin{subfigure}{\textwidth}
		\includegraphics[width=0.485\textwidth]{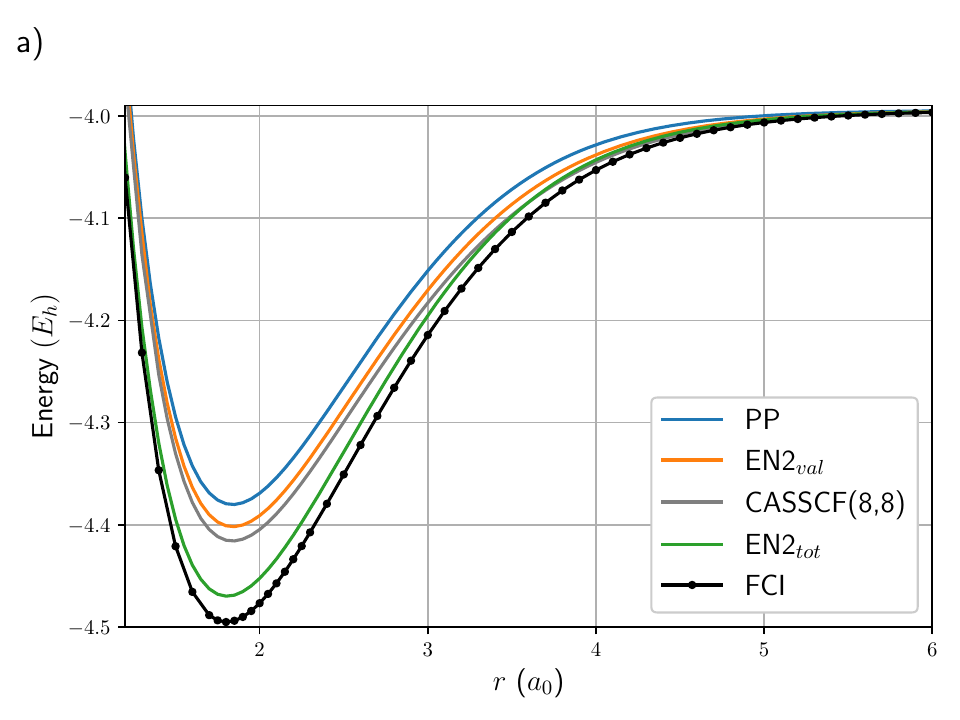}
		\hfill
		\includegraphics[width=0.485\textwidth]{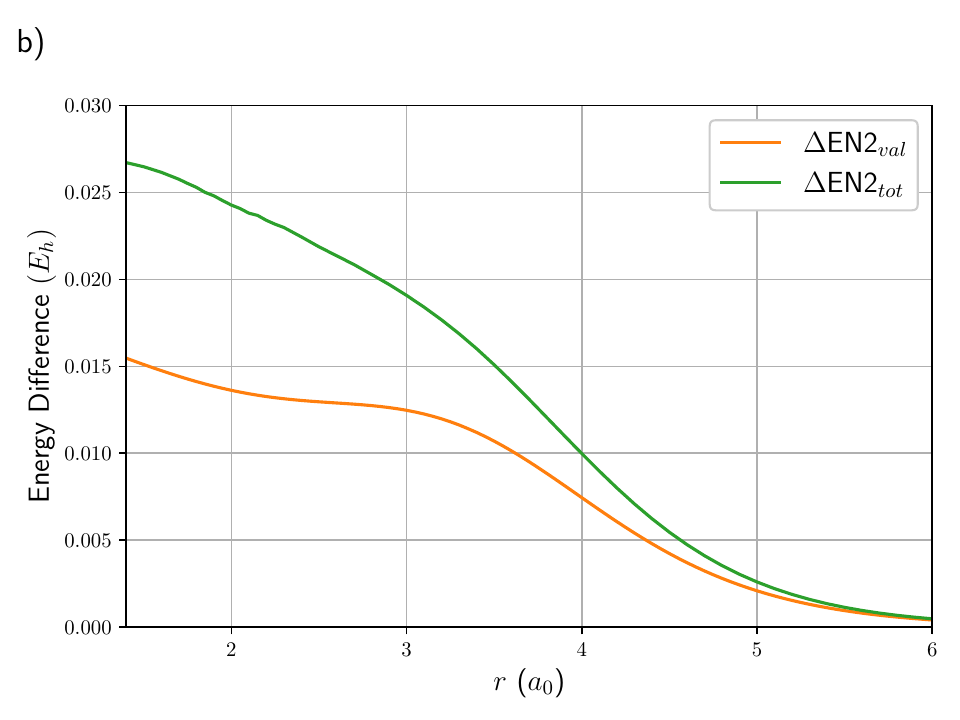}
	\end{subfigure}
	\caption[]{Symmetric bond dissociation of linear H$_{8}$ in the cc-pVDZ basis: PP and EN2 results computed in the PP orbitals. FCI results are from ref. \citenum{kossoski:2022}. (a) Absolute energies: EN2$_{val}$ is the correction in the valence only, EN2$_{tot}$ is the complete correction in the semi-canonical basis. (b) $\Delta$EN2$_{val}$ is the difference between EN2$_{val}$ and CASSCF(8,8) while $\Delta$EN2$_{tot}$ is the difference between EN2$_{tot}$ and FCI.}
	\label{fig:h8_ccpvdz}
\end{figure}

Symmetric H-chain dissociations are usually computed in a minimal basis to isolate the effects of strong correlation. There is no difficulty in computing these systems in more realistic bases. The symmetric dissociation of linear H$_8$ in the cc-pVDZ basis is shown in Figure \ref{fig:h8_ccpvdz}.
Here the EN2 correction includes external excitations involving the virtual orbitals. PP is invariant to core-core and virtual-virtual rotations, whereas the EN2 correction is not. Directly computing the EN2 correction therefore does not produce a smooth result. This is easily fixed with the use of a semi-canonical basis. The core orbitals are chosen to diagonalize the core-core block of the generalized Fock matrix $f_{ij}$ \eqref{eq:genfock}. The virtual-virtual block of the generalized Fock matrix is identically zero, so the virtuals are chosen to diagonalize
\begin{align}
	\tilde{f}_{ab} = 2 h_{ab} + \sum_p 2 (2 V_{abpp} - V_{appb} ) n_p.
\end{align}
An orbital-invariant PT2 correction could be constructed, however, as EN2 is not perfect even in the valence, a better option would be to forget perturbation theory and build an analogue of coupled cluster. For the present purpose, the point to emphasize is that weak correlation missing from the PP reference \emph{can} be extracted from its excitations. It is clear that the EN2$_{val}$ correction is a reasonable approximation to CASSCF(8,8) which would be the standard treatment of strong correlation. CASSCF is an SCF procedure with an eigenvalue solver employed at each iteration. EN2 is a one-shot correction to the already computed PP reference. The complete EN2 correction in a semi-canonical basis, EN2$_{tot}$, is much closer to FCI, but the curves remain discernible, and thus a better treatment would be required. One can see in Figure \ref{fig:h8_ccpvdz} (b) that the difference between EN2$_{tot}$ and FCI is larger than between EN2$_{val}$ and CASSCF(8,8).

\begin{figure} 
	\begin{subfigure}{\textwidth}
		\includegraphics[width=0.485\textwidth]{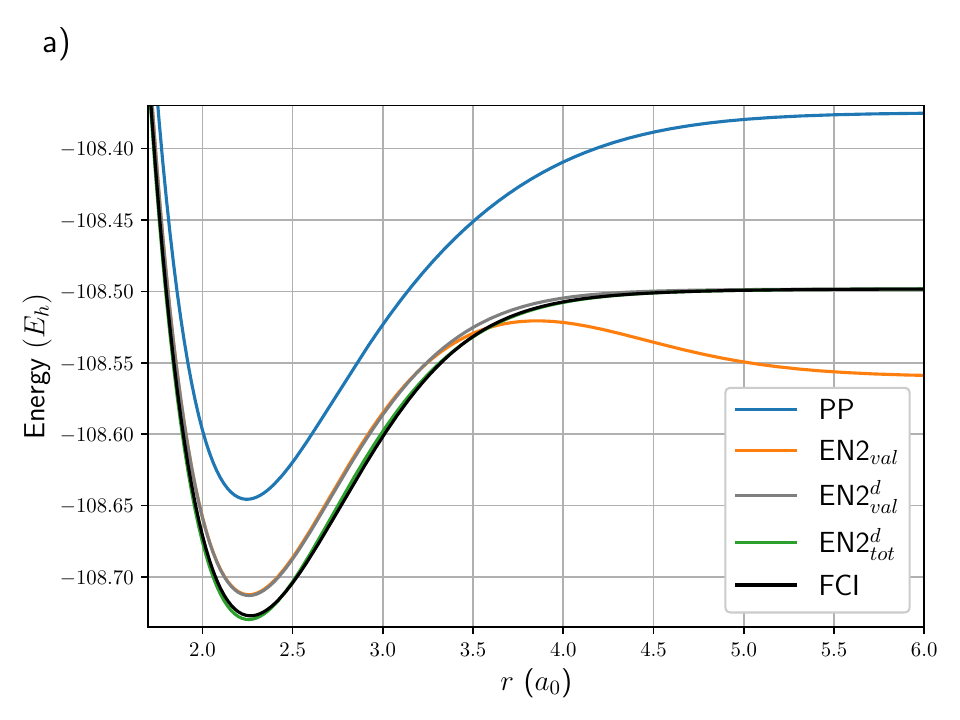}
		\hfill
		\includegraphics[width=0.485\textwidth]{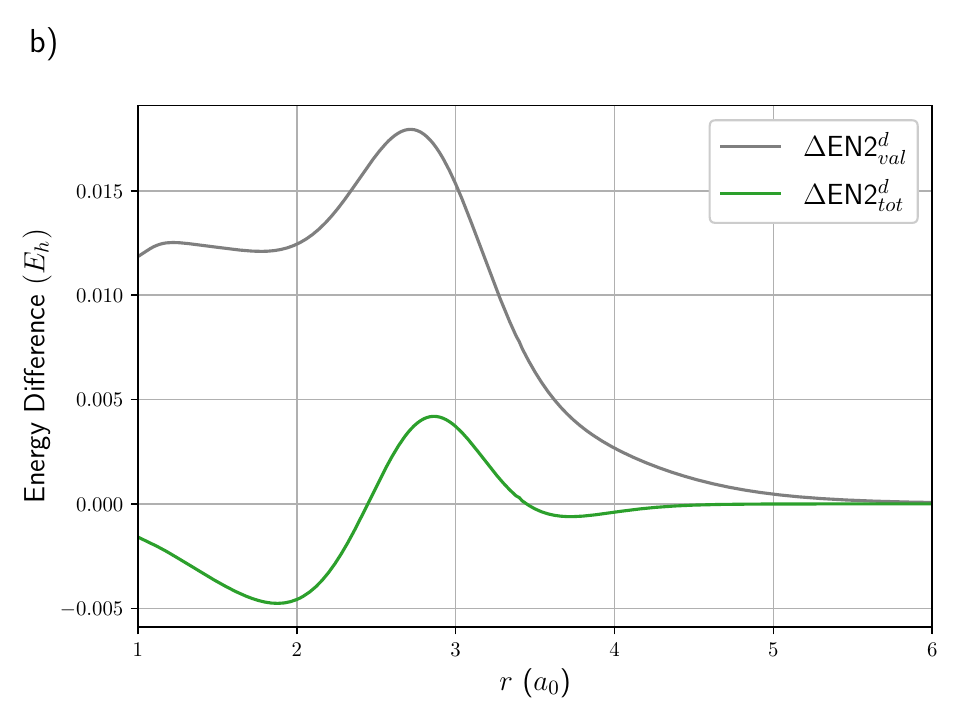}
	\end{subfigure}
	\caption[]{Dissociation of N$_2$ in the STO-6G basis: PP and EN2 results computed in the PP orbitals. (a) EN2$_{val}$ is the correction in the valence only, EN2$^d_{val}$ is the intruder-free correction in the valence (see text), and EN2$^d_{tot}$ is the complete intruder-free correction in the semi-canonical basis. (b) Differences between EN2$^d_{val}$ and FCI, and EN2$^d_{tot}$ and FCI.}
	\label{fig:n2_sto6g}
\end{figure}

\begin{figure} 
	\begin{subfigure}{0.5\textwidth}
		\includegraphics[width=\textwidth]{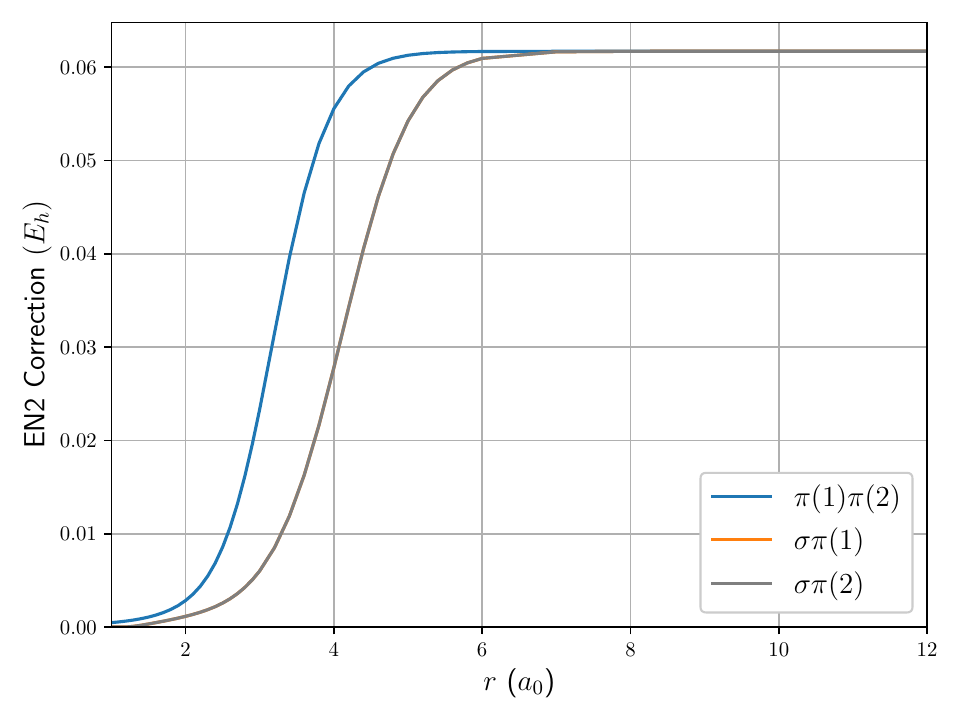}
	\end{subfigure}
	\caption[]{Dissociation of N$_2$ in the STO-6G basis: EN2 corrections from complementary double-splits (intruder states). The two $\sigma\pi$ contributions are equivalent.}
	\label{fig:n2_intruder}
\end{figure}

PP is usually considered a method for dealing with several single bonds. It is much more capable than that. The dissociation of N$_2$ with the STO-6G basis is shown in Figure \ref{fig:n2_sto6g}.
First, PP is strictly a seniority-zero state and cannot dissociate the three bonds correctly on its own. If one shifted the PP curve downward so that the dissociation limit agreed with FCI, the curves would then be quite similar, with the disagreement largest near the minimum. This is the usual approach which is \emph{not} the purpose here. As the core EN2 correction is not smooth, it will not be discussed further. Adding the EN2 correction in the valence, one can see that the result is much better near the minimum, but breaks, and goes below FCI at dissociation. Notice that even here the result levels off to a constant at large distances. This is not a size-consistency problem. It is an intruder-state problem: the three complementary double-splits $\ket{\ftwo{\alpha\beta}{\alpha\beta}}$ have small, but non-zero, excitation energies and large couplings with the PP reference. The EN2 corrections from these three states are shown in Figure \ref{fig:n2_intruder}. 
Two of the complementary double-splits are equivalent, involving the $\sigma$ VBS and one $\pi$ VBS, while the third involves both $\pi$ VBS. Notice that the EN2 contributions from the intruders reach limiting values. They do not become degenerate with the PP reference and thus the EN2 curve is smooth and does not diverge. The intruder state problem is resolved by including the intruders in the reference: build the CI matrix of $\ket{\PP}$ along with the three complementary double-splits and take its lowest eigenvalue. This requires computing the couplings
\begin{align}
	\begin{split}
	\braket{\ftwo{\alpha \beta}{\alpha \beta} | \hH_C | \ftwo{\alpha \gamma}{ \alpha \gamma }} &= 
		- \frac{1}{2} ( \sqrt{n_{\ger{\beta}} n_{\ung{\gamma}} } + \sqrt{ n_{\ung{\beta}} n_{\ger{\gamma}} } )
			V_{ \ger{\gamma} \ung{\beta} \ger{\beta} \ung{\gamma} } \\
		&\quad - \frac{1}{2} ( \sqrt{ n_{\ger{\beta}} n_{\ger{\gamma}} } + \sqrt{ n_{\ung{\beta}} n_{\ung{\gamma}} } )
			V_{ \ger{\gamma} \ger{\beta} \ung{\beta} \ung{\gamma} }.
	\end{split}
\end{align}
Couplings between complementary double-splits can only be non-zero if one index is shared. Adding the EN2 valence correction for all of the other valence states gives the curve labelled EN2$^d_{val}$, while adding all the corrections for all the other states gives the curve EN2$^d_{tot}$ in Figure \ref{fig:n2_sto6g} (a). This is a shortcut, as strictly speaking the couplings between the double-splits and the remainder of the states \emph{does} contribute, but one can see that these effects are very small. 

\begin{figure} 
	\begin{subfigure}{\textwidth}
		\includegraphics[width=0.485\textwidth]{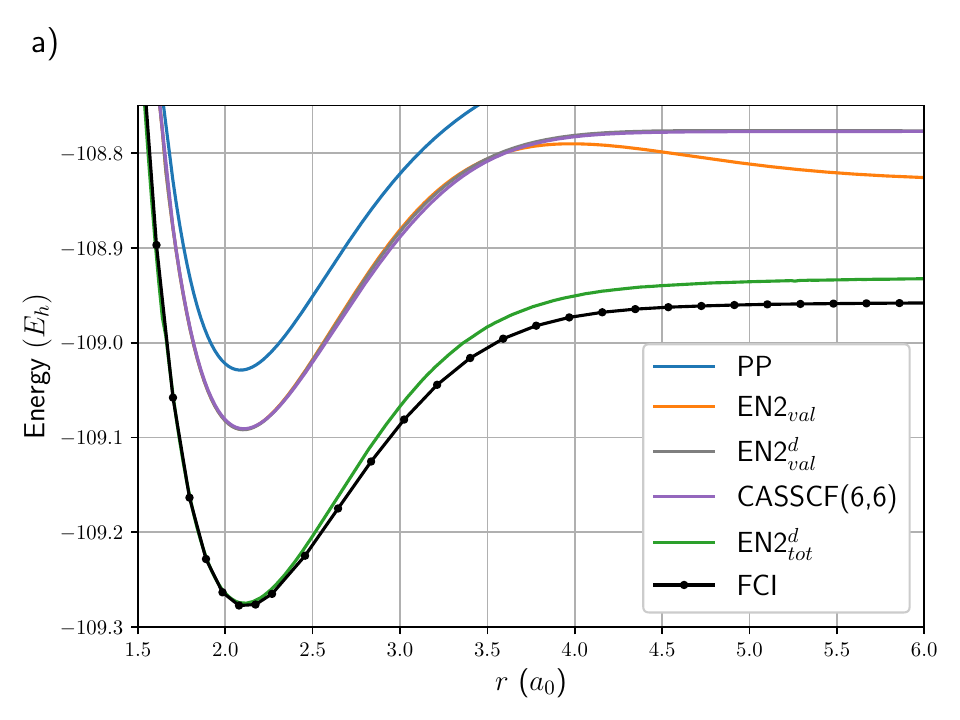}
		\hfill
		\includegraphics[width=0.485\textwidth]{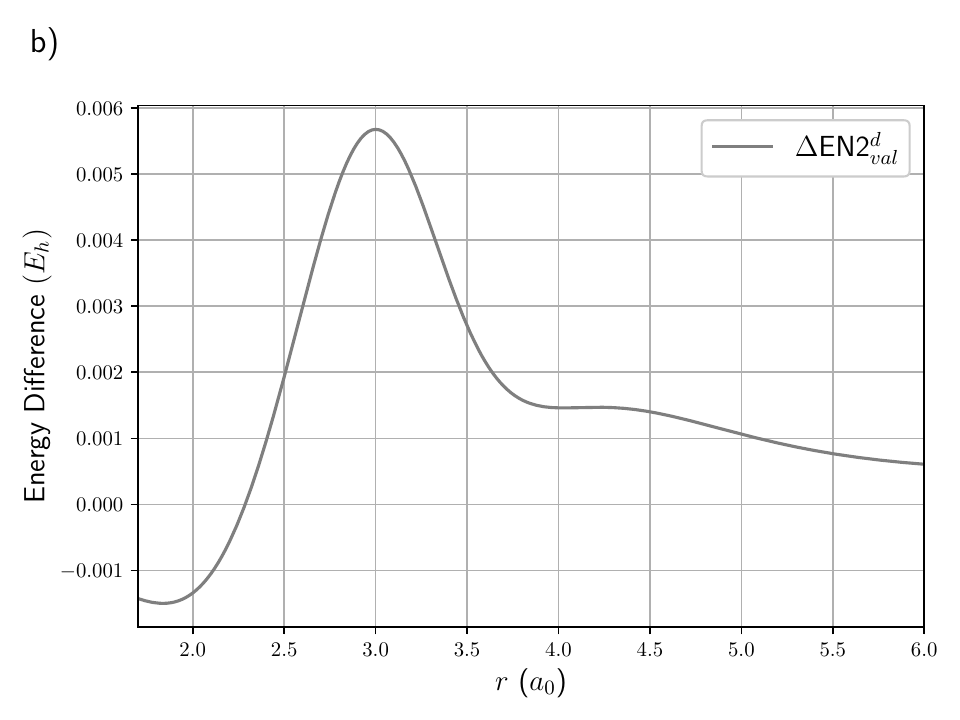}
	\end{subfigure}
	\caption[]{Dissociation of N$_2$ in the cc-pVDZ basis. (a) PP and EN2 results computed in the PP orbitals. EN2$_{val}$ is the correction in the valence only, EN2$^d_{val}$ is the intruder-free correction in the valence, while EN2$^d_{tot}$ is the complete intruder-free correction in the semi-canonical basis. FCI results are from ref. \citenum{kossoski:2022}. (b) Difference between EN2$^d_{val}$ and CASSCF(6,6).}
	\label{fig:n2_ccpvdz}
\end{figure}

These results carry forward to real basis sets. In cc-pVDZ, there are still three intruder states, and the results are shown in Figure \ref{fig:n2_ccpvdz}.
It is clear that CASSCF(6,6) is very well approximated as EN2$^d_{val}$, while FCI is well approximated as EN2$^d_{tot}$ at equilibrium with noticeable disagreement at dissociation. The agreement between CASSCF and EN2$^d_{val}$ is much better for N$_2$ than for H$_8$. For N$_2$, the entire (6,6) active space is required to dissociate the triple bond correctly, whereas for H$_8$, the (8,8) active space describes the H--H single bonds breaking in addition to weak correlation between them. This dissociation only requires four copies of CASSCF(2,2) which is essentially PP. At equilibrium ($r=1.10\;\text{\AA} \approx 2.08\;a_0$), EN2$^d_{tot}$ and FCI agree to 3 m$E_h$ but this disagreement grows to 27 m$E_h$ at dissociation ($r=4.00\;\text{\AA} \approx 7.56\;a_0$). Quantitative agreement at dissociation would be cleaned up in a CI or coupled cluster approach in terms of PP states.

\begin{figure} 
	\begin{subfigure}{\textwidth}
		\includegraphics[width=0.485\textwidth]{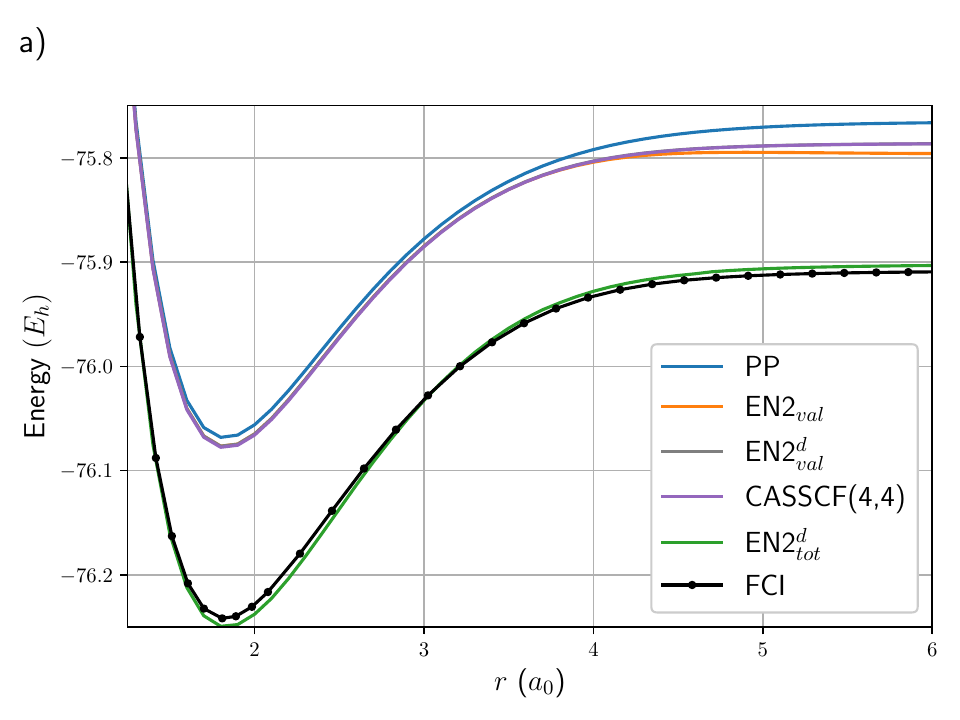}
		\hfill
		\includegraphics[width=0.485\textwidth]{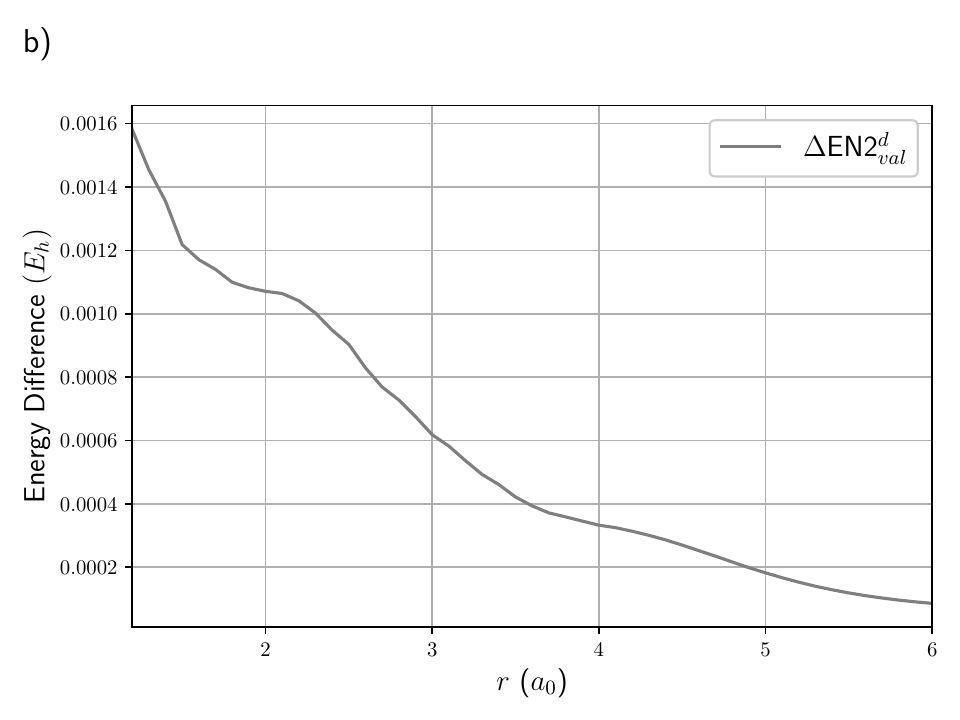}
	\end{subfigure}
	\caption[]{Symmetric dissociation of H$_2$O in the cc-pVDZ basis: PP and EN2 results computed in the PP orbitals. (a) PP, EN2 corrections, CASSCF(4,4) and FCI. EN2$_{val}$ is the correction in the valence only, EN2$^d_{val}$ is the intruder-free correction in the valence, while EN2$^d_{tot}$ is the complete intruder-free correction in the semi-canonical basis. FCI results are from ref. \citenum{kossoski:2022}. (b) Energy difference between EN2$^d_{val}$ and CASSCF(4,4).}
	\label{fig:h2o_ccpvdz}
\end{figure}

\begin{figure} 
	\begin{subfigure}{0.5\textwidth}
		\includegraphics[width=\textwidth]{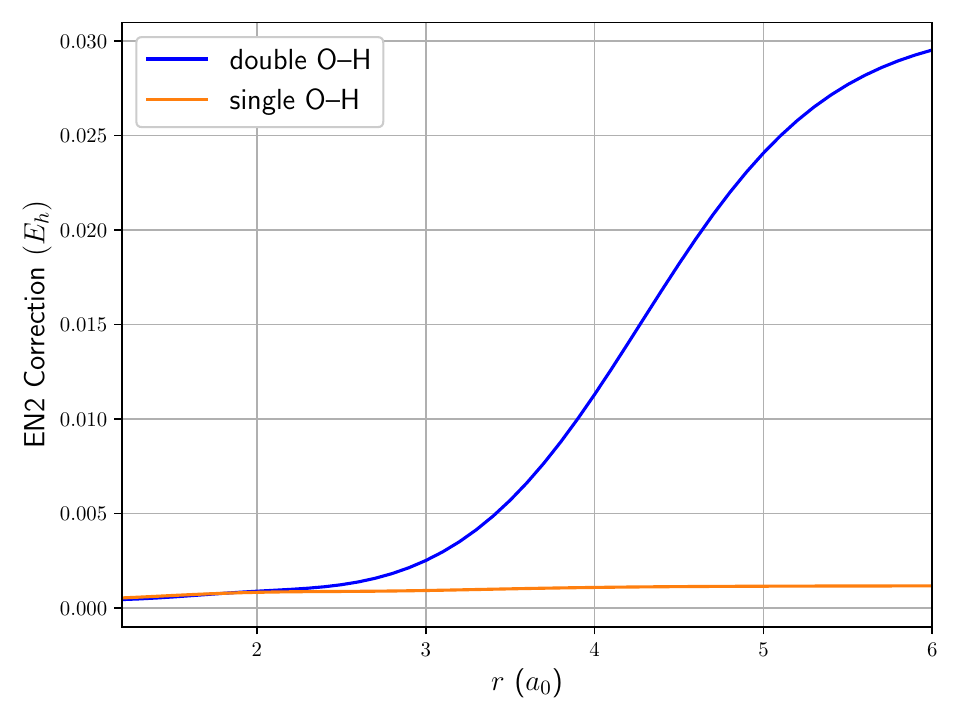}
	\end{subfigure}
	\caption[]{EN2 correction from the complementary double-split intruder states: simultaneous two O--H bond dissociations (blue), single O--H bond dissociation(orange).}
	\label{fig:h2o_intruder}
\end{figure}

The dissociation of a triple bond thus leads to three specific intruder states. Similarly, the dissociation of doubly-bonded O$_2$ leads to \emph{one} intruding complementary double-split, though the curve is omitted. So, multiple dissociating bonds between the same set of 2 atoms leads to intruder states. What happens if only one atom is shared amongst two dissociating bonds? This case is described by the symmetric dissociation of H$_2$O. Results are computed in the cc-pVDZ basis and presented in Figure \ref{fig:h2o_ccpvdz}.
One can again see that there is an intruder state problem, and analysing the EN2 correction reveals the one intruder is the complementary double-split. The observed effect is also much weaker than in N$_2$ as seen in Figure \ref{fig:h2o_intruder}. 
The second, much smaller curve in Figure \ref{fig:h2o_intruder} represents the complementary double-split in H$_2$O where one of the O--H bonds is held fixed at the equilibrium distance of $r_e = 1.8099\;a_0$ while the other is stretched. Technically this is an intruder state, though the difference between EN2$_{val}$ and EN2$^d_{val}$ is on the order of $2\times 10^{-6} E_h$ at dissociation. Once again, the agreement between CASSCF(4,4) and EN2$^d_{val}$ is excellent. The agreement between FCI and EN2$^d_{tot}$ appears good, with an overcorrection of 8 m$E_h$ at equilibrium ($r=0.9572\;\text{\AA} \approx 1.8088\;a_0$) and an undercorrection of 7 m$E_h$ at dissociation.

From these observations it seems that in general dissociating bonds sharing at least one atomic centre see intruder complementary double-splits in EN2. The effect is greater if both atoms are shared. The localized nature of the PP orbitals prevents these intruders for bonds that do not share any atomic centres as the couplings \eqref{eq:conjugate_dble_split} depend on exchange-type integrals that will vanish.

\section{Discussion}
The pretext for the present contribution was to study how the PP limit of RG states\cite{johnson:2025d} carried forward to seniorities two and four. It was expected that the PP states would be a reasonable approximation to the RG states at a reduced cost. Each RG state required a solution of Richardson's equations, and a matrix inversion to compute its RDM elements,\cite{faribault:2022} a cost on the order of $\mathcal{O}(N^4)$ \emph{for each state}. For PP, the VBS gaps \emph{are} the variables, which once optimized directly define all the states. As it turns out, the PP states even outperform the RG states in EN2. 

For the EN2 correction of PP states, the scaling is the integral transformation. In the present case, the orbitals are transformed at each iteration of the PP reference optimization, which could be improved.\cite{limacher:2026} However, the complete set of integrals must be transformed once before the EN2 correction is computed. This is also the case for MP2. With a basis of $N$ orbitals, this step costs $\mathcal{O}(N^5)$ floating-point operations. There are $\mathcal{O}(N^4)$ seniority-four pair-transfer excitations, whose EN2 corrections each require a fixed number of floating-point operations. If one wanted to be aggressive about cost-cutting, the dominant contributions to EN2 are the single electron-transfers, double-splits, and complementary double-splits. There are only $\mathcal{O}(N^2)$ of these states.

Other researchers have employed PP and APSG as starting points in perturbation theory,\cite{rosta:2002,beran:2006,pernal:2014,pastorczak:2015,margocsy:2018,pastorczak:2019} block-correlated coupled cluster,\cite{wang:2020,ren:2024,han:2025,han:2026} and CI type treatments.\cite{cullen:1999,cullen:2007} However, all of these treatments fundamentally consider PP as a multireference wavefunction ansatz rather than an eigenvector of an exactly-solvable Hamiltonian. The clear benefit of the current picture is a complete set of low-lying excitations to which PP couples through $\hH_C$. Ref. \citenum{rosta:2002} in particular considers the possibility of an EN partitioning, but chooses not to pursue the idea as it is not additively separable, and hence is not necessarily size-consistent. This gets immediately to the next point.

While the EN2 correction is not necessarily size-consistent, the numerical results \emph{in the valence} certainly seem to be. This has been discussed for the seniority-zero correction at length.\cite{johnson:2024b} The localized nature of the optimal PP orbitals eliminates many of the would-be offending terms, while others are weighted by roots of occupation numbers or VBS gaps. Even when intruder states poison the EN2 correction, the energies level-off to constant values. The EN2 corrections computed with semi-canonical core and virtuals are not size-consistent. This is to be expected and must be corrected with a coupled cluster type treatment. The present purpose is again only to demonstrate the simplest possible correction in terms of PP states.

The intruding complementary double-splits are dealt with by diagonalizing a small Hamiltonian matrix composed of the PP reference and the intruders. It is natural to ask how quickly the size of this matrix grows. Main group atoms following the octet rule tend to have at maximum four bonds apiece which severely limits the number of potentially intruding complementary double-splits. Intruder states only appear for bonding pairs that share at least one atomic centre. In methane, there are six potential intruders: the four C--H bonds share the common carbon centre giving $\binom{4}{2}=6$ potential intruders. In ethane, there are twelve: the C--C bond shares one carbon centre with each of the six C--H bonds, while on each carbon centre there are three C--H bonds $6 + 2 \binom{3}{2} = 12$. In general, a saturated alkane with $N_C$ carbon centres has $6 N_C$ couples of bonds sharing at least one centre. Double and triple bonds decrease this number. Thus, the Hamiltonian matrix required for EN2$^d_{val}$ grows only linearly with the number of carbon centres. In addition, most of the resulting complementary double-splits will not end up being intruders. Hypervalent elements and transition metals will of course grow faster than alkanes, but still have limited connectivity and hence the number of potential intruder states will grow linearly with the number of atomic centres.

As the energy expression \eqref{eq:pp_pair_energy} can be understood as independent electrons in different orbitals stabilised by a pairing interaction, there is an obvious question as to how far this viewpoint can be taken. It was certainly productive to think about the PP excited states qualitatively from this viewpoint. However, trying to use effective electrons directly does not lead very far, which again serves to emphasize that pairs of electrons and individual electrons are very different, and trying to understand strong correlation in terms of Slater determinants is not productive. To be clear, it is possible to use two sets of electrons $c^{\dagger}_{\alpha_{\mu}}$ and $d^{\dagger}_{\alpha_{\mu}}$ to define an auxiliary Hamiltonian 
\begin{align}
\begin{split}
	\hH_F &= \sum_{i} \varepsilon_i \hat{n}_i 
	+ \sum_{\alpha}\sum_{\mu} \varepsilon_{\alpha_{\mu}} 
	\left(
	   c^{\dagger}_{\alpha_{\mu}} c_{\alpha_{\mu}} 
	+  d^{\dagger}_{\alpha_{\mu}} d_{\alpha_{\mu}} \right) \\
	& - \sum_{\alpha} \eg{\alpha} L_{\ger{\alpha}\ung{\alpha}}
	\left(
	   c^{\dagger}_{\ger{\alpha}} d^{\dagger}_{\ung{\alpha}} d_{\ung{\alpha}} c_{\ger{\alpha}} 
	-  d^{\dagger}_{\ger{\alpha}} c^{\dagger}_{\ung{\alpha}} c_{\ung{\alpha}} d_{\ger{\alpha}}
	\right).
\end{split}
\end{align}
Each spatial orbital can hold a $c$ electron and a $d$ electron, so that they could be understood as some rotation of up- and down-spin. The Slater determinant
\begin{align}
	\ket{\tilde{\PP}} = \prod_{\alpha} c^{\dagger}_{\ger{\alpha}} d^{\dagger}_{\ung{\alpha}}
	\ket{\Phi}
\end{align}
is an eigenvector of $\hH_F$ with eigenvalue
\begin{align}
	E_F[\tilde{\PP}] = 2 \sum_i \varepsilon_i 
	+ \sum_{\alpha} [ \varepsilon_{\ger{\alpha}} + \varepsilon_{\ung{\alpha}}
	- \eg{\alpha} L_{\ger{\alpha} \ung{\alpha}} ],
\end{align}
which is the same as the PP energy \eqref{eq:pp_pair_energy} $E[\PP]$ without the double-counting correction. The seniority-conserving excitations are clear in this language. In particular the Slater determinant,
\begin{align}
	\ket{\tilde{0}^{\alpha}_{\alpha}} = d^{\dagger}_{\ger{\alpha}} c^{\dagger}_{\ung{\alpha}}
	\prod_{\beta (\neq \alpha)} c^{\dagger}_{\ger{\beta}} d^{\dagger}_{\ung{\beta}}
	\ket{\Phi}
\end{align}
reproduces the swap excitation
\begin{align}
	E_F[\tilde{0}^{\alpha}_{\alpha}] - E_F[\tilde{\PP}] =
	2 \eg{\alpha} L_{\ger{\alpha} \ung{\alpha}},
\end{align}
while the analogue for the double-swap
\begin{align}
	\ket{\tilde{0}^{\alpha\beta}_{\alpha\beta}} = 
	d^{\dagger}_{\ger{\alpha}} c^{\dagger}_{\ung{\alpha}}
	d^{\dagger}_{\ger{\beta}}  c^{\dagger}_{\ung{\beta}}
	\prod_{\gamma (\neq \alpha, \beta)}
	c^{\dagger}_{\ger{\gamma}} d^{\dagger}_{\ung{\gamma}} \ket{\Phi}
\end{align}
is correct up to the double-counting correction
\begin{align}
	E_F[\tilde{0}^{\alpha\beta}_{\alpha\beta}] - E_F[\tilde{\PP}] =
	2 \eg{\alpha} L_{\ger{\alpha}\ung{\alpha}} + 2 \eg{\beta} L_{\ger{\beta}\ung{\beta}}.
\end{align}
Likewise, the pair transfer can be understood with the Slater determinant
\begin{align}
	\ket{\tilde{0}^{\beta\beta}_{\alpha\alpha}} =
	c^{\dagger}_{\ger{\beta}} d^{\dagger}_{\ung{\beta}}
	d^{\dagger}_{\ger{\beta}} c^{\dagger}_{\ung{\beta}}
	\prod_{\gamma (\neq \alpha, \beta)}
	c^{\dagger}_{\ger{\gamma}} d^{\dagger}_{\ung{\gamma}} \ket{\Phi}
\end{align}
giving an excitation energy
\begin{align}
\begin{split}
	E_F[\tilde{0}^{\beta\beta}_{\alpha\alpha}] - E_F[\tilde{\PP}] &= 
	  \varepsilon_{\ger{\beta}}  + \varepsilon_{\ung{\beta}} 
	- \varepsilon_{\ger{\alpha}} - \varepsilon_{\ung{\alpha}} \\
	&\quad + \eg{\alpha} L_{\ger{\alpha} \ung{\alpha}}
	+ \eg{\beta}  L_{\ger{\beta}  \ung{\beta}},
\end{split}
\end{align}
which is $E[0^{\beta\beta}_{\alpha\alpha}] - E[\PP]$ up to the double-counting correction. Unfortunately, this is as far as the idea can be taken. One can choose either $c^{\dagger}_{\ger{\alpha}}c^{\dagger}_{\ung{\alpha}}$ \emph{or} $d^{\dagger}_{\ger{\alpha}}d^{\dagger}_{\ung{\alpha}}$ to represent a split in VBS $\alpha$, and introduce more terms in the auxiliary Hamiltonian to count $t_{\ger{\alpha}\ung{\alpha}}$, which is arbitrary, but works. Trying to do the same for single electron transfers does not work, and this line of reasoning is thus to be discouraged. 

Even though the above effective electrons are not particularly useful, the connection between PP and UHF is understood and actively exploited as a guess.\cite{wang:2019} In short, with $\bm{S}$ the atomic orbital (AO) overlap matrix, the up/down overlap of the occupied UHF orbitals is constructed
\begin{align}
	\bm{M} = \bm{C}^{\uparrow}_{occ} \bm{S} \bm{C}^{\downarrow}_{occ},
\end{align}
and the singular value decomposition (SVD) is taken $\bm{M} = \bm{U} \bm{\varSigma} \bm{V}^T$. Most of the singular values are identically equal to one, indicating identical up- and down-spin orbitals, while the remainder will be $0 < \sigma_{\alpha} < 1$. The index $\alpha$ is employed as it will end up labelling the VBS $\alpha$. It is not to be confused with the usual spin label convention. The SVD is used to build the \emph{corresponding orbitals}
\begin{subequations}
\begin{align}
	\tilde{\bm{C}}^{\uparrow}_{occ} &= \bm{C}^{\uparrow}_{occ} \bm{U} \\
	\tilde{\bm{C}}^{\downarrow}_{occ} &= \bm{C}^{\downarrow}_{occ} \bm{V}
\end{align}
\end{subequations}
which separate into clean pairs $\tilde{\phi}^{\uparrow}_{\alpha}$ and $\tilde{\phi}^{\downarrow}_{\alpha}$. Finally, the unrestricted natural orbitals (UNOs) are constructed
\begin{subequations}
\begin{align}
	\phi_{\ger{\alpha}} &= \frac{\tilde{\phi}^{\uparrow}_{\alpha} + \tilde{\phi}^{\downarrow}_{\alpha}}{\sqrt{2 (1 + \sigma_{\alpha})}} \\
	\phi_{\ung{\alpha}} &= \frac{\tilde{\phi}^{\uparrow}_{\alpha} - \tilde{\phi}^{\downarrow}_{\alpha}}{\sqrt{2 (1 - \sigma_{\alpha})}},
\end{align}
\end{subequations}
with occupation numbers
\begin{align}
	n_{\alpha_{\mu}} &= 1 + (-1)^{\mu} \sigma_{\alpha}.
\end{align}
UNOs are understood to provide a good guess for the PP orbitals, while the singular values provide a guess for the VBS gaps
\begin{align}
	\omega_{\alpha} = \frac{\sigma_{\alpha}}{ \sqrt{1 - \sigma_{\alpha}^2 } }.
\end{align}
In ref. \citenum{wang:2019}, the authors localize the bonding and antibonding UNOs separately before matching them together by maximizing the transition dipole. This guess emphasizes that the workflow
\begin{align*}
	\text{UHF} \rightarrow \text{PP} \rightarrow \text{post-PP}
\end{align*}
is feasible and productive.

While it is not the optimal way to understand the energies of excited PP states, it is worth summarizing the occupation-number point of view in terms of energy cumulants. These are presented for the valence excitations in Table \ref{tab:energy_cumulants}.
\begin{center}
	\begin{table}[h]
		
		\begin{tabular}{c|c}
			$\ket{\Psi}$ & $\Delta [\Psi]$ \\
			\toprule
			
			$\ket{0^{\alpha}_{\alpha}}$	&
			$\frac{2}{\eg{\alpha}} L_{\ger{\alpha}\ung{\alpha}}$ \\
			
			$\ket{2^{\alpha}_{\alpha}}$ &
			$\frac{1}{\eg{\alpha}} L_{\ger{\alpha}\ung{\alpha}} + \twamp{\ger{\alpha}\ung{\alpha}}$ \\
			
			$\ket{2^{\beta_{\nu}}_{\alpha_{\mu}} }$ &
			$\frac{1}{\eg{\alpha}} L_{\ger{\alpha}\ung{\alpha}} +
			\frac{1}{\eg{\beta}}  L_{\ger{\beta} \ung{\beta} } +
			\twamp{\alpha_{\mu} \beta_{\nu}} +
			G_{ \ger{\beta} \ung{\beta} }$ \\
			
			\midrule
			
			$\ket{0^{\alpha\beta}_{\alpha\beta}}$ &
			$\Delta [0^{\alpha}_{\alpha}] + \Delta[0^{\beta}_{\beta}] $ \\
			
			$\ket{2^{\alpha\beta}_{\alpha\beta}}$ &
			$\Delta [0^{\alpha}_{\alpha}] + \Delta[2^{\beta}_{\beta}] $ \\
			
			$\ket{4^{\alpha\beta}_{\alpha\beta}}$ &
			$\Delta [2^{\alpha}_{\alpha}] + \Delta[2^{\beta}_{\beta}] $ \\
			
			$\ket{2^{\alpha \gamma_{\lambda}}_{\alpha \beta_{\nu}} } $ &
			$\Delta [0^{\alpha}_{\alpha}] + \Delta [2^{\gamma_{\lambda}}_{\beta_{\nu}}]$\\
			
			$\ket{4^{\alpha \gamma_{\lambda}}_{\alpha \beta_{\nu}} } $ &
			$\Delta [2^{\alpha}_{\alpha}] + \Delta [2^{\gamma_{\lambda}}_{\beta_{\nu}}]$\\
			
			\midrule
			
			$\ket{0^{\beta\beta}_{\alpha\alpha}}$ &
			$\frac{1}{\eg{\alpha}} L_{\ger{\alpha}\ung{\alpha}} +
			\frac{1}{\eg{\beta}}  L_{\ger{\beta} \ung{\beta }} +
			2 G_{\ger{\beta} \ung{\beta}}$ \\
			
			$\ket{2^{\gamma\gamma}_{\alpha_{\mu} \beta_{\nu}}} $ &
			$\frac{1}{\eg{\alpha}} L_{\ger{\alpha} \ung{\alpha}} +
			\frac{1}{\eg{\beta}}  L_{\ger{\beta}  \ung{\beta }} +
			\frac{1}{\eg{\gamma}} L_{\ger{\gamma} \ung{\gamma}} +
			\twamp{\alpha_{\mu} \beta_{\nu}} + 2 G_{\ger{\gamma} \ung{\gamma}}$ \\
			
			$\ket{2^{\alpha_{\mu} \beta_{\nu}}_{\gamma \gamma} }$ &
			$\frac{1}{\eg{\alpha}} L_{\ger{\alpha} \ung{\alpha}} +
			\frac{1}{\eg{\beta}}  L_{\ger{\beta}  \ung{\beta }} +
			\frac{1}{\eg{\gamma}} L_{\ger{\gamma} \ung{\gamma}} +
			\twamp{\alpha_{\mu} \beta_{\nu}} 
			+ G_{\ger{\alpha} \ung{\alpha}} + G_{\ger{\beta} \ung{\beta}}$ \\
			
			$\ket{4^{\gamma_{\lambda} \delta_{\kappa}}_{\alpha_{\mu}\beta_{\nu}}}$ &
			$\frac{1}{\eg{\alpha}} L_{\ger{\alpha} \ung{\alpha}} +
			\frac{1}{\eg{\beta}}  L_{\ger{\beta}  \ung{\beta }} +
			\frac{1}{\eg{\gamma}} L_{\ger{\gamma} \ung{\gamma}} +
			\frac{1}{\eg{\delta}} L_{\ger{\delta} \ung{\delta}} +
			\twamp{\alpha_{\mu}\beta_{\nu}} + \twamp{\gamma_{\lambda} \delta_{\kappa}} +
			G_{\ger{\gamma} \ung{\gamma}} + G_{\ger{\delta} \ung{\delta}}$
			
		\end{tabular}
		\caption{Valence excitation energy cumulants.}
		\label{tab:energy_cumulants}
	\end{table}
\end{center}
The shapes of the energy cumulants are unsurprisingly quite similar to the already computed excitation energies. A simple recipe suffices to obtain energy cumulants: take the excitation energy and discard any terms related to double-counting and orbital energies, then replace the pair amplitudes $\eg{}$ with $\frac{1}{\eg{}}$. One advantage of this viewpoint is that double excitations separate additively to single excitations. In the end, the results are identical so that there is no particular computational advantage to this point of view.

Finally, while it has been mentioned several times in the text, it bears emphasis that Slater determinants and second-quantized operators are the wrong way to understand strong electronic correlation. The PP reference and low-lying excitations are simple and should be understood on their own without relying on a Slater determinant expansion. CASSCF, the standard treatment in terms of Slater determinants, scales very poorly but is competitive with a straightforward one-shot perturbative correction in terms of PP states. 

\section{Conclusion}
Slater determinant CI partitioned in terms of seniority is understood to describe strong correlation at exponential cost. Parts I and II in this series established that RG states can achieve the same result at a (large) polynomial cost. Here, a particular limit of the RG state construction leads to PP and its natural excitations. In perturbation theory, these PP states are just as effective as RG states but are much cheaper to employ. In particular, an EN2 correction in terms of PP states experiences particular intruder states, which when removed leads to a simple computation in the valence that is competitive with CASSCF in the cc-pVDZ basis. The next contribution will establish the couplings between all PP states to build a CI or coupled cluster type approach.

\section{Acknowledgments}
The author thanks the Natural Sciences and Engineering Research Council of Canada (Grant No. RGPIN-2024-05610) for funding. This research was made possible in part by the Digital Research Alliance of Canada. The author is also indebted to W. Poelmans, P.-F. Loos, F. Kossoski, and Y. Damour for sharing reference curves for H$_8$, N$_2$ and H$_2$O.

\section{Conflict of Interest}
The author has no conflicts to disclose.

\section{Data Availability}
The data that support the findings of this study are available from the author upon reasonable request.

\appendix

\section{External Excitation Energies}

\subsection{Singles}
The only external single excitations are electron-transfers between spaces. The orbital optimization necessarily decouples these states from $\ket{\PP}$, and the excitation energies are positive.
\begin{align}
	E[2^{\alpha_{\mu}}_i] - E[\PP] &= \eg{\alpha} L_{\ger{\alpha} \ung{\alpha}}
		+ \twamp{i \alpha_{\mu}} 
		+ \varepsilon_{\alpha_{1 - \mu}} - \varepsilon_i
		+ G_{\ger{\alpha} \ung{\alpha}} 
		- \frac{1}{2} G_{i \alpha_{1 - \mu}}
		+ \gone{\alpha}{i} \\
	E[2^a_{\alpha_{\mu}}] - E[\PP] &= \eg{\alpha} L_{\ger{\alpha} \ung{\alpha}}
		+\twamp{\alpha_{\mu} a}
		+ \varepsilon_a - \varepsilon_{\alpha_{1 - \mu}}
		- \frac{1}{2} G_{\alpha_{1 - \mu} a}
		- \gone{\alpha}{a} \\
	E[2^a_i] - E[\PP] &= \twamp{ia} 
		+ \varepsilon_a - \varepsilon_i
		- \frac{1}{2} G_{ia}
\end{align}

\subsection{Doubles}
External doubles all involve a valence single, either a swap or a split, plus an external electron transfer. The swap plus electron transfer excitation energies are
\begin{align}
	\begin{split}
		E[2^{\alpha \beta_{\nu}}_{\alpha i}] - E[\PP] &= 
		2 \eg{\alpha} L_{\ger{\alpha} \ung{\alpha}} + \eg{\beta} L_{\ger{\beta} \ung{\beta}}
		+ \twamp{i \beta_{\nu}} 
		+ \varepsilon_{\beta_{1 - \nu}} - \varepsilon_i 
		+ G_{\ger{\beta} \ung{\beta}} \\
		&\quad + 2 \gtwo{\alpha}{\beta}
		+ \gone{\beta}{i} + 2 \gone{\alpha}{i} - 2 \gone{\alpha}{\beta_{1 - \nu}}
		- \frac{1}{2} G_{i \beta_{1 - \nu}}
	\end{split}
	\\
	&= ( E[0^{\alpha}_{\alpha}] - E[\PP] ) + ( E[2^{\beta_{\nu}}_i] - E[\PP] ) 
	+ 2 \gtwo{\alpha}{\beta} + 2 \gone{\alpha}{i} - 2 \gone{\alpha}{\beta_{1 - \nu}}
	\\
	\begin{split}
		E[2^{\alpha a}_{\alpha \beta_{\nu}}] - E[\PP] &=
		2 \eg{\alpha} L_{\ger{\alpha} \ung{\alpha}} + \eg{\beta} L_{\ger{\beta} \ung{\beta}} 
		+ \twamp{\beta_{\nu} a} + \varepsilon_a - \varepsilon_{\beta_{1 - \nu}} \\
		&\quad + 2 \gtwo{\alpha}{\beta}
		+ 2 \gone{\alpha}{\beta_{1 - \nu}} - 2 \gone{\alpha}{a} - \gone{\beta}{a}
		- \frac{1}{2} G_{\beta_{1 - \nu} a}
	\end{split}
	\\
	&= ( E[0^{\alpha}_{\alpha}] - E[\PP] ) + ( E[2_{\beta_{\nu}}^a] - E[\PP] ) 
	+ 2 \gtwo{\alpha}{\beta} + 2 \gone{\alpha}{\beta_{1 - \nu}} - 2 \gone{\alpha}{a}
	\\
	\begin{split}
		E[2^{\alpha a}_{\alpha i}] - E[\PP] &= 
		2 \eg{\alpha} L_{\ger{\alpha} \ung{\alpha}}
		+ \twamp{ia} + \varepsilon_a - \varepsilon_i
		+ 2 \gone{\alpha}{i} - 2 \gone{\alpha}{a} 
		- \frac{1}{2} G_{ia}
	\end{split}
	\\
	&= ( E[0^{\alpha}_{\alpha}] - E[\PP] ) + ( E[2^a_i] - E[\PP] )
	+ 2 \gone{\alpha}{i} - 2 \gone{\alpha}{a},
\end{align}
while the split plus electron-transfer excitation energies are
\begin{align}
	\begin{split}
		E[ \fone{\alpha \beta_{\nu}}{\alpha i}] - E[\PP] &=
		\eg{\alpha} L_{\ger{\alpha} \ung{\alpha}} + \eg{\beta} L_{\ger{\beta} \ung{\beta}}
		+ \twamp{\ger{\alpha} \ung{\alpha}} + \twamp{i \beta_{\nu}}
		+ \varepsilon_{\beta_{1 - \nu}} - \varepsilon_i
		+ G_{\ger{\beta} \ung{\beta}} \\
		&\quad + \gtwo{\alpha}{\beta}
		- \gone{\alpha}{\beta_{1 - \nu}} + \gone{\alpha}{i} + \gone{\beta}{i}
		- \frac{1}{2} G_{i \beta_{1 - \nu}}
	\end{split}
	\\
	&= ( E[2^{\alpha}_{\alpha}] - E[\PP]) + ( E[2^{\beta_{\nu}}_i ] - E[\PP] )
	+ \gtwo{\alpha}{\beta} - \gone{\alpha}{\beta_{1 - \nu}} + \gone{\alpha}{i}
	\\
	\begin{split}
		E[ \fone{\alpha a}{\alpha \beta_{\nu}}] - E[\PP] &= 
		\eg{\alpha} L_{\ger{\alpha} \ung{\alpha}} + \eg{\beta} L_{\ger{\beta} \ung{\beta}}
		+ \twamp{\ger{\alpha} \ung{\alpha}} + \twamp{\beta_{\nu} a}
		+ \varepsilon_a - \varepsilon_{\beta_{1 - \nu}} \\
		&\quad + \gtwo{\alpha}{\beta}
		- \gone{\alpha}{a} - \gone{\beta}{a} + \gone{\alpha}{\beta_{1 - \nu}} 
		- \frac{1}{2} G_{\beta_{1 - \nu} a}
	\end{split}
	\\
	&= (E[2^{\alpha}_{\alpha}] - E[\PP]) + (E[2^a_{\beta_{\nu}}] - E[\PP] ) 
	+ \gtwo{\alpha}{\beta} - \gone{\alpha}{a} + \gone{\alpha}{\beta_{1 - \nu}}
	\\
	\begin{split}
		E[ \fone{\alpha a}{\alpha i}] - E[\PP] &=
		\eg{\alpha} L_{\ger{\alpha} \ung{\alpha}}
		+ \twamp{\ger{\alpha} \ung{\alpha}}	+ \twamp{ia}  
		+ \varepsilon_a - \varepsilon_i 
		- \gone{\alpha}{a} + \gone{\alpha}{i}
		- \frac{1}{2} G_{ia}
	\end{split}
	\\
	&= ( E[2^{\alpha}_{\alpha}] - E[\PP]) + ( E[2^{a}_{i}] - E[\PP] ) - \gone{\alpha}{a} + \gone{\alpha}{i}.
\end{align}
Excitation energies for the complementary seniority-four excitations are obtained as updates involving only the exchange integrals of the blocked levels.

\subsection{Pair-Transfers}
Seniority-conserving pair-transfers occur between spaces with excitation energies
\begin{align}	
	E[0^{\alpha\alpha}_{ii}] - E[\PP] &= \eg{\alpha} L_{\ger{\alpha} \ung{\alpha}} 
		+ \varepsilon_{\ger{\alpha}} + \varepsilon_{\ung{\alpha}} - 2 \varepsilon_i 
		+ 2 G_{\ger{\alpha} \ung{\alpha}} 
		- G_{i \ger{\alpha}} - G_{i \ung{\alpha}} + 2 \gone{\alpha}{i}	\\
	E[0^{aa}_{\alpha\alpha}] - E[\PP] &= \eg{\alpha} L_{\ger{\alpha} \ung{\alpha}} 
		+ 2 \varepsilon_a - \varepsilon_{\ger{\alpha}} - \varepsilon_{\ung{\alpha}}
		- G_{\ger{\alpha} a} - G_{\ung{\alpha} a} - 2 \gone{\alpha}{a} \\
	E[0^{aa}_{ii}] - E[\PP] &= 2 \varepsilon_a - 2 \varepsilon_i - 2 G_{ia}.
\end{align}
The last is the excitation energy of a doubly-excited Slater determinant in an EN partitioning of the Hamiltonian.

Seniority-two pairs can be scattered to seniority-zero pairs with excitation energies
\begin{align}
	\begin{split}
		E[2^{\alpha \alpha}_{i j}] - E[\PP] &= \eg{\alpha} L_{\ger{\alpha} \ung{\alpha}} 
			+ \twamp{i j}
			+ \varepsilon_{\ger{\alpha}} + \varepsilon_{\ung{\alpha}}
			- \varepsilon_{i} - \varepsilon_{j}
			+ 2 G_{\ger{\alpha} \ung{\alpha}} \\
		&\quad + \gone{\alpha}{i}  + \gone{\alpha}{j}
			- \frac{1}{2} G_{i \ger{\alpha}} - \frac{1}{2} G_{j \ger{\alpha}} 
			- \frac{1}{2} G_{i \ung{\alpha}} - \frac{1}{2} G_{j \ung{\alpha}}
	\end{split}
	\\
	\begin{split}
		E[2^{aa}_{\alpha_{\mu} \beta_{\nu}}] - E[\PP] &= \eg{\alpha} L_{\ger{\alpha} \ung{\alpha}} + \eg{\beta} L_{\ger{\beta} \ung{\beta}}
			+ \twamp{\alpha_{\mu} \beta_{\nu}} + 2 \varepsilon_a
			- \varepsilon_{\alpha_{1 - \mu}} - \varepsilon_{ \beta_{1 - \nu}} \\
		&\quad + \gtwo{\alpha}{\beta} 
			- 2 \gone{\alpha}{a} - 2 \gone{\beta}{a}
			+ \gone{\beta}{\alpha_{1 - \mu}}
			+ \gone{\alpha}{\beta_{1 - \nu}} \\
		&\quad + \frac{1}{2} G_{\alpha_{1 - \mu} \beta_{1 - \nu}}
			- G_{\alpha_{1 - \mu} a} - G_{\beta_{1 - \nu} a}
	\end{split}
	\\
	E[2^{aa}_{i j}] - E[\PP] &=	\twamp{i j}
		+ 2 \varepsilon_a - \varepsilon_{i} - \varepsilon_{j}
		- G_{i a} - G_{j a}
	\\
	\begin{split}
		E[2^{aa}_{i \alpha_{\mu}}] - E[\PP] &= \eg{\alpha} L_{\ger{\alpha} \ung{\alpha}}
			+ \twamp{i \alpha_{\mu}}
			+ 2\varepsilon_a - \varepsilon_i - \varepsilon_{\alpha_{1 - \mu}} \\
		&\quad - 2 \gone{\alpha}{a} + \gone{\alpha}{i} 
			+ \frac{1}{2} G_{i \alpha_{1 - \mu}}
			- G_{ia} - G_{\alpha_{1 - \mu} a}
	\end{split}
	\\
	\begin{split}
		E[2^{\beta \beta}_{i \alpha_{\mu}}] - E[\PP] &= 
			  \eg{\alpha} L_{\ger{\alpha} \ung{\alpha}} + \eg{\beta} L_{\ger{\beta} \ung{\beta}}
			+ \twamp{i \alpha_{\mu}}
			+ \varepsilon_{\ger{\beta}} + \varepsilon_{\ung{\beta}}
			- \varepsilon_{\alpha_{1 - \mu}} - \varepsilon_i
			+ 2 G_{\ger{\beta} \ung{\beta}} \\
		&\quad + \gtwo{\alpha}{\beta}
			- \gone{\alpha}{\ger{\beta}} - \gone{\alpha}{\ung{\beta}}
			+ \gone{\beta}{\alpha_{1 - \mu}} + \gone{\alpha}{i} + \gone{\beta}{i} \\
		&\quad + \frac{1}{2} G_{i \alpha_{1 - \mu}}
			- \frac{1}{2} G_{\alpha_{1 - \mu} \ger{\beta}} - \frac{1}{2} G_{\alpha_{1 - \mu} \ung{\beta}}
			- \frac{1}{2} G_{i \ger{\beta}} - \frac{1}{2} G_{i \ung{\beta}},
	\end{split} 
\end{align}
while seniority-zero pairs can be scattered to seniority-two pairs with energies
\begin{align}
	\begin{split}
		E[2^{\alpha_{\mu} \beta_{\nu}}_{ii} ] - E[\PP] &= \eg{\alpha} L_{\ger{\alpha} \ung{\alpha}} + \eg{\beta} L_{\ger{\beta} \ung{\beta}}
			+ \twamp{\alpha_{\mu} \beta_{\nu}} 
			+ \varepsilon_{\alpha_{1 - \mu}} + \varepsilon_{ \beta_{1 - \nu}} - 2 \varepsilon_i
			+ G_{\ger{\alpha} \ung{\alpha}} + G_{\ger{\beta} \ung{\beta}} \\
		&\quad + \gtwo{\alpha}{\beta} - \gone{\beta}{\alpha_{1 - \mu}} - \gone{\alpha}{\beta_{1 - \nu}}
			+ 2 \gone{\alpha}{i} + 2 \gone{\beta}{i} \\
		&\quad + \frac{1}{2} G_{\alpha_{1 - \mu} \beta_{1 - \nu}}
			- G_{i \alpha_{1 - \mu}} - G_{i \beta_{1 - \nu}}
	\end{split}
	\\
	\begin{split}
		E[2^{a b}_{\alpha \alpha}] - E[\PP] &= \eg{\alpha} L_{\ger{\alpha} \ung{\alpha}}
			+ \twamp{a b} + \varepsilon_{a} + \varepsilon_{b}
			- \varepsilon_{\ger{\alpha}} - \varepsilon_{\ung{\alpha}} \\
		&\quad  - \gone{\alpha}{a}  - \gone{\alpha}{b}
			- \frac{1}{2} G_{\ger{\alpha} a} - \frac{1}{2} G_{\ger{\alpha} b}
			- \frac{1}{2} G_{\ung{\alpha} a} - \frac{1}{2} G_{\ung{\alpha} b}
	\end{split}
	\\
	E[2^{a b}_{ii}] - E[\PP] &=	\twamp{a b}
		+ \varepsilon_{a} + \varepsilon_{b}
		- 2 \varepsilon_i
		- G_{i a} - G_{i b}
	\\
	\begin{split}
		E[2^{\alpha_{\mu} a}_{ii} ] - E[\PP] &= \eg{\alpha} L_{\ger{\alpha} \ung{\alpha}}
			+ \twamp{\alpha_{\mu} a}
			+ \varepsilon_{\alpha_{1 - \mu}} + \varepsilon_{a}
			- 2 \varepsilon_i
			+ G_{\ger{\alpha} \ung{\alpha }} \\
		&\quad - \gone{\alpha}{a} + 2 \gone{\alpha}{i}
			+ \frac{1}{2} G_{\alpha_{1 - \mu} a}
			- G_{ia} - G_{i \alpha_{1 - \mu}}
	\end{split}
	\\
	\begin{split}
		E[2^{\alpha_{\mu} a}_{\beta \beta}] - E[\PP] &=
			  \eg{\alpha} L_{\ger{\alpha} \ung{\alpha}} + \eg{\beta} L_{\ger{\beta} \ung{\beta}}
			+ \twamp{\alpha_{\mu} a}
			+ \varepsilon_{\alpha_{1 - \mu}} + \varepsilon_a
			- \varepsilon_{\ger{\beta}} - \varepsilon_{\ung{\beta}}
			+ G_{\ger{\alpha} \ung{\alpha}} \\
		&\quad + \gtwo{\alpha}{\beta}
			- \gone{\beta}{\alpha_{1 - \mu}} - \gone{\alpha}{a} - \gone{\beta}{a}
			+ \gone{\alpha}{\ger{\beta}} + \gone{\alpha}{\ung{\beta}} \\
		&\quad + \frac{1}{2} G_{\alpha_{1 - \mu} a}
			- \frac{1}{2} G_{\alpha_{1 - \mu} \ger{\beta}} - \frac{1}{2} G_{\alpha_{1 - \mu} \ung{\beta}}
			- \frac{1}{2} G_{a \ger{\beta}} - \frac{1}{2} G_{a \ung{\beta}}.
	\end{split}
\end{align}

Finally, seniority-two pairs scatter to seniority-two pairs
\begin{align}
	\begin{split}
		E[\fone{\alpha_{\mu} \beta_{\nu}}{i j}] - E[\PP] &=
			  \eg{\alpha} L_{\ger{\alpha} \ung{\alpha}} + \eg{\beta} L_{\ger{\beta} \ung{\beta}}
			+ \twamp{i j} + \twamp{\alpha_{\mu} \beta_{\nu}} \\
		&\quad
			+ \varepsilon_{\alpha_{1 - \mu}} + \varepsilon_{ \beta_{1 - \nu}}
			- \varepsilon_{i} - \varepsilon_{j}
			+ G_{\ger{\alpha} \ung{\alpha}} + G_{\ger{\beta} \ung{\beta}} \\
		&\quad + \gtwo{\alpha}{\beta} 
			- \gone{\beta}{\alpha_{1 - \mu}} - \gone{\alpha}{\beta_{1 - \nu}}
			+ \gone{\alpha}{i} + \gone{\beta}{i} + \gone{\alpha}{j} + \gone{\beta}{j} \\
		&\quad + \frac{1}{2} G_{\alpha_{1 - \mu} \beta_{1 - \nu}}
			- \frac{1}{2} G_{i \alpha_{1 - \mu}} - \frac{1}{2} G_{i \beta_{1 - \nu}}
			- \frac{1}{2} G_{j \alpha_{1 - \mu}} - \frac{1}{2} G_{j \beta_{1 - \nu}}
	\end{split}
	\\
	\begin{split}
		E[\fone{a b}{\alpha_{\mu} \beta_{\nu} }] - E[\PP] &=
			  \eg{\alpha} L_{\ger{\alpha} \ung{\alpha}} + \eg{\beta} L_{\ger{\beta} \ung{\beta}}
			+ \twamp{\alpha_{\mu} \beta_{\nu}} + \twamp{a b} \\
		&\quad
			+ \varepsilon_{a} + \varepsilon_{b}
			- \varepsilon_{\alpha_{1 - \mu}} - \varepsilon_{ \beta_{1 - \nu}} \\
		&\quad + \gtwo{\alpha}{\beta}
			- \gone{\alpha}{a} - \gone{\beta}{a}
			- \gone{\alpha}{b} - \gone{\beta}{b}
			+ \gone{\beta}{\alpha_{1 - \mu}} + \gone{\alpha}{\beta_{1 - \nu}} \\
		&\quad + \frac{1}{2} G_{\beta_{1 - \nu} \alpha_{1 - \mu}}
			- \frac{1}{2} G_{\alpha_{1 - \mu} a} - \frac{1}{2} G_{\alpha_{1 - \mu} b}
			- \frac{1}{2} G_{\beta_{1 - \nu}  a} - \frac{1}{2} G_{\beta_{1 - \nu}  b}
	\end{split}
\end{align}
\begin{align}
	\begin{split}
		E[4^{a b}_{i j}] - E[\PP] &= \twamp{i j} + \twamp{a b}
			+ \varepsilon_{a} + \varepsilon_{b}
			- \varepsilon_{i} - \varepsilon_{j} 
		 - \frac{1}{2} G_{i a} - \frac{1}{2} G_{i b}
			   - \frac{1}{2} G_{j a} - \frac{1}{2} G_{j b}
	\end{split}
	\\
	\begin{split}
		E[\fone{\alpha_{\mu} a}{i j}] - E[\PP] &=
			  \eg{\alpha} L_{\ger{\alpha} \ung{\alpha}}
			+ \twamp{i j} + \twamp{\alpha_{\mu} a}
			+ \varepsilon_{\alpha_{1 - \mu}} + \varepsilon_a
			- \varepsilon_{i} - \varepsilon_{j}
			+ G_{\ger{\alpha} \ung{\alpha}} \\
		&\quad - \gone{\alpha}{a} + \gone{\alpha}{i}  + \gone{\alpha}{j} \\
		&\quad + \frac{1}{2} G_{\alpha_{1 - \mu} a}
			- \frac{1}{2} G_{i \alpha_{1 - \mu}} - \frac{1}{2} G_{j \alpha_{1 - \mu}} 
			- \frac{1}{2} G_{i a} - \frac{1}{2} G_{j a}
	\end{split}
	\\
	\begin{split}
		E[\fone{a b}{i \alpha_{\mu}}] - E[\PP] &=
			  \eg{\alpha} L_{\ger{\alpha} \ung{\alpha}}
			+ \twamp{i \alpha_{\mu}} + \twamp{a b}
			+ \varepsilon_{a} + \varepsilon_{b}
			- \varepsilon_{\alpha_{1 - \mu}} - \varepsilon_i
		 - \gone{\alpha}{a} - \gone{\alpha}{b} + \gone{\alpha}{i} \\
		&\quad + \frac{1}{2} G_{i \alpha_{1 - \mu}}
			- \frac{1}{2} G_{\alpha_{1 - \mu} a} 
			- \frac{1}{2} G_{\alpha_{1 - \mu} b}
			- \frac{1}{2} G_{i a}
			- \frac{1}{2} G_{i b}
	\end{split}
	\\
	\begin{split}
		E[\fone{a\beta_{\nu}}{i \alpha_{\mu}}] - E[\PP] &=
			  \eg{\alpha} L_{\ger{\alpha} \ung{\alpha}} + \eg{\beta} L_{\ger{\beta} \ung{\beta}} 
			+ \twamp{c \alpha_{\mu}} + \twamp{v \beta_{\nu}} 
			+ \varepsilon_{\beta_{1 - \nu}} + \varepsilon_a
			- \varepsilon_{\alpha_{1 - \mu}} - \varepsilon_i
			+ G_{\ger{\beta} \ung{\beta}} \\
		&\quad + \gtwo{\alpha}{\beta}
			- \gone{\alpha}{\beta_{1 - \nu}} 
			- \gone{\alpha}{a} - \gone{\beta}{a}
			+ \gone{\beta}{\alpha_{1 - \mu}}
			+ \gone{\alpha}{i} + \gone{\beta}{i} \\
		&\quad + \frac{1}{2} G_{i \alpha_{1 - \mu}} + \frac{1}{2} G_{a \beta_{1 - \nu}} \\
		&\quad
			- \frac{1}{2} G_{ia} - \frac{1}{2} G_{\alpha_{1 - \mu} a}
			- \frac{1}{2} G_{i \beta_{1 - \nu}} - \frac{1}{2} G_{\alpha_{1 - \mu} \beta_{1 - \nu}},
	\end{split}
\end{align}
and
\begin{align}
	\begin{split}
		E[\fone{\gamma_{\lambda} a}{\alpha_{\mu} \beta_{\nu}}] - E[\PP] &=
			  \eg{\alpha} L_{\ger{\alpha} \ung{\alpha}}	
			+ \eg{\beta}  L_{\ger{ \beta} \ung{ \beta}}
			+ \eg{\gamma} L_{\ger{\gamma} \ung{\gamma}}
			+ \twamp{\alpha_{\mu} \beta_{\nu}}
			+ \twamp{\gamma_{\lambda} a} \\
		&\quad
			+ \varepsilon_{\gamma_{1 - \lambda}} + \varepsilon_a
			- \varepsilon_{\alpha_{1 - \mu}} - \varepsilon_{\beta_{1 - \nu}}
			+ G_{\ger{\gamma} \ung{\gamma}} \\
		&\quad + \gtwo{\alpha}{\beta} + \gtwo{\alpha}{\gamma} + \gtwo{\beta}{\gamma}
			- \gone{\alpha}{a} - \gone{\beta}{a} - \gone{\gamma}{a} \\
		&\quad - \gone{\alpha}{\gamma_{1 - \lambda}} - \gone{\beta }{\gamma_{1 - \lambda}}
			+ \gone{\beta}{\alpha_{1 - \mu}} + \gone{\gamma}{\alpha_{1 - \mu}}
			+ \gone{\alpha}{\beta_{1 - \nu}} + \gone{\gamma}{\beta_{1 - \nu}} \\
		&\quad + \frac{1}{2} G_{\alpha_{1 - \mu} \beta_{1 - \nu}}
			+ \frac{1}{2} G_{\gamma_{1 - \lambda} a} \\
		&\quad
			- \frac{1}{2} G_{\alpha_{1 - \mu} \gamma_{1 - \lambda}}
			- \frac{1}{2} G_{\beta_{1 - \nu} \gamma_{1 - \lambda}}
			- \frac{1}{2} G_{\alpha_{1 - \mu} a}
			- \frac{1}{2} G_{\beta_{1 - \nu}  a}
	\end{split}
	\\
	\begin{split}
		E[\fone{\beta_{\nu} \gamma_{\lambda} }{i \alpha_{\mu}}] - E[\PP] &=
			  \eg{\alpha} L_{\ger{\alpha} \ung{\alpha} } 
			+ \eg{ \beta} L_{\ger{\beta}  \ung{\beta}}
			+ \eg{\gamma} L_{\ger{\gamma} \ung{\gamma}} 
			+ \twamp{i \alpha_{\mu}} + \twamp{\beta_{\nu}\gamma_{\lambda}} \\
		&\quad
			+ \varepsilon_{\beta_{1 - \nu}} + \varepsilon_{\gamma_{1 - \lambda}}
			- \varepsilon_{\alpha_{1 - \mu}} - \varepsilon_i 
			+ G_{\ger{\beta} \ung{\beta}} + G_{\ger{\gamma} \ung{\gamma}} \\
		&\quad + \gtwo{\alpha}{\beta} + \gtwo{\alpha}{\gamma} + \gtwo{\beta}{\gamma}
			+ \gone{\alpha}{i} + \gone{\beta}{i} + \gone{\gamma}{i} \\
		&\quad - \gone{\alpha}{\beta_{1 - \nu}} - \gone{\gamma}{\beta_{1 - \nu}}
			- \gone{\alpha}{\gamma_{1 - \lambda}}
			- \gone{\beta }{\gamma_{1 - \lambda}}
			+ \gone{\beta }{\alpha_{1 - \mu}}
			+ \gone{\gamma}{\alpha_{1 - \mu}} \\
		&\quad + \frac{1}{2} G_{i \alpha_{1 - \mu}} + \frac{1}{2} G_{\beta_{1 - \nu} \gamma_{1 - \lambda}} \\
		&\quad
			- \frac{1}{2} G_{\alpha_{1 - \mu} \beta_{1 - \nu}} 
			- \frac{1}{2} G_{\alpha_{1 - \mu} \gamma_{1 - \lambda}}
			- \frac{1}{2} G_{i \beta_{1 - \nu}}
			- \frac{1}{2} G_{i \gamma_{1 - \lambda}}.
	\end{split}
\end{align}
Again, an update involving exchange integrals of the blocked levels gives the excitation energies of the complementary seniority-four states.

\bibliography{non_zero_sen_3}

\end{document}